\documentclass[a4paper]{article}
\usepackage{graphicx,amsmath,amssymb}

\def\d{\mathrm{d}}

\let\oldcaption=\caption
\def\caption#1{\oldcaption{\small #1}}

\def\figs#1#2{%
\resizebox{\linewidth}{!}{%
\includegraphics[height=0.5\linewidth]{#1}%
\includegraphics[height=0.5\linewidth]{#2}%
}}

\begin{document}

\title{Centrality control of hadron-nucleus interactions by detection of
slow nucleons}

\author{Ferenc Sikl\'er \\
  {\small\sl KFKI Research Institute for Particle
       and Nuclear Physics, Budapest, Hungary} \\
  {\small\it sikler@rmki.kfki.hu}}
\date{March 31, 2003}

\maketitle

\begin{abstract}

Slow nucleons emitted during a hadron-nucleus interaction can give
information on the centrality, impact parameter of the collision. The aim
of this note is to provide the reader with the important characteristics
of the slow nucleons, focusing on their spectra, correlations. This study
tries to build on evidences, hence more weight is put on experimental
results than on models.

\end{abstract}

%%%%%%%%%%%%%%%%
% Introduction
%%%%%%%%%%%%%%%%

\section{Introduction}

High energy hadron-nucleus collisions have a long history and an enormous
literature: a very complete review contains more than thousand references
\cite{Fredriksson:1987nb}. Though studied by many experiments, the main
questions are still unsolved, making the subject a still open field.

The terminology of slow particles comes from pioneering emulsion work. The
emitted slow particles have been classified according to their grain
density left in the detection material: "black" or "gray". They are called
in a word "heavy". The lighter-colored particles are concentrated forward
and called "shower". These names can be converted to corresponding ranges
in $\beta$, momentum or energy (Table~\ref{tab:class}).

\begin{table}[h]
\begin{center}
 \begin{tabular}{cccc}
 \hline\hline
 name   & $\beta$ & $p$ [MeV/$c$] & $E_{kin}$ [MeV] \\
 \hline
 black \\
 \rule[0.5ex]{10ex}{0.4pt} & 0.25 & 250 & 30 \\
 gray \\
 \rule[0.5ex]{10ex}{0.4pt} & 0.7 & 1000 & 400 \\
 shower \\
 \hline\hline
 \end{tabular}

 \caption{\label{tab:class} Classification of particles produced in
hadron-nucleus collisions with the borders of the ranges. Momentum and
energy ranges are given for nucleons.}

\end{center}
\end{table}

%%%%%%%%%%%%%%%%%%%%%%%%
% Experimental history
%%%%%%%%%%%%%%%%%%%%%%%%

\section{Experimental history}

\begin{figure}
 \figs{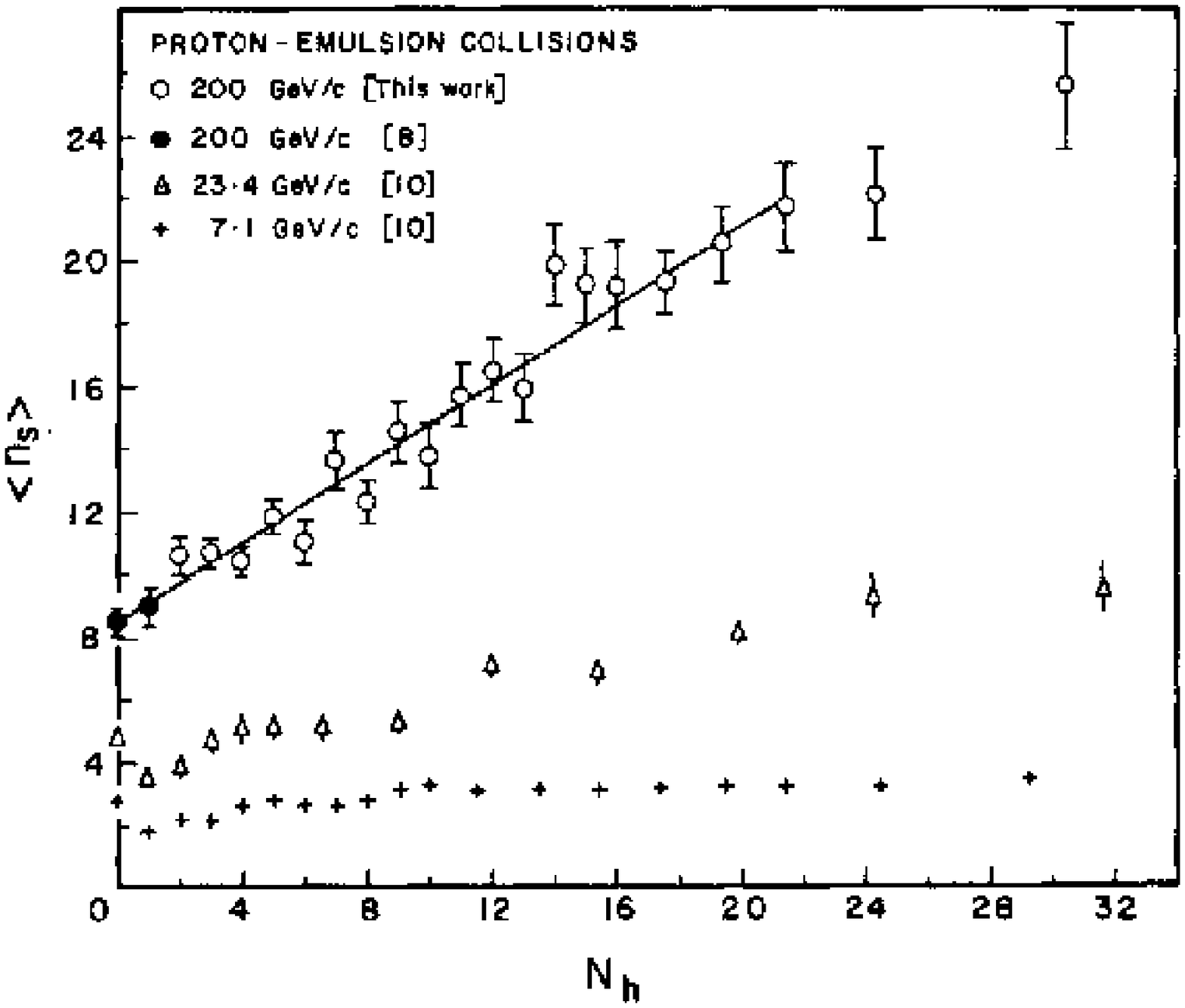}
      {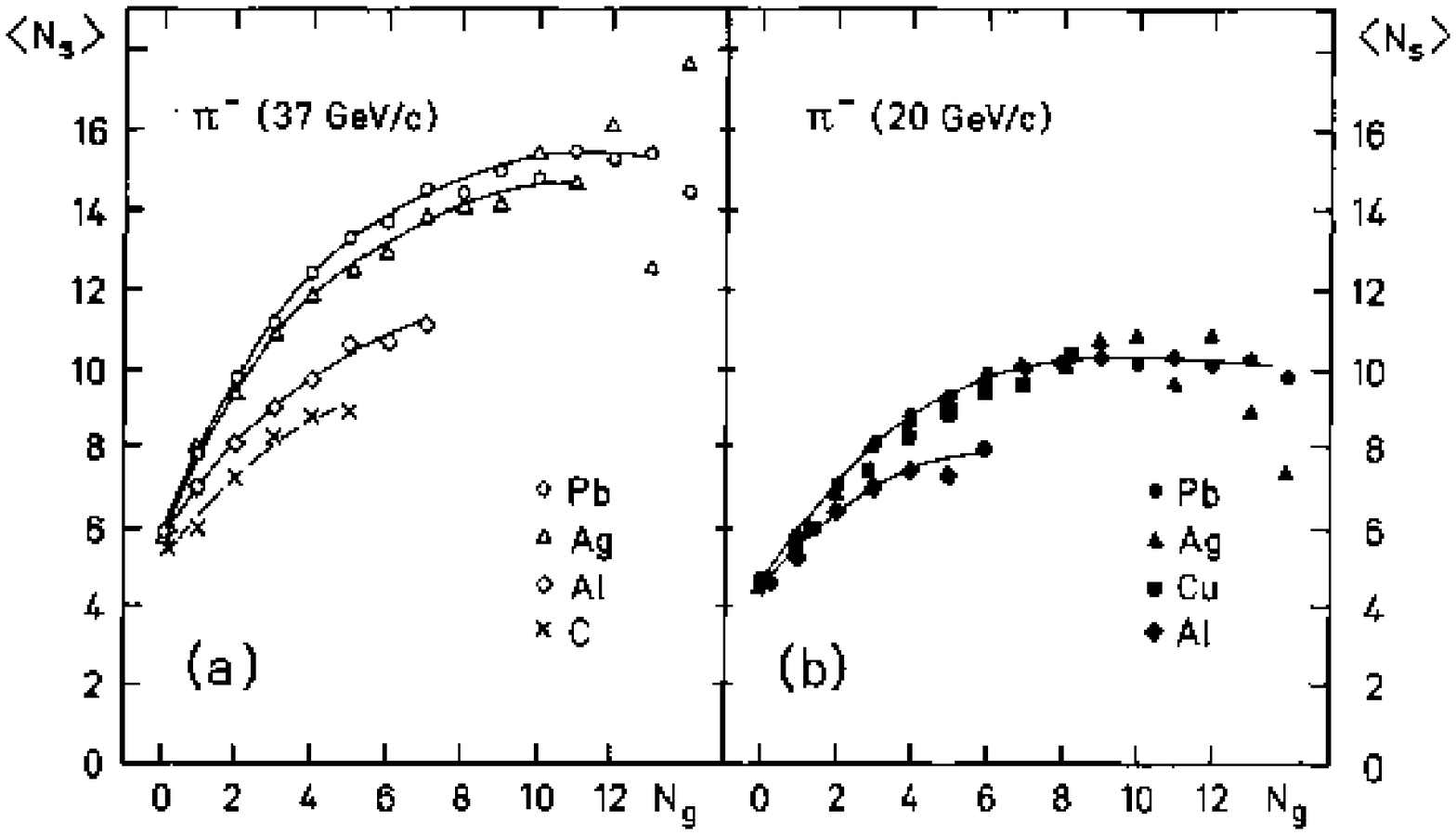}

 \caption{\label{fig:gurtu} a) Dependence of $\langle n_S \rangle$ on the
number $N_h$ of heavy particles at 200, 23.4 and 7.1 GeV/$c$, from
Ref.~\cite{Gurtu:1974ei}. Line shown is the best linear fit. b) Dependence
of the mean number $\langle N_s \rangle$ of fast particles on the number
$N_g$ of slow particles, from Ref.~\cite{Faessler:1979sn}.}

\end{figure}

The general observation is that the multiplicities of produced shower
particles increase with increasing the number of slow particles: for heavy
prongs a closely linear dependence (representative example in
Fig.~\ref{fig:gurtu}a), for grey ones a more curved relation is observed
(representative example in Fig.~\ref{fig:gurtu}b).

The following review is not meant to be exhaustive, but shows the most
important steps and developments. The experiments are grouped around
measuring techniques, in the end a summary of recent results is given.

Early cosmic ray experiments already indicated that collisions with nuclei
produce all kinds of particles: pions, kaons, protons and antiprotons.

%%%%%%%%%%%%%%%%%%%%%%%%%%%%%%%%%%
% Spectrometers

\subsection{Spectrometers}

\paragraph{Cocconi et al.} At the CERN-PS 25 GeV protons interacted with
Al and Pt targets \cite{Cocconi:1960}. Secondary particles at
$15.9^{\circ}$ were identified essentially by a mass spectrometer which
used magnetic field and the measurement of time of flight between two
scintillator counters. The copious production of deuterons and mass-three
nuclei was discovered.

\begin{figure}
 \figs{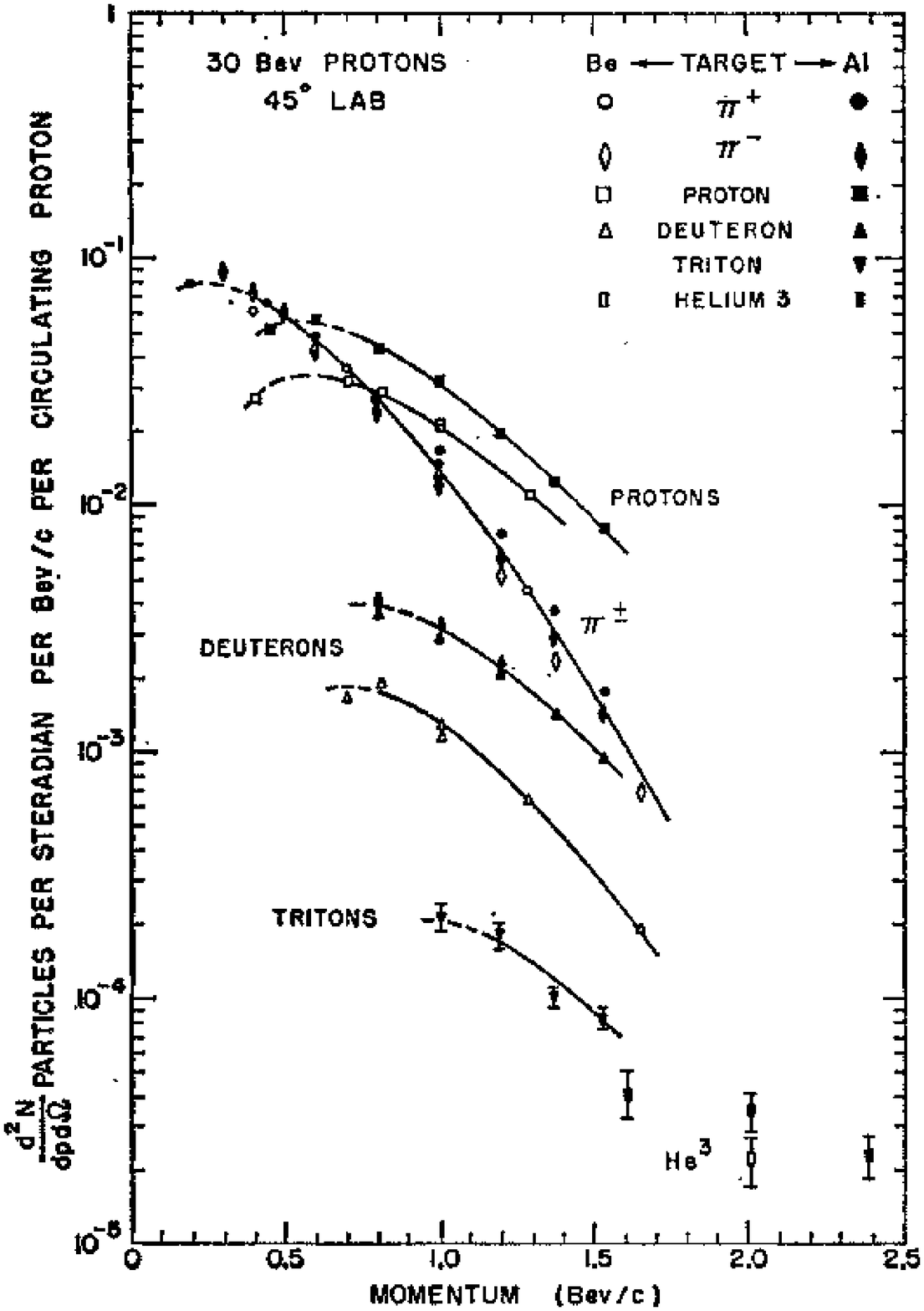}
      {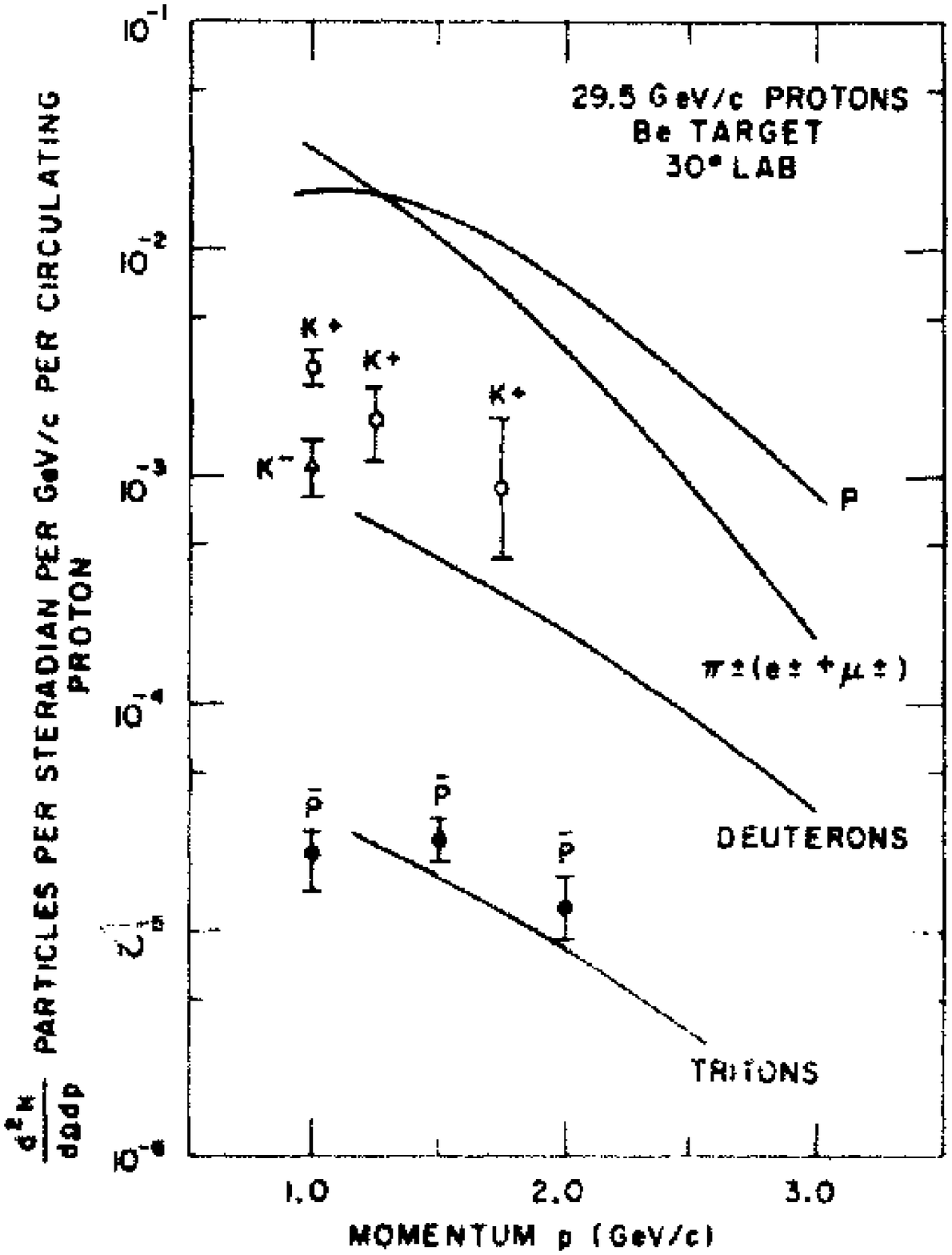} 

 \caption{\label{fig:fitch} a) Momentum spectra of particles emitted at
$45^{\circ}$ from Au and Be targets when struck by 30 GeV protons, from
Ref.~\cite{Fitch:1962}. b) Momentum distribution of particles from
Ref.~\cite{Schwarzschild:1963}.}

\end{figure}

\paragraph{Fitch et al.} Following this line, similar mass analysis was
done at Brookhaven AGS by mass analysis of particles emitted from Al and
Be targets when struck by 30 and 33 GeV protons \cite{Fitch:1962}. Here
several angle settings ($14\frac{1}{4}^{\circ}$, $45^{\circ}$ and $90^{\circ}$)
were used (Fig.~\ref{fig:fitch}a). They speculate that most of
the $\pi$ mesons originated in secondary or cascade processes. while
low-momentum deuterons could be products of "pick-up".

\paragraph{Schwarzschild et al.} Another group at Brookhaven AGS performed
beam survey at $30^{\circ}$ using 30 GeV protons on Al, Be and Fe targets
with same technique \cite{Schwarzschild:1963} (Fig.~\ref{fig:fitch}b).
They state that production of light nuclei involves cooperative phenomena
involving several nucleons of the target nucleus, namely by coalescing
shower nucleons.

\begin{figure}
 \figs{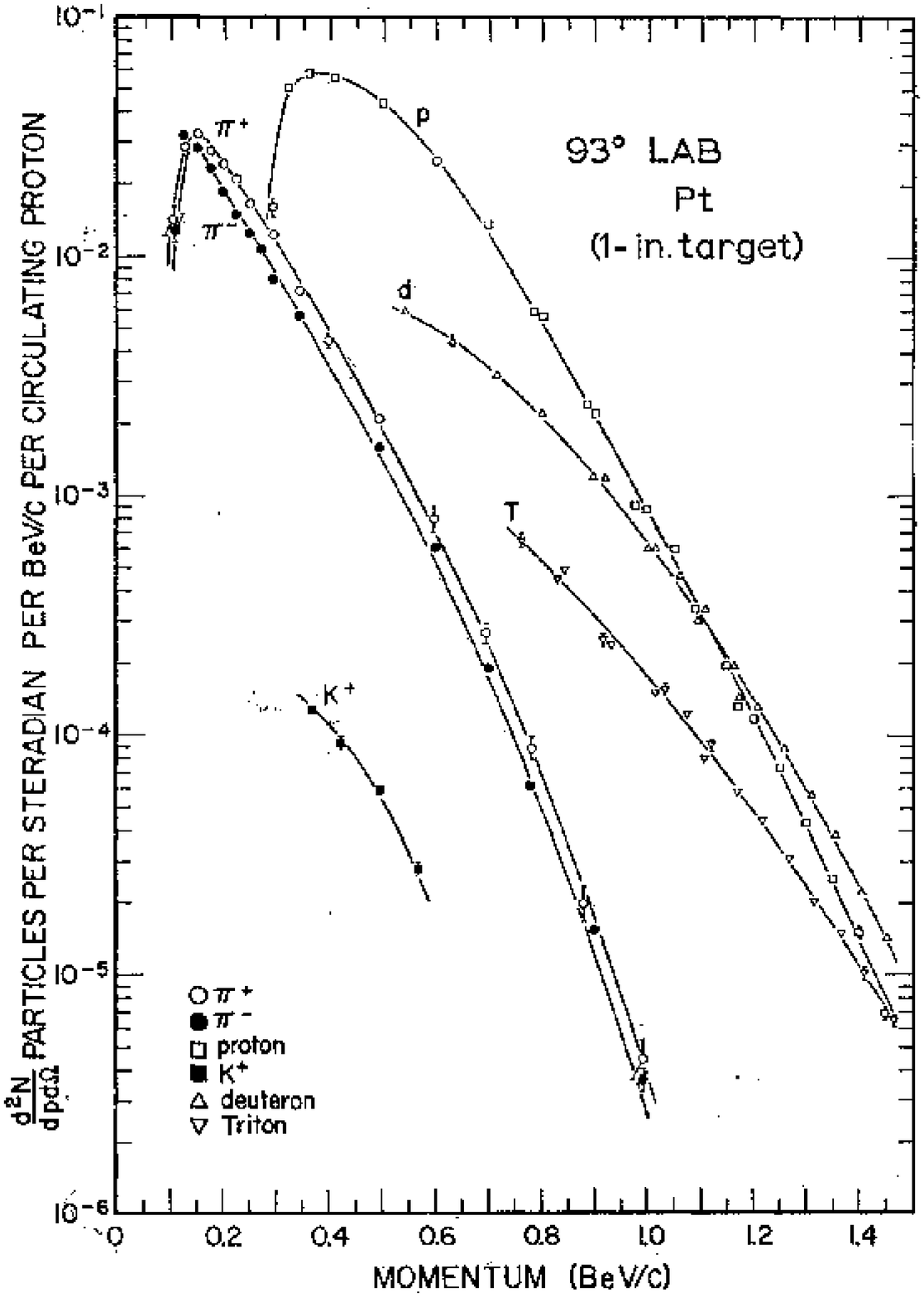}
      {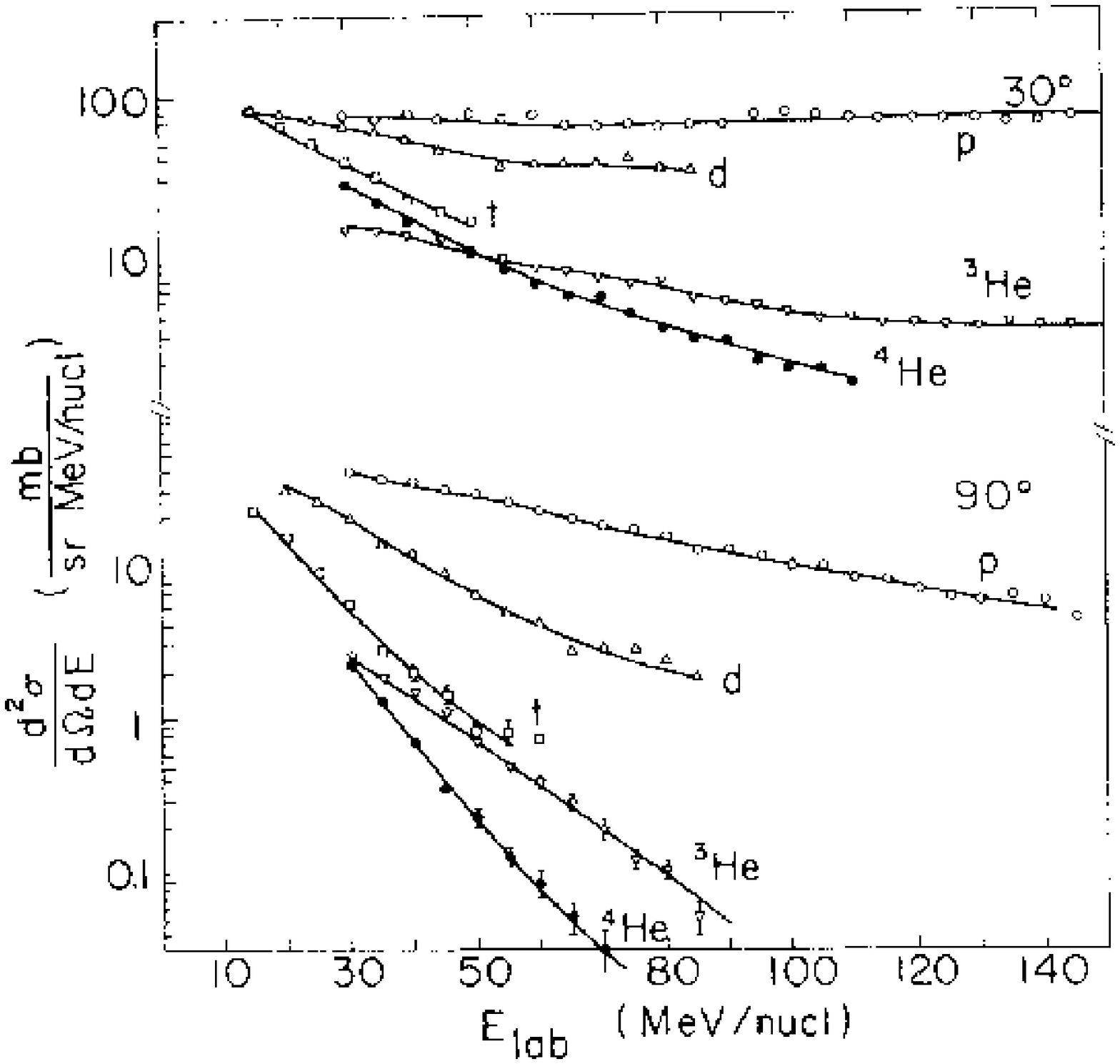}

 \caption{\label{fig:piroue} a) Momentum spectra of particles emitted at
$93^\circ$ from Pt target struck by 2.9 GeV protons, from
Ref.~\cite{Piroue:1966}. b) Double-differential cross-sections for
fragments from the irradiation of U by 400 MeV/nucleon $^{20}$Ne ions,
from Ref.~\cite{Gutbrod:1976gt}.}

\end{figure}

\paragraph{Pirou\'e et al.} Particle production by 2.9 GeV protons on Be
and Pt targets have been studied at the Princeton-Pennsylvania Accelerator
at various laboratory angles by mass analysis \cite{Piroue:1966}
(Fig.~\ref{fig:piroue}a). It is found that proton production results both
from nucleon-nucleon encounters and processes involving nuclear matter.
This latter appears to be necessary to account for observed deuteron,
triton, etc., yields.

\paragraph{Poskanzer et al.} The energy spectra of nuclear fragments
produced by the interaction of 5.5 GeV protons with U have been determined
at several laboratory angles at Berkeley Bevatron \cite{Poskanzer:1971rz}.
The measurement was done by means of $dE/dx$ -- $E$ measurements with
semiconductor-detector telescopes. By integration angular distributions
are obtained and fitted with curves based on the isotropic emission of
fragments from a system moving along the beam axis. Being mostly sensitive
to black protons the average velocity of the moving system is about
$\beta=0.006$.

\paragraph{Gutbrod et al.} Later particles emitted from U targets
irradiated with $^{20}$Ne ions at 400 MeV/nucleon energies were measured
with the same method \cite{Gutbrod:1976gt} (Fig.~\ref{fig:piroue}b).
Strong evidence for final-state interactions in the production of
high-energy fragments is found. It is suggested that observation of larger
composite particles might be a way of selecting central collisions.

\begin{figure}
 \figs{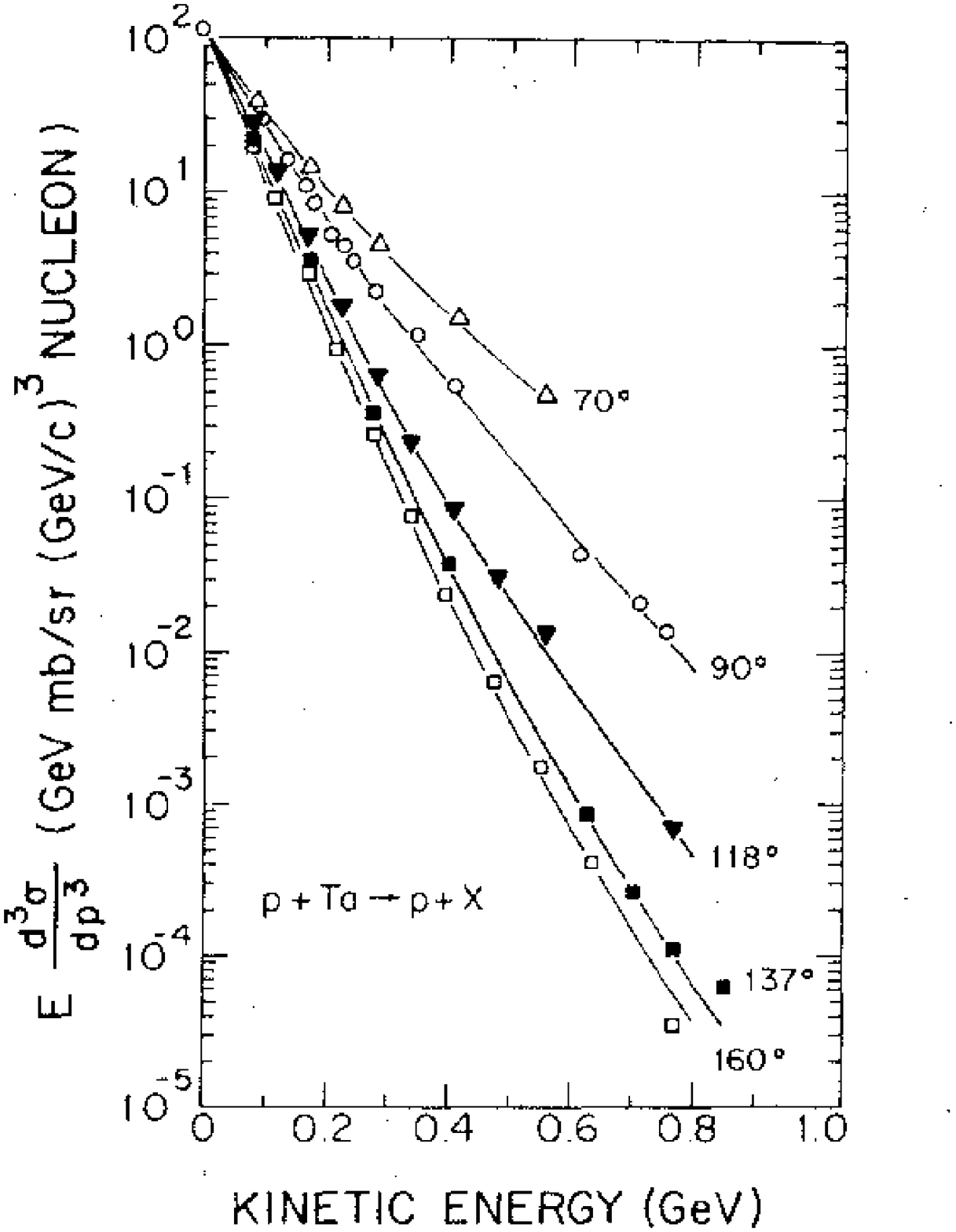}
      {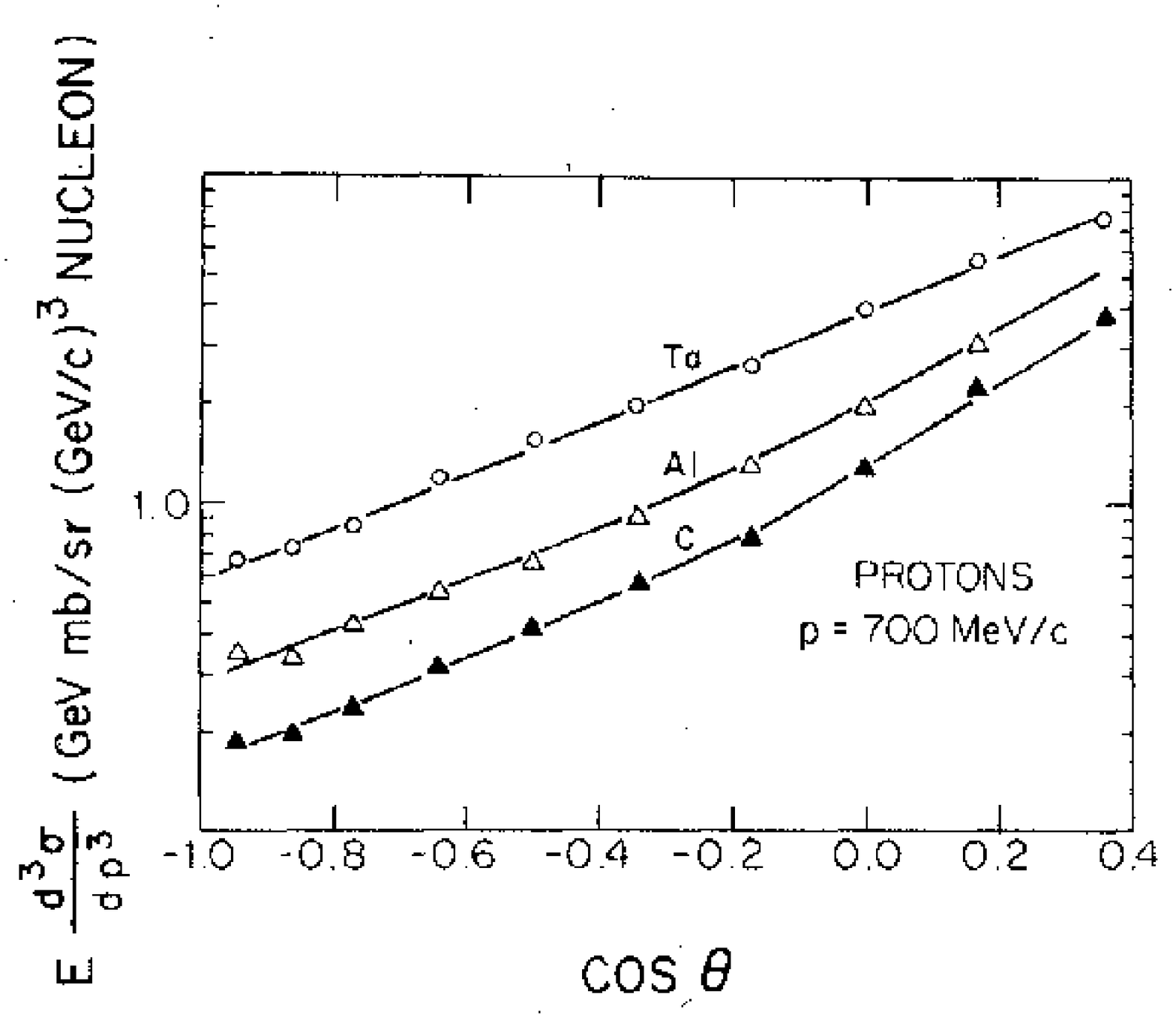}

 \caption{\label{fig:bayukov} a). Invariant cross-section per nucleon vs
kinetic energy at $70^\circ$, $90^\circ$, $118^\circ$, $137^\circ$, and
$160^\circ$ (laboratory) for Ta. b) Invariant cross-section per nucleon vs
angle at 0.7 GeV/$c$ for C, Al, and Ta. Both from
Ref.~\cite{Bayukov:1979vf}.}

\end{figure}

\begin{figure}
 \figs{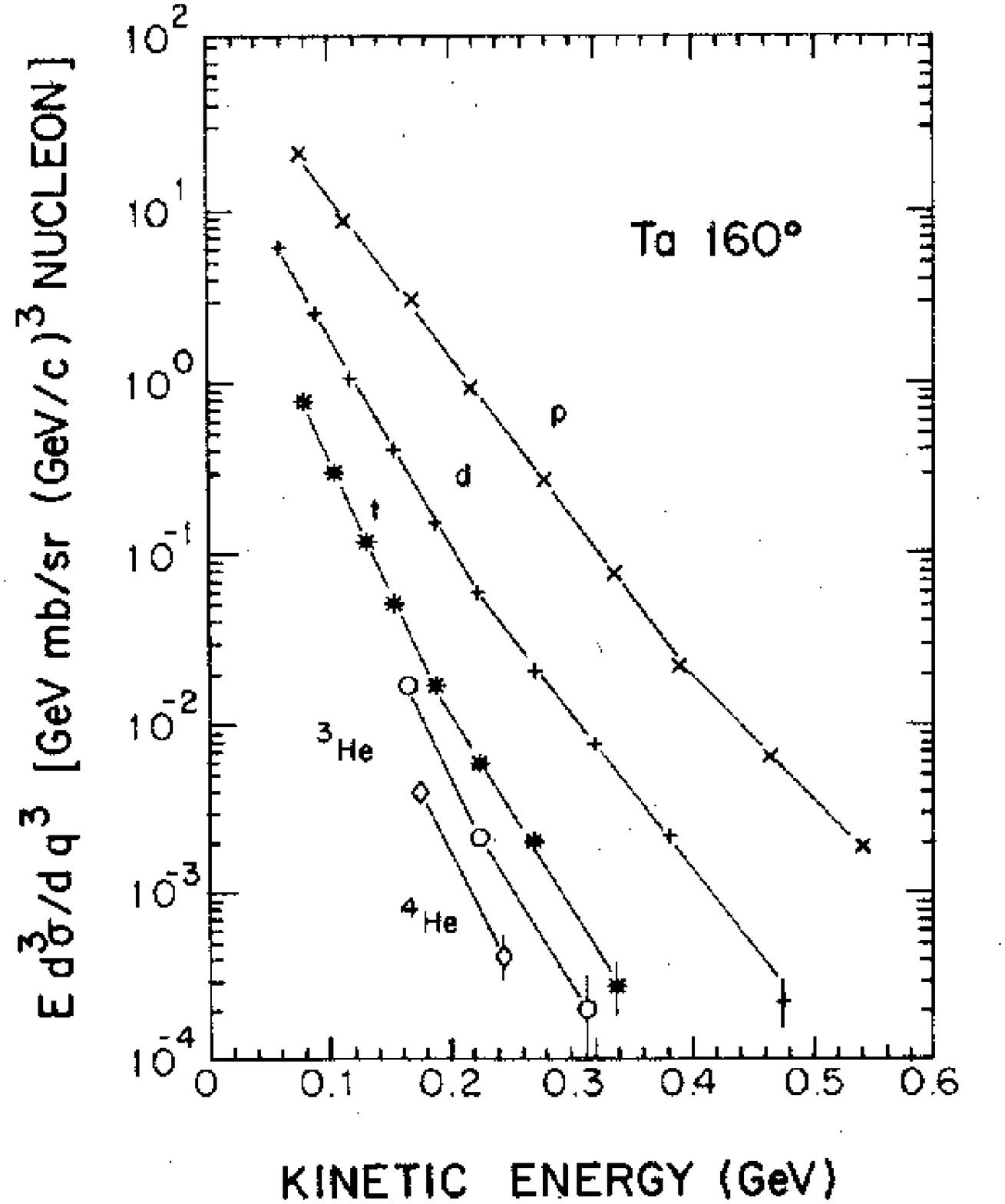}
      {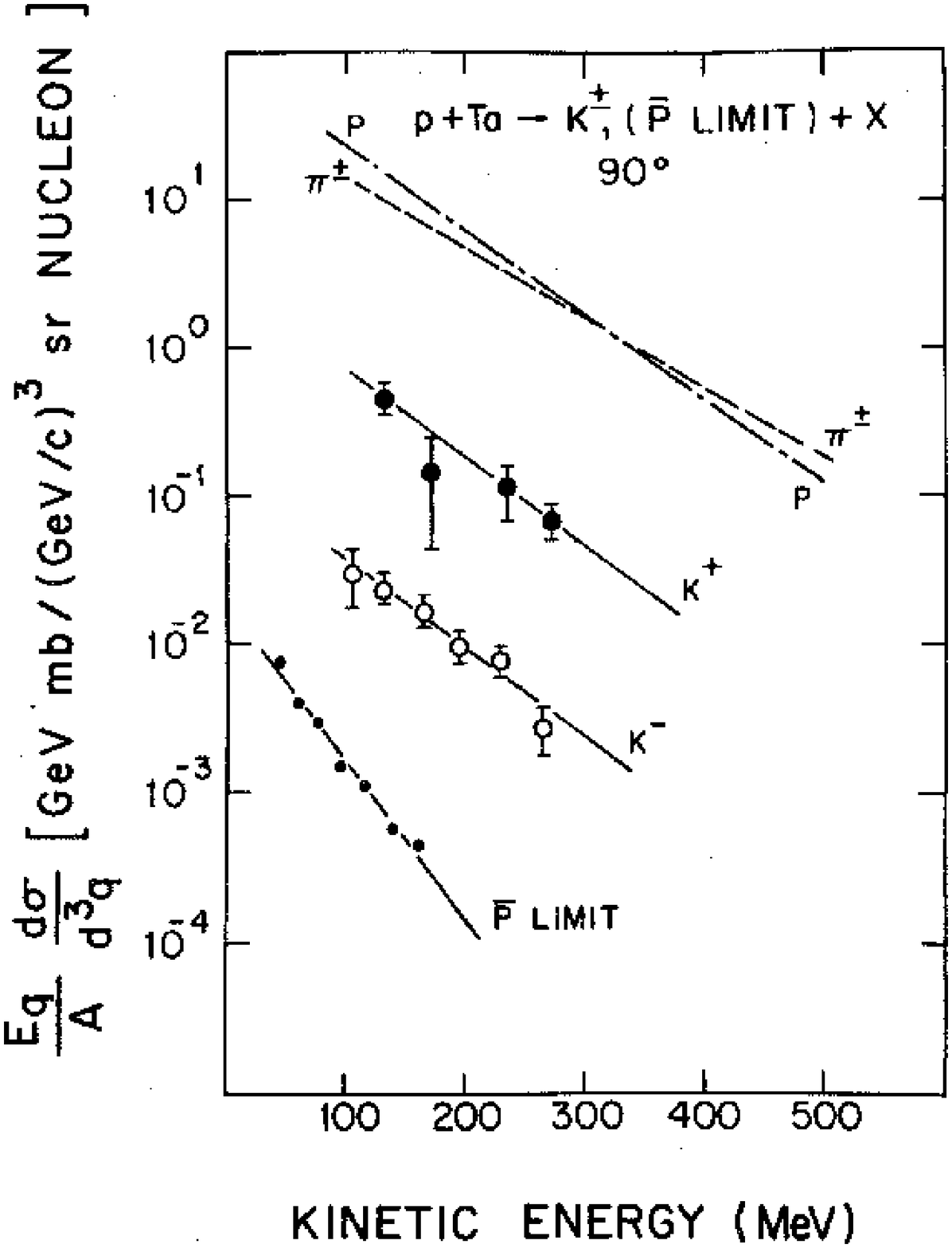}

 \caption{\label{fig:frankel} a) Invariant cross-sections per nucleon vs
kinetic energy for d, t, $^3$He, and $^4$He: Ta target at $160^\circ$
laboratory angle, from Ref.~\cite{Frankel:1979uq}. b) Invariant cross
sections per nucleon vs kinetic energy $T_q$ at $90^\circ$, from
Ref.~\cite{Nikiforov:1980gd}.}

\end{figure}

\paragraph{FNAL-E592.} Very complete, high precision measurements of
invariant cross-sections for the production of protons, deuterons,
tritons, $^3$He, $^4$He, pions and kaons by 400 GeV protons from a variety
of nuclear targets -- $^6$Li, Be, C, Al, Cu, Ta -- have been performed by
the FNAL-E592 collaboration, at angles $70^\circ$, $90^\circ$,
$118^\circ$, $137^\circ$ and $160^\circ$, in the momentum range of 0.1 to
1.4 GeV/$c$
\cite{Bayukov:1979nr,Bayukov:1979vf,Frankel:1979uq,Nikiforov:1980gd}. The
experiment was performed using independent measurements of the time of
flight in the telescope, as well as $dE/dx$ measurements in the
scintillators traversed by the particles. Although they stick to the mere
presentation of their data, they note that the empirical function
$\exp(-E_{kin}/E_0)$ is a useful parametrization at a fixed angle
(Figs.~\ref{fig:bayukov}a~and~\ref{fig:frankel}); there is a clear need
for exponential angular dependence of the cross-section on $\cos\theta$
(Fig.~\ref{fig:bayukov}b), which arises naturally in several single
scattering models via the scaling variable $E_{kin}-p\cos\theta$.

%%%%%%%%%%%%%%%%%%%%%%%%%%%%%%%%%%%%%%%%%%%%%%%%%%%%
% Emulsions

\subsection{Emulsions}

\begin{figure}
 \resizebox{2.25\linewidth}{!}{
 \includegraphics[height=0.5\linewidth]
                 {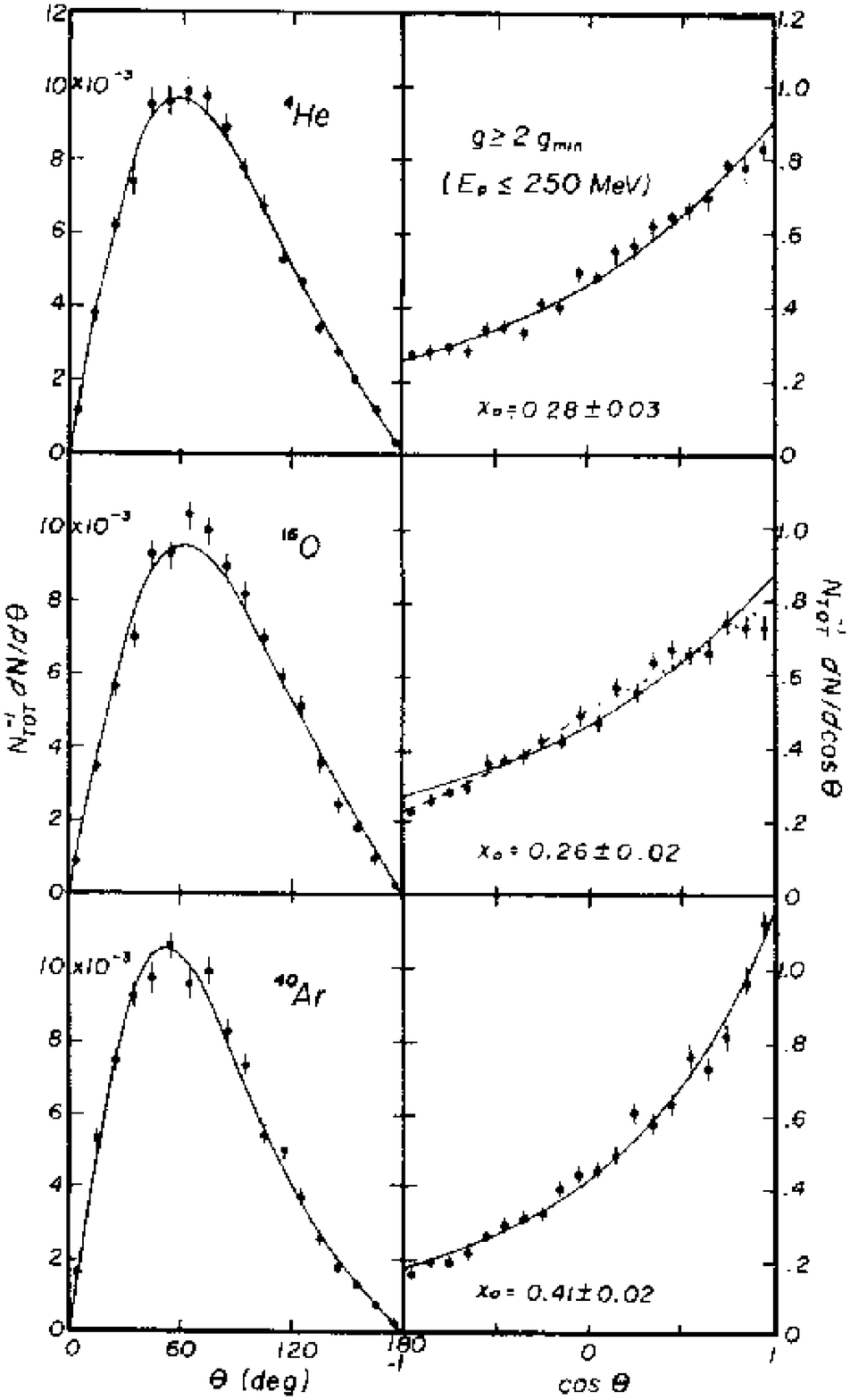}
 \vbox{\includegraphics[height=0.25\linewidth]
                       {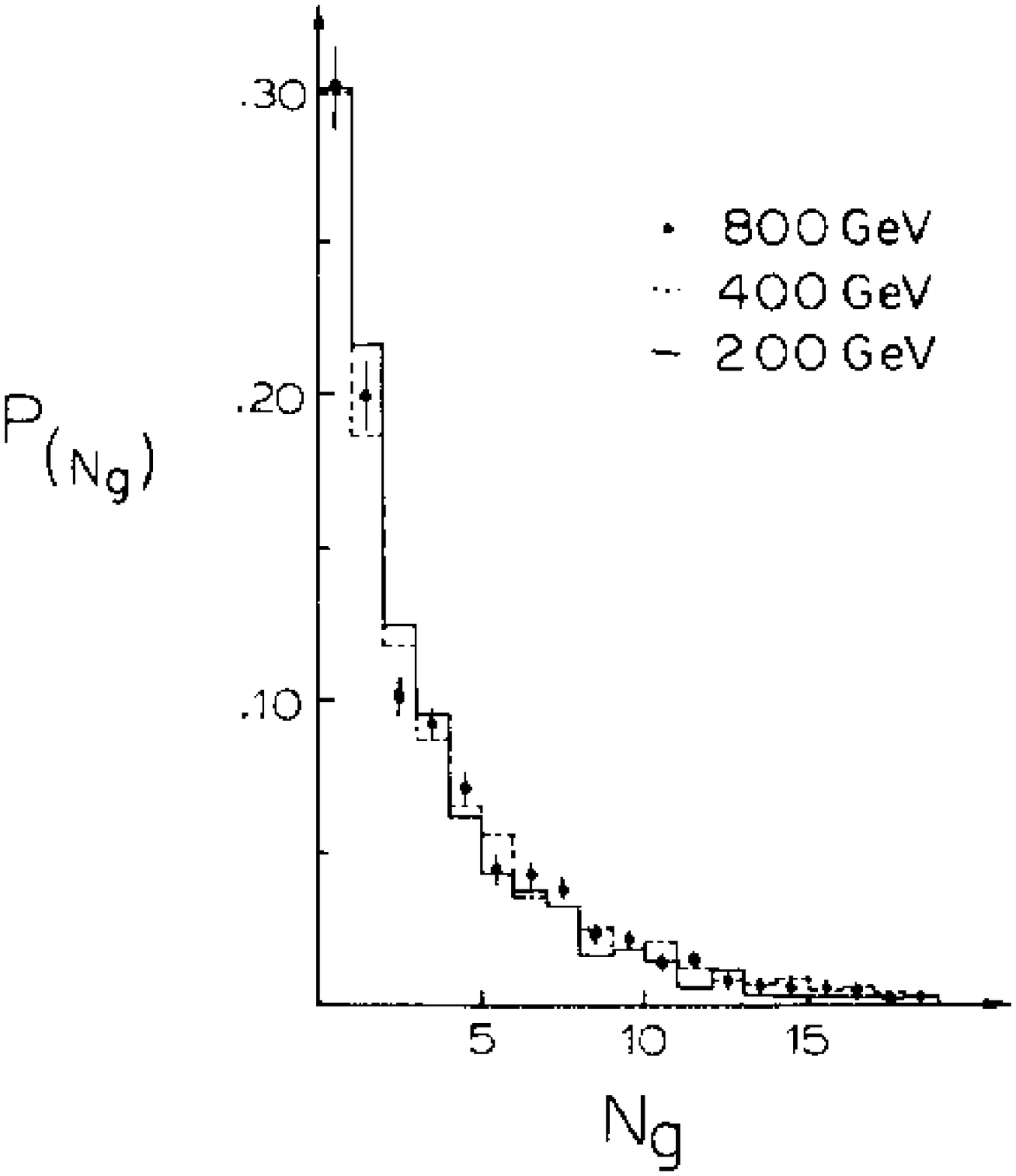}
   
    \includegraphics[height=0.25\linewidth]
                       {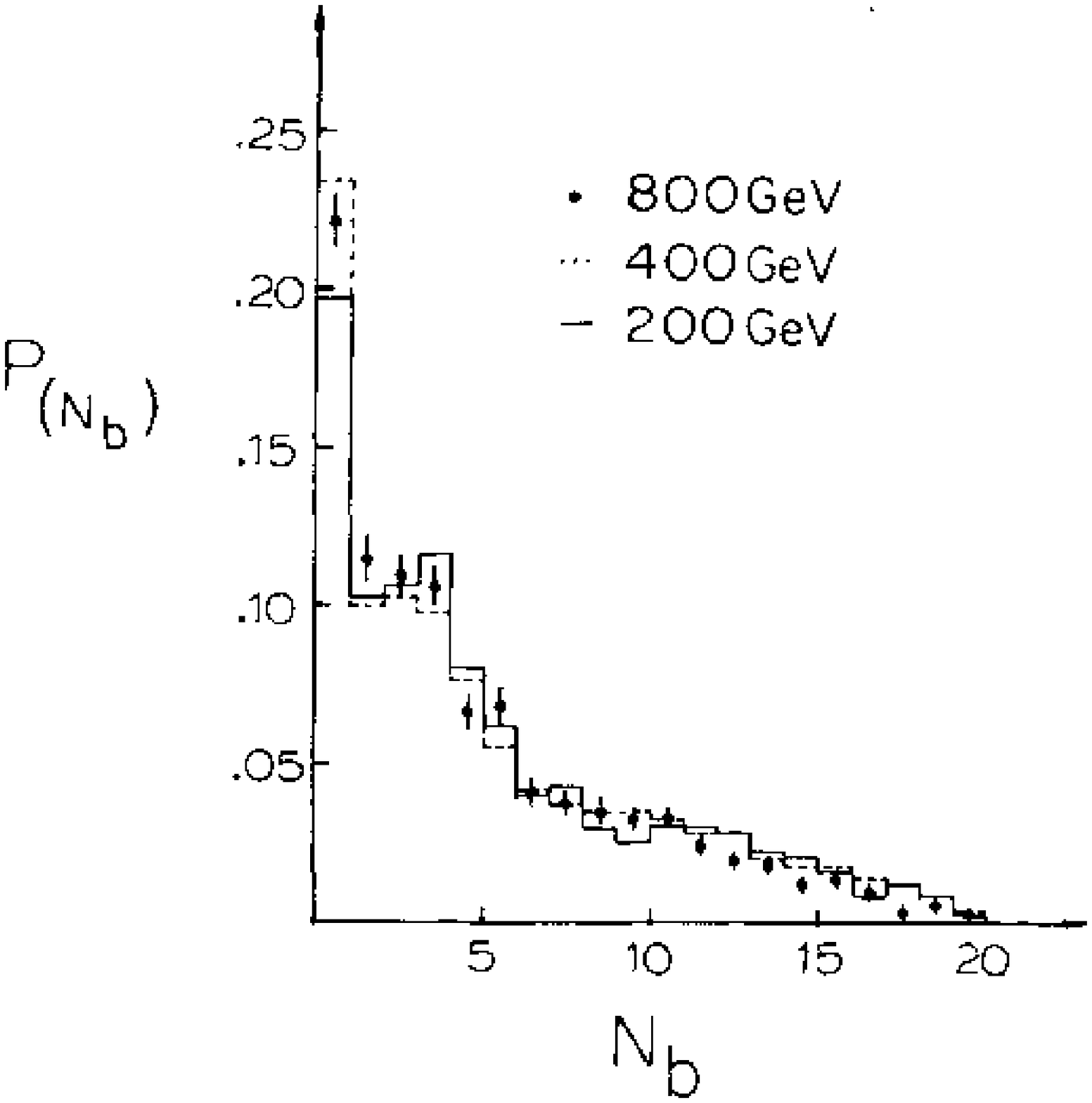}}}

 \caption{\label{fig:heckman} a) Angular distributions for fragments with
$E<250$ MeV, emitted from central collisions observed in nuclear emulsion,
from Ref.~\cite{Heckman:1978}. b) Distribution of the number of gray and
black tracks in proton-emulsion interactions at 200, 400, and 800 GeV,
from Ref.~\cite{Abduzhamilov:1989fr}.}

\end{figure}

\paragraph{Heckman et al.} Angular and momentum distributions of fragments
emitted from central collisions between emulsion nuclei (AgBr) and heavy
ion projectiles ${}^4$He, $^{16}$O and $^{40}$Ar have been studied at
Bevatron-Bevalac at an energy of around 2 GeV/$A$ \cite{Heckman:1978}.
Production angles and ranges of fragments having grain density $g\ge
2g_{min}$ corresponding to protons of $E\le$ 250 MeV were measured. The
data are successfully fitted and analyzed in terms of a modified
Maxwell-Boltzmann distribution (see
Sec.~\ref{sec:modified_maxwell_boltzmann}) from which estimates of the
longitudinal velocity $\beta_{\parallel}$ and characteristic spectral
velocity $\beta_0$ of the particle-emitting systems are obtained
(Fig.~\ref{fig:heckman}a). (It is assumed that the fragments are dominated
by one species and the effective Coulomb barrier for emission is small.)

For the subclass of fragments having $E<30$ MeV (black protons) they find
that the angular distributions are independent of the mass of the
projectile. Longitudinal velocities of the particle-emitting systems are
low, typically $0.014\pm0.002$, other representative experiments showed
$0.01<\beta_{\parallel}<0.03$. The "temperature" $\tau = m\beta_0^2/2$ is
typically 6-7 MeV, independent of projectile.

\paragraph{Fujioka et al.} Japanese group investigated particles emitted
in backward hemisphere with momenta around the kinematic limit of single
proton-nucleon collisions, using 205 GeV protons on nuclear emulsion at
FNAL \cite{Fujioka:1975wc}. They find that backward pions cannot come from
the simple superposition of elementary processes, but may originate from very
highly excited states of residual nucleus which took part in the reaction.

\paragraph{FNAL-E668.} The interaction of 800 GeV protons in nuclear
emulsion has been investigated by FNAL-E668 \cite{Abduzhamilov:1989fr}.
The distribution of heavily ionizing particles, thus target excitation, is
found to be independent of energy when comparing to collisions at 67, 200
and 400 GeV (Fig.~\ref{fig:heckman}b). This supports the hypothesis that
the number of heavy particles measure the impact parameter of the
collision and is related to the number of nucleon-nucleon collisions in
the target nucleus.

The superposition models assume that each collision of the projectile
yields the same distribution of gray particles and that consecutive
collisions contribute independently. Hence the angular distribution of
gray particles should not depend on primary energy or the number of
collisions $\nu$. Their results on angular distributions demonstrate both
the energy and $\nu$ independence.

\begin{figure}
 \figs{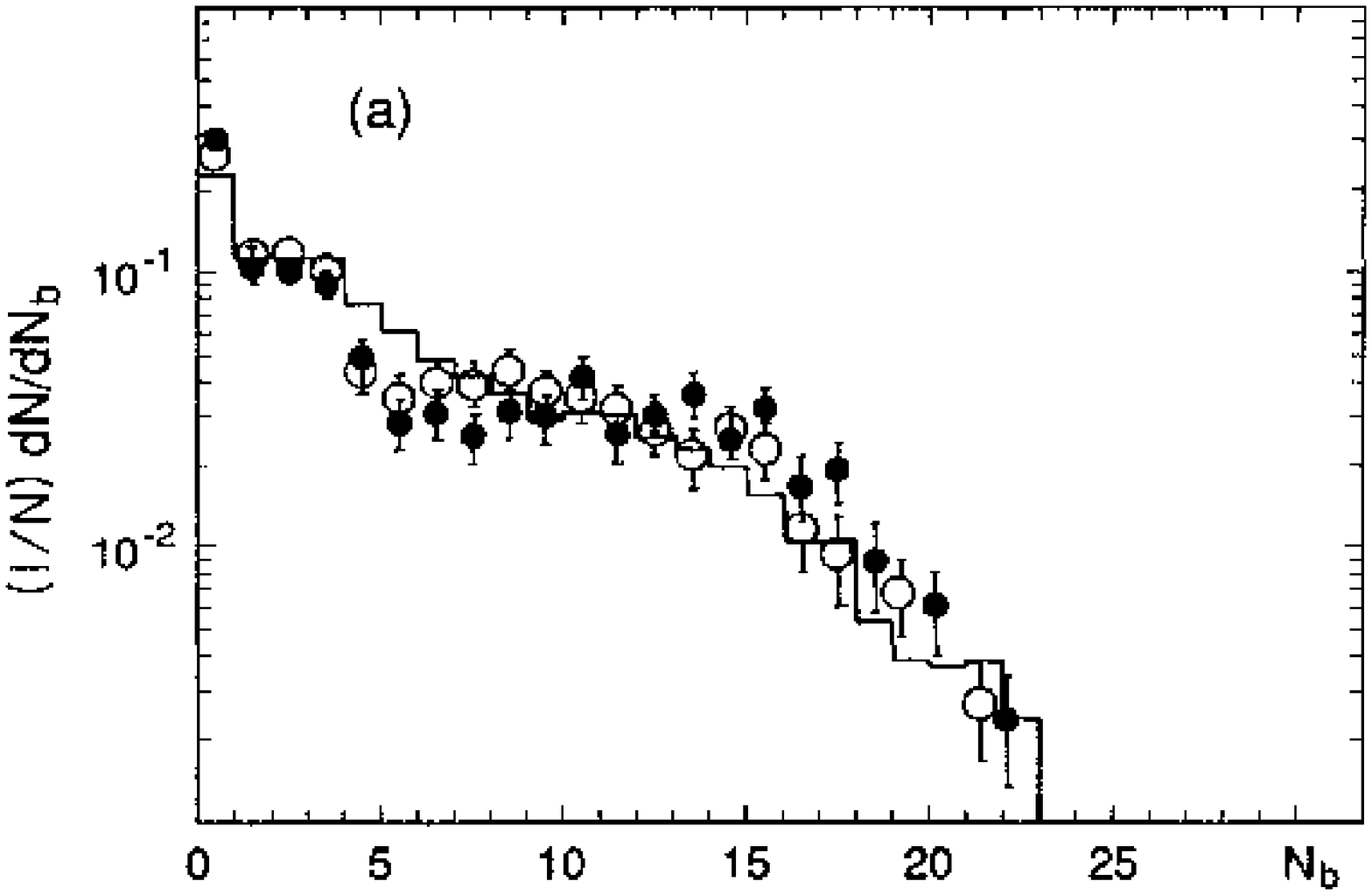}
      {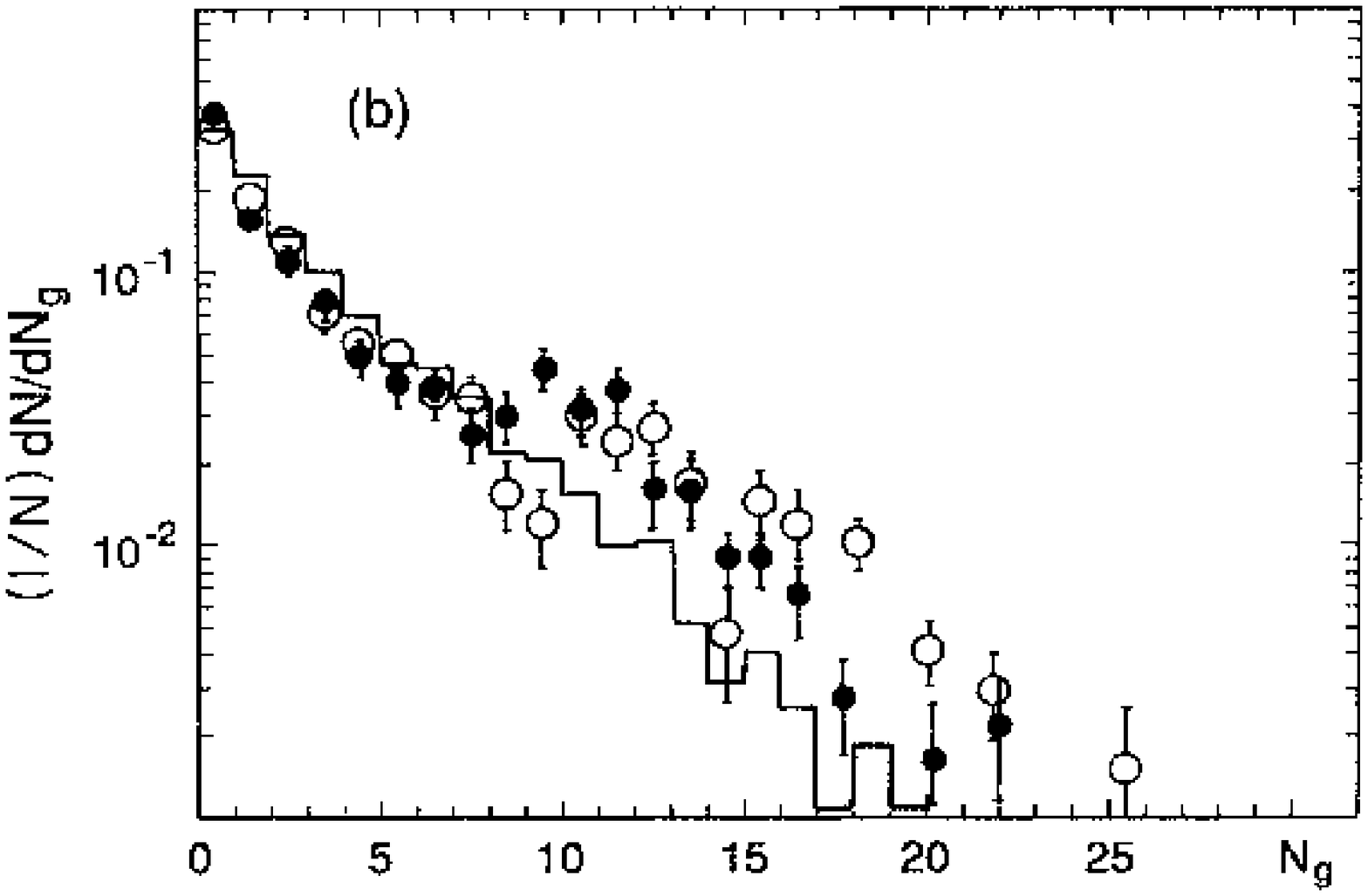}

 \caption{\label{fig:dabrowska1} The multiplicity distribution of black
and grey tracks for O and S interactions, from
Ref.~\cite{Dabrowska:1993pj}.}

\end{figure}

\begin{figure}
 \figs{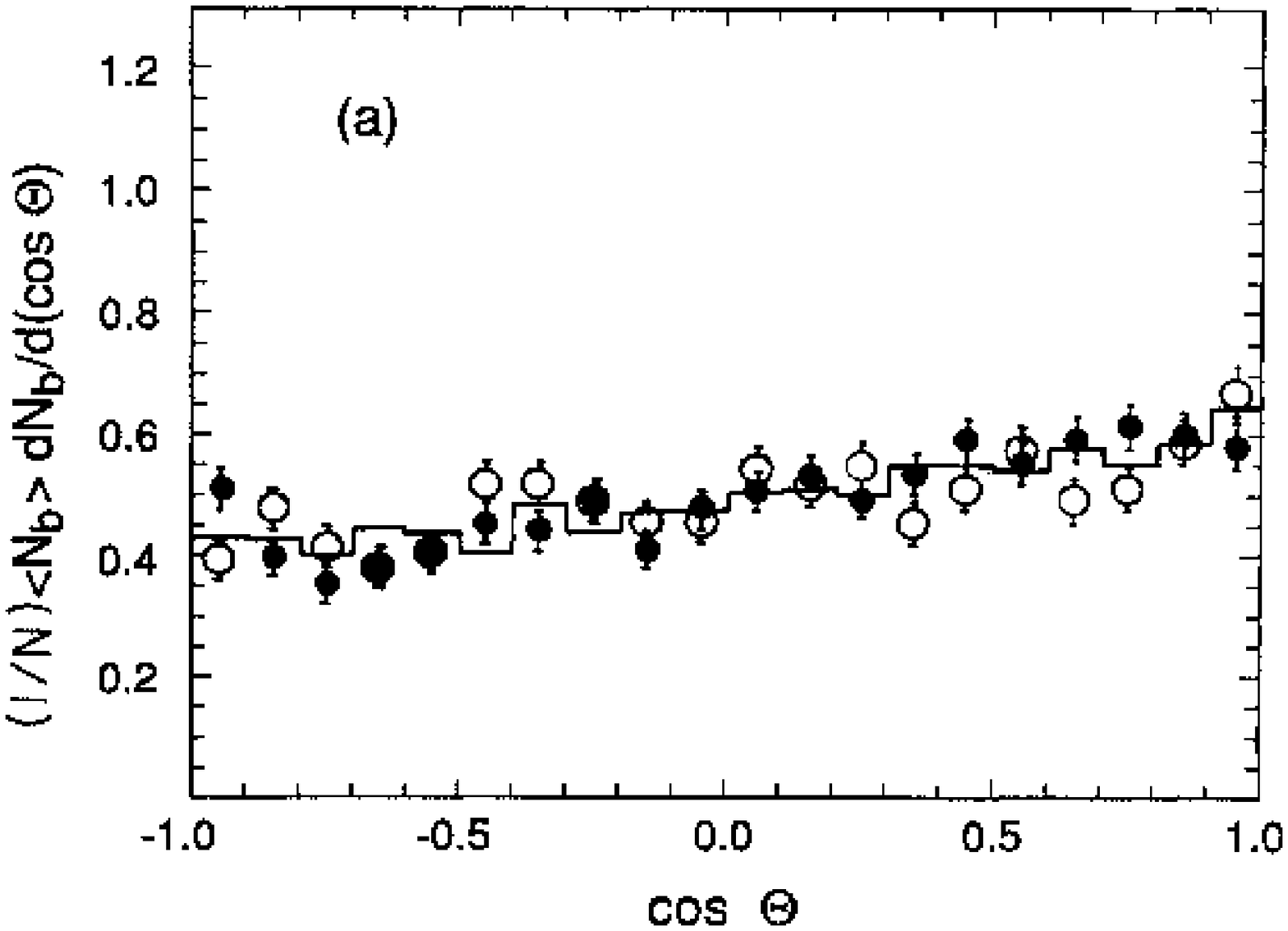}
      {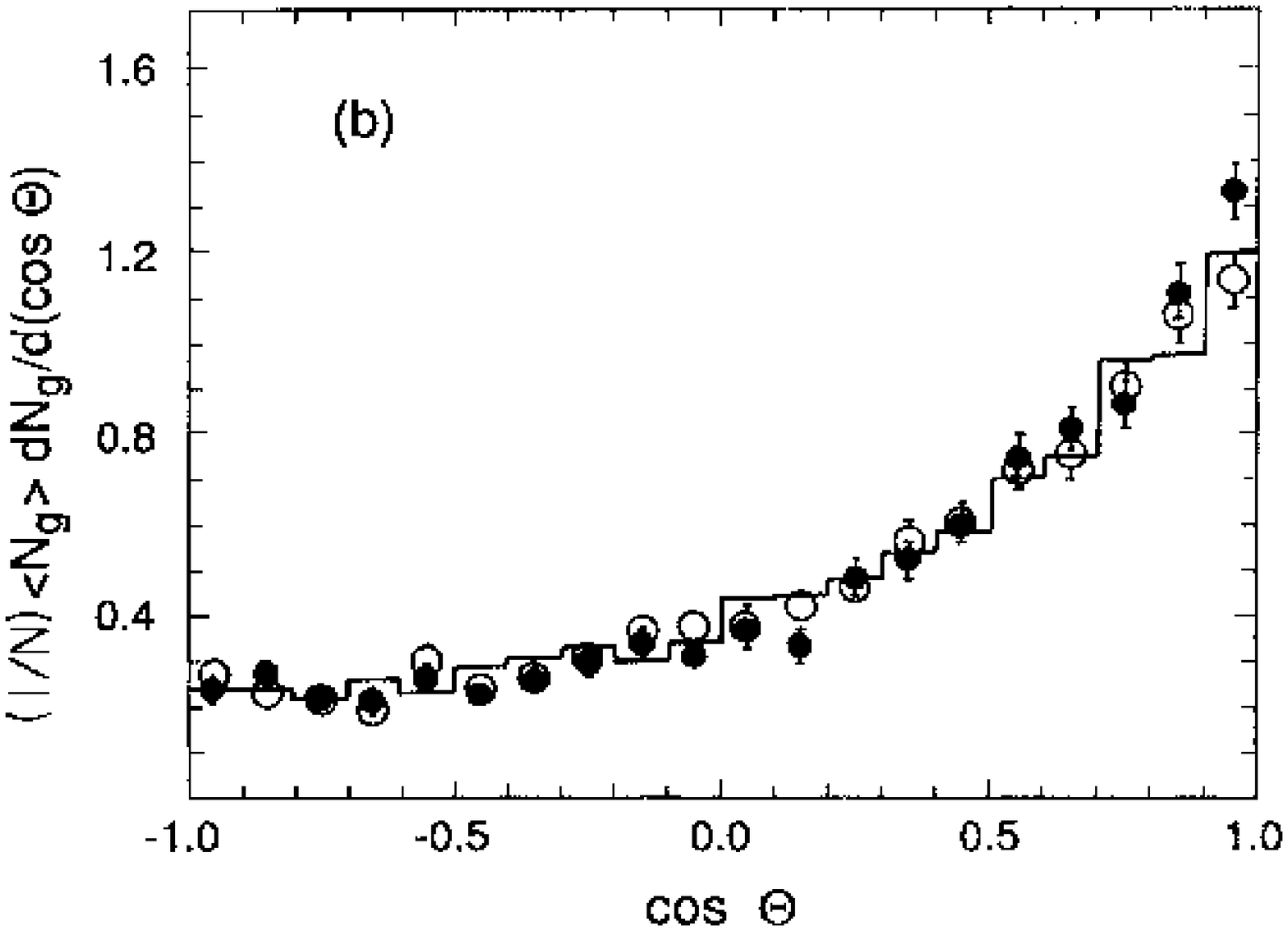}

 \caption{\label{fig:dabrowska2} The angular distribution of black and
grey tracks from O and S interactions, from Ref.~\cite{Dabrowska:1993pj}.
For comparison the distribution for proton interactions is shown by the
histogram.}

\end{figure}

\begin{figure}
\figs{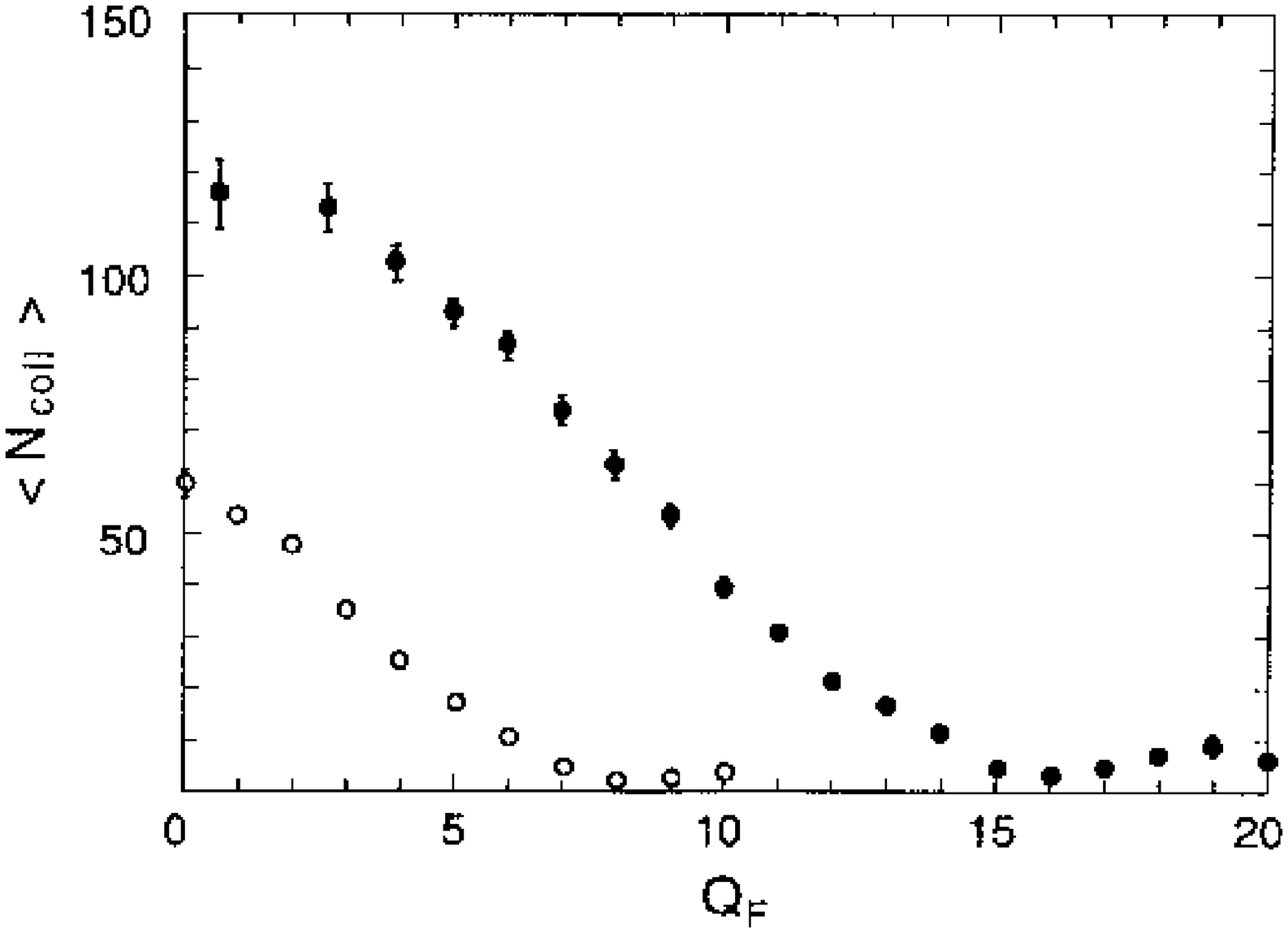}
      {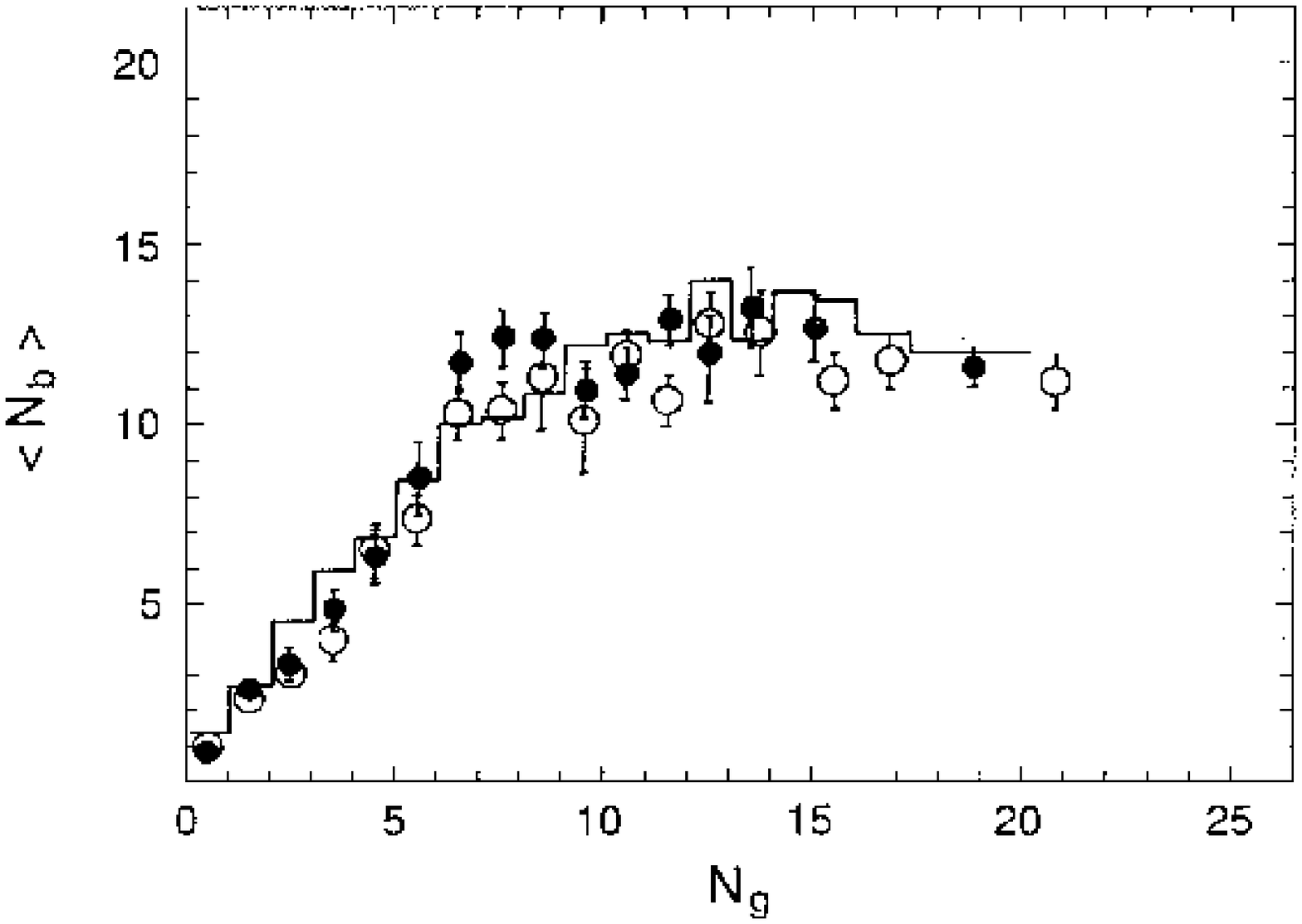}

 \caption{\label{fig:dabrowska3} a) The relation derived from the Venus
model between the mean number $\langle N_{coll} \rangle$ of intranuclear
collisions and the charge $Q_F$ for O (open) and S (closed) interactions
in emulsion. b) The dependence of the mean number $\langle N_b \rangle$ of
black tracks on the number $N_g$ of grey tracks, for O (open) and S
(closed) interactions in emulsion. For comparison the dependence for
proton interactions is shown by the histogram. Both from
Ref.~\cite{Dabrowska:1993pj}.}

\end{figure}

\paragraph{CERN-EMU07.} Interactions of O and S nuclei of 200 GeV/$A$ in
nuclear emulsion have been observed by CERN-EMU07 experiment (KLM
collaboration) \cite{Dabrowska:1993pj}. They do not find significant
deviations from models, such as Venus, describing the interactions as
being the superposition of individual nucleon-nucleon collisions.

The value of the total charge $Q_F(\theta)$ emitted within the very
forward cone of angle $\theta$ can be measured. When $\theta$ is chosen
such that all spectator protons but only a small number of produced
particles are contained, $Q_F$ measures the total charge carried by the
noninteracting projectile nucleons. Comparing experimental probability
distributions of $Q_F$ with predictions from Venus model there is
generally good agreement. This justifies the use of Venus model to relate
the experimentally measured values with the calculated mean number of
intranuclear collisions $\langle N_{coll} \rangle$
(Fig.~\ref{fig:dabrowska3}a). The latter represents the centrality of the
collision. It is shown that $\langle N_g \rangle$ increases almost
linearly with decreasing $Q_F$ (increasing centrality of the collision),
while the $\langle N_b \rangle$ value for central collisions becomes
constant.

The slight angular dependence for the black tracks suggests that they
could be emitted isotropically from a moving frame of reference with the
velocity $\beta=0.01$
(Figs.~\ref{fig:dabrowska1}~and~\ref{fig:dabrowska2}). They note that
differences in the multiplicity and angular distributions of the grey and
black tracks point to their different creation processes: initial
interaction and evaporation. The mechanism is independent of the number of
collisions $\nu$. The correlation between the numbers of black and grey
tracks is strong for $N_g\le 7$, but for larger values of $N_g$, the mean
$N_b$ levels off at a constant value of 12 (Fig.~\ref{fig:dabrowska3}b).

\begin{figure}
 \figs{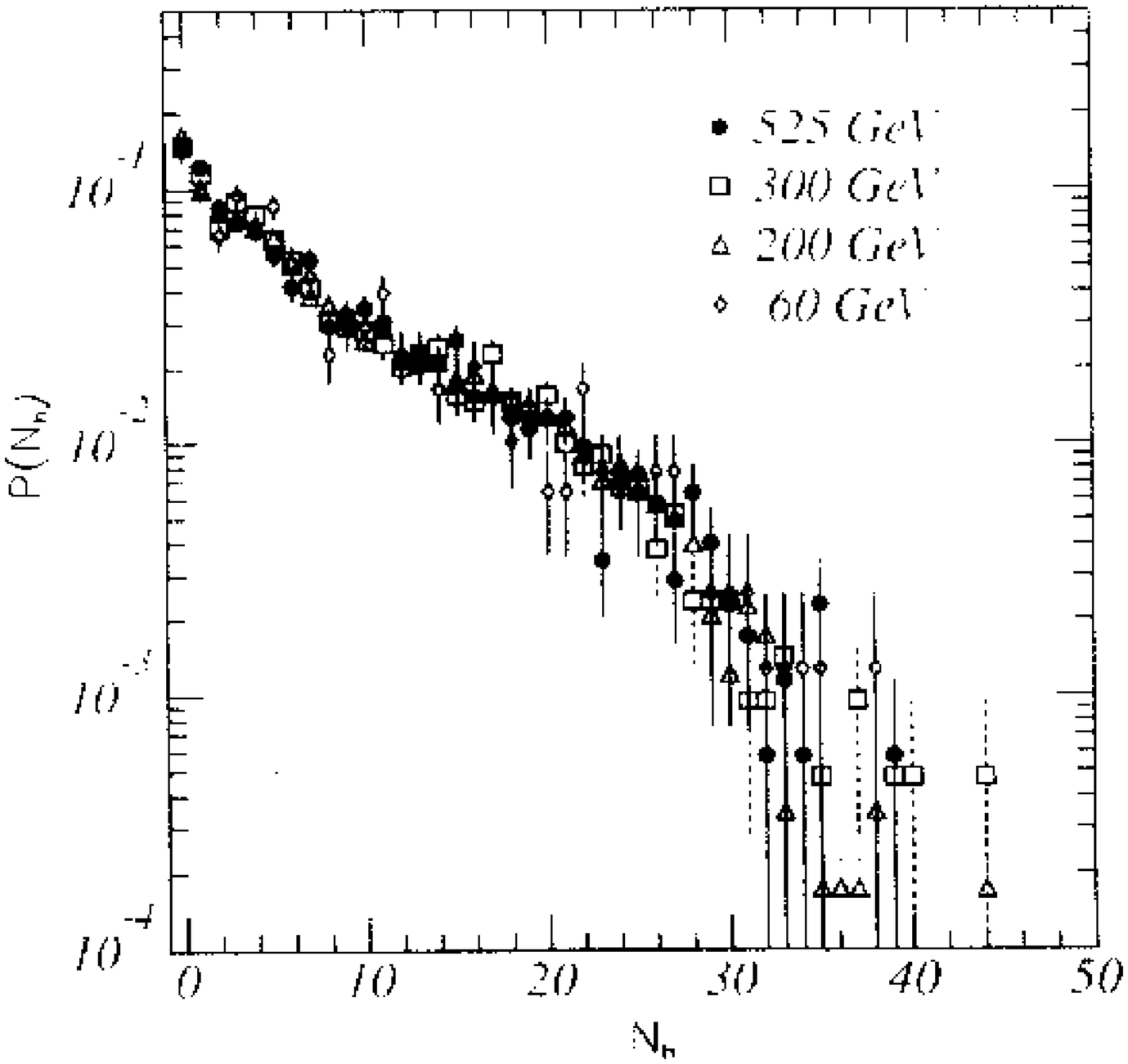}
      {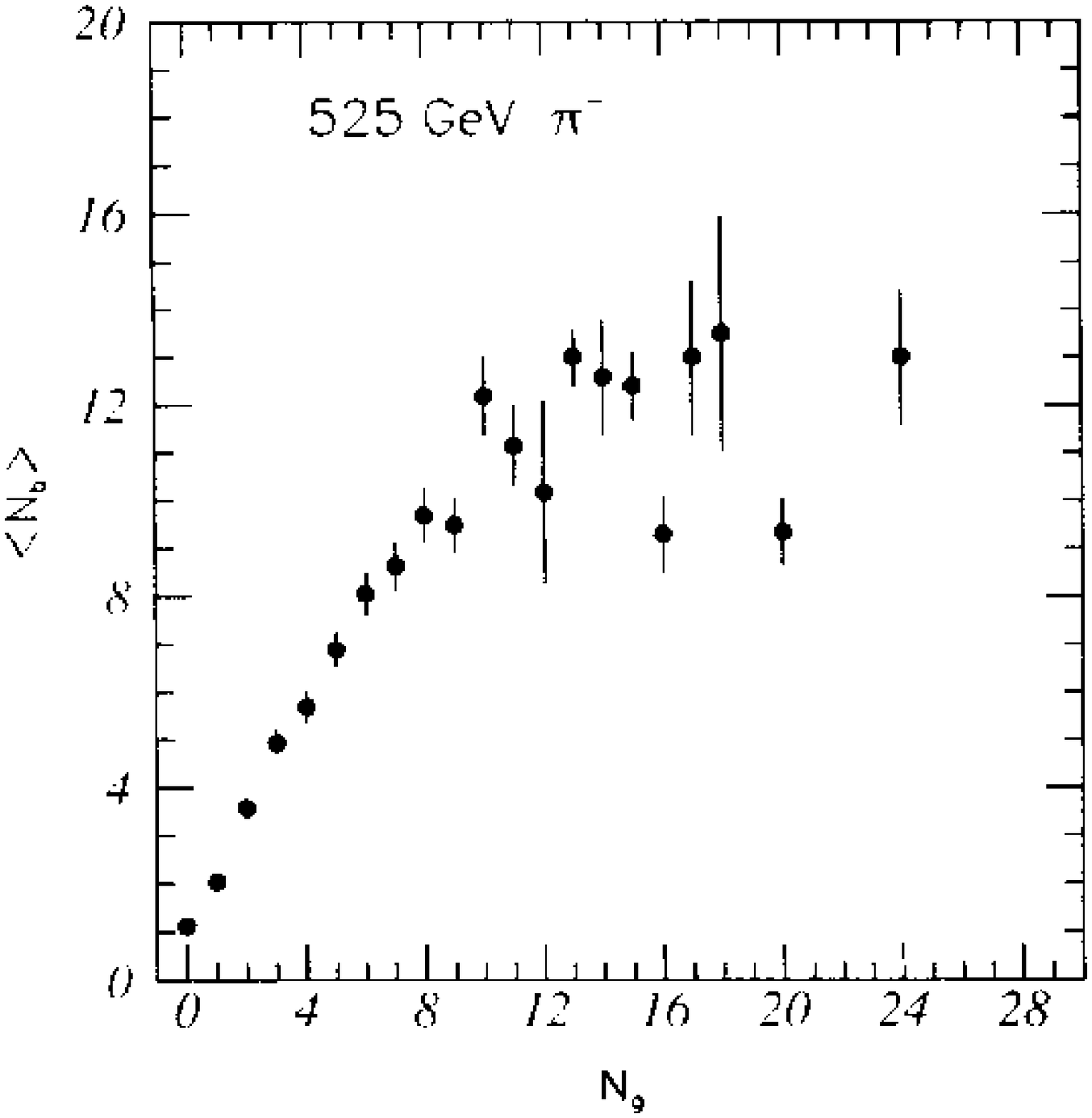}

 \caption{\label{fig:cherry1} a) $P(N_h)$ vs $N_h$ for 60-525 GeV pions.  
b) $\langle N_b \rangle$ vs $N_g$ for 200 GeV $\pi$ and p and 200
GeV/nucleon $^{16}$O and $^{32}$S interactions. Both from
Ref.~\cite{Cherry:1994}.}

\end{figure}

\begin{figure}
 \figs{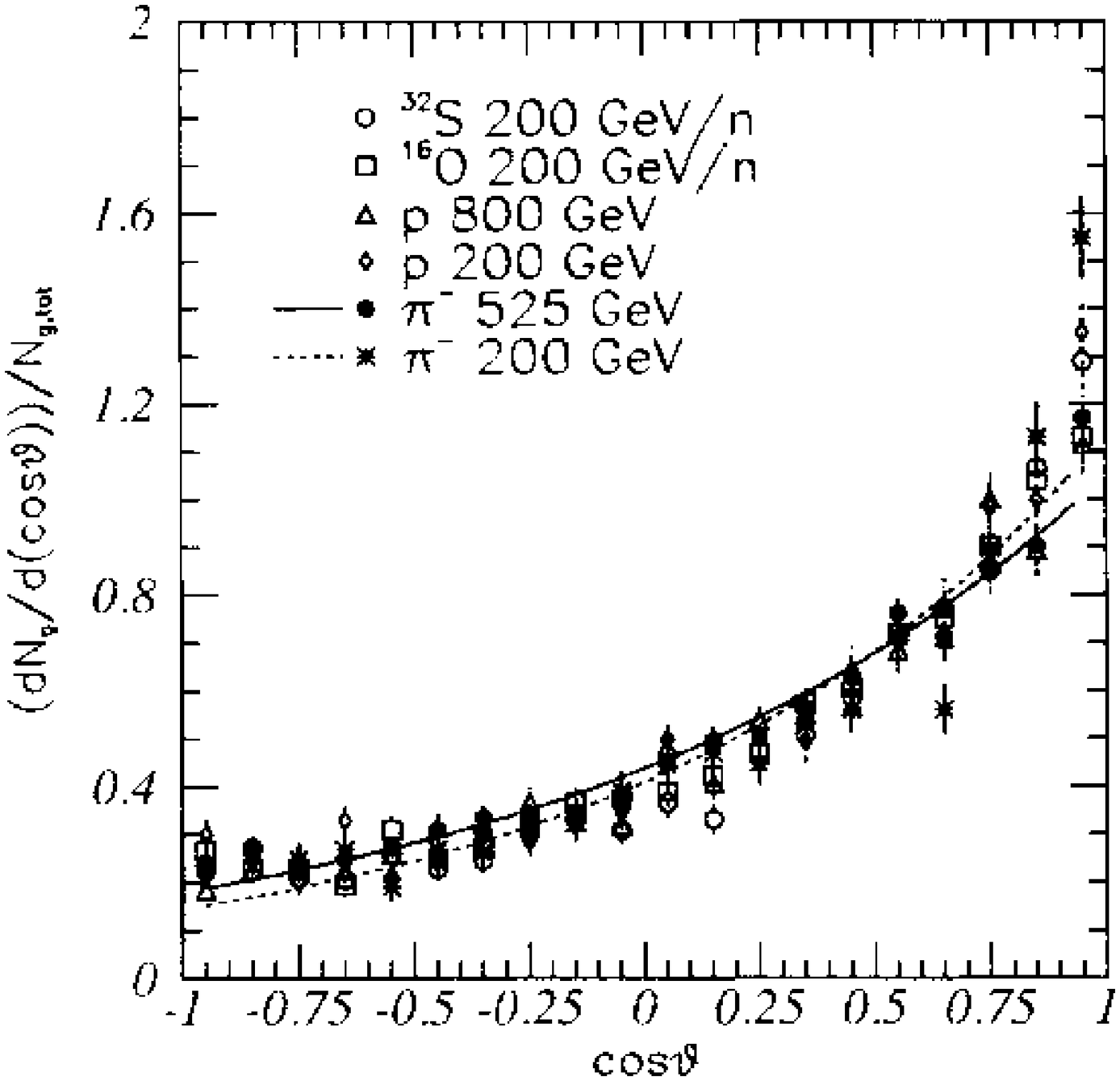}    
      {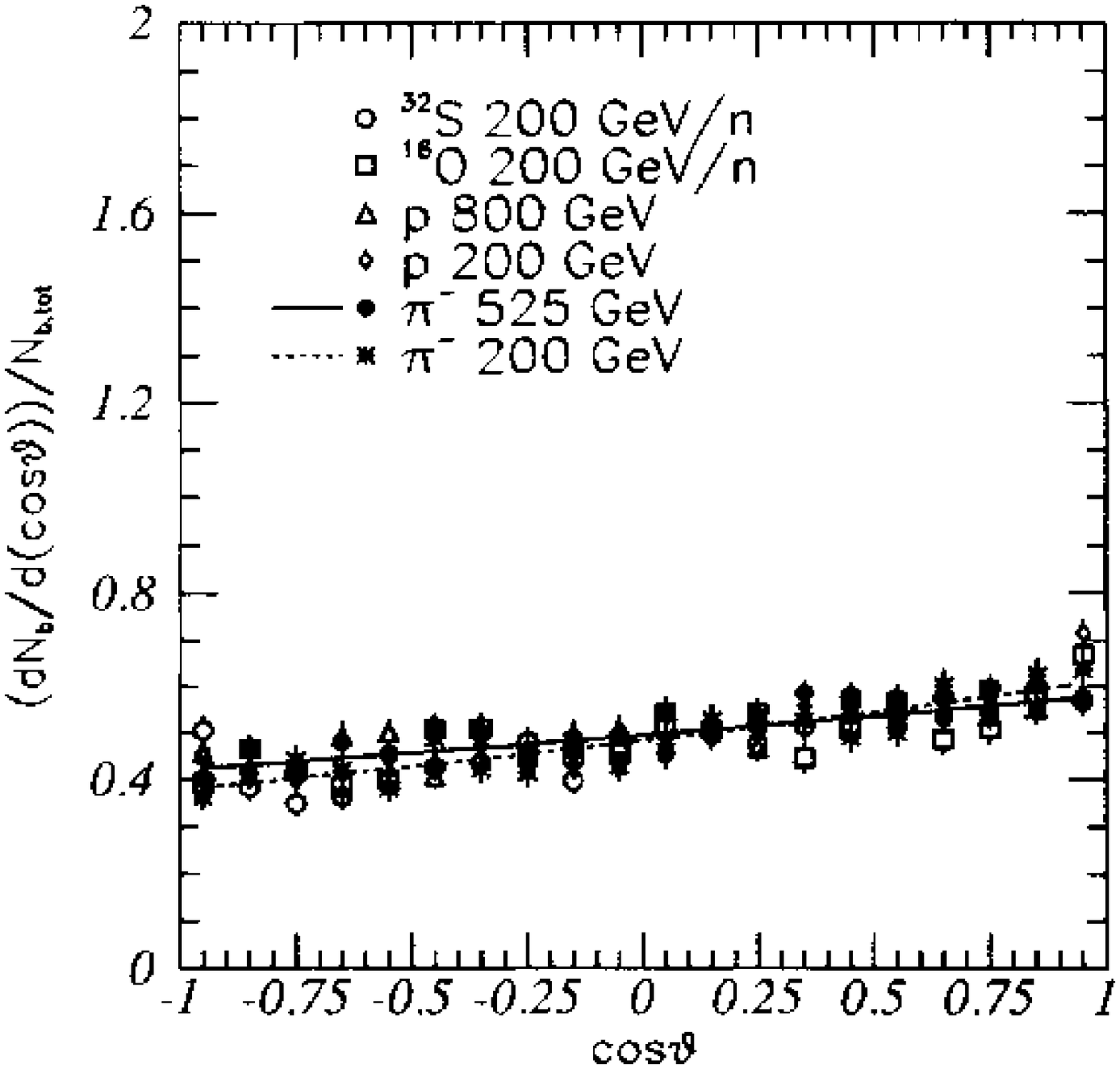}
 
 \caption{\label{fig:cherry2} a). Angular distribution of grey tracks for
$\pi$, p and heavy ion interactions in emulsion. Smooth curves are fit to
the two-step vector model. b) Same for black protons. Both from
Ref.~\cite{Cherry:1994}.}

\end{figure}

\paragraph{FNAL-E667.} Inclusive 525 GeV $\pi^-$ interactions in emulsion have
been measured by FNAL-E667 \cite{Cherry:1994}. Mentioning the systematic
differences between grey and black track identification using range and
ionization measurements, only comparison of the number of heavily ionizing
track $N_h$ is considered. They find that multiplicity distributions of
heavy tracks do not vary significantly with the energy of the incoming
pion, similarly to proton projectile (Fig.~\ref{fig:cherry1}a). Saturation
is observed when $\langle N_b \rangle$ is plotted against $N_g$, similarly
to the experiment described above (Fig.~\ref{fig:cherry1}b). It appears
that same number of gray tracks produces a similar degree of target
excitation ($N_b$) independent of the kind of projectile and incident
energy. The gray track angular distributions are strongly forward-peaked,
while the black ones exhibit little asymmetry (Fig.~\ref{fig:cherry2}).
The distributions do not depend significantly on the type of projectile or
its energy. The forward-backward ratios decrease slightly with increasing
projectile energy, but this can be due to the uncertainties in track
identification. If the typical fragmentation energy per nucleon is $E_0
\sim 6-8$ MeV then $\beta_\parallel \le 0.01$ for the residual system
producing black tracks.

%%%%%%%%%%%%%%%%%%%
% Bubble chambers

\subsection{Bubble chambers}

\paragraph{Yeager et al.} Detailed studies of $\pi^+$Ne and $\pi^-$Ne
interactions at 10.5 GeV/$c$ have been carried out at SLAC, using bubble
chamber \cite{Yeager:1977ym}. Curvature, range and track density
information have been used to separate pions from protons, plus the help
of isospin symmetry to untangle the ambiguities. The enhancement of pion
production in the target-fragmentation region arises mainly from events
with large values of $N_h$ heavy tracks. $N_h$ appears to be the measure
of the number of struck nucleons inside the nucleus.

\paragraph{Hayashino et al.} Protons and pions emitted in the backward
hemisphere in collisions of 28.5 GeV/$c$ protons with Ta have been studied
using hydrogen bubble chamber at BNL \cite{Hayashino:1976dc}. The kind of
particles was identified by the usual method from ionization and momentum.
Longitudinal momentum distributions show that protons and pions have
different average momenta, thus the contribution from isobar decay is
small. On the other hand kinetic energy spectra of both particles can be
expressed by simple exponential form, suggesting that they come from some
kind of thermodynamically equilibrium-like state. Similar results have
been obtained using 12.6 GeV/$c$ $K^-$ beam \cite{Fukushima:1978ed}.

\begin{figure}
%bailly_zpc35_301_1987_fig2r
 \figs{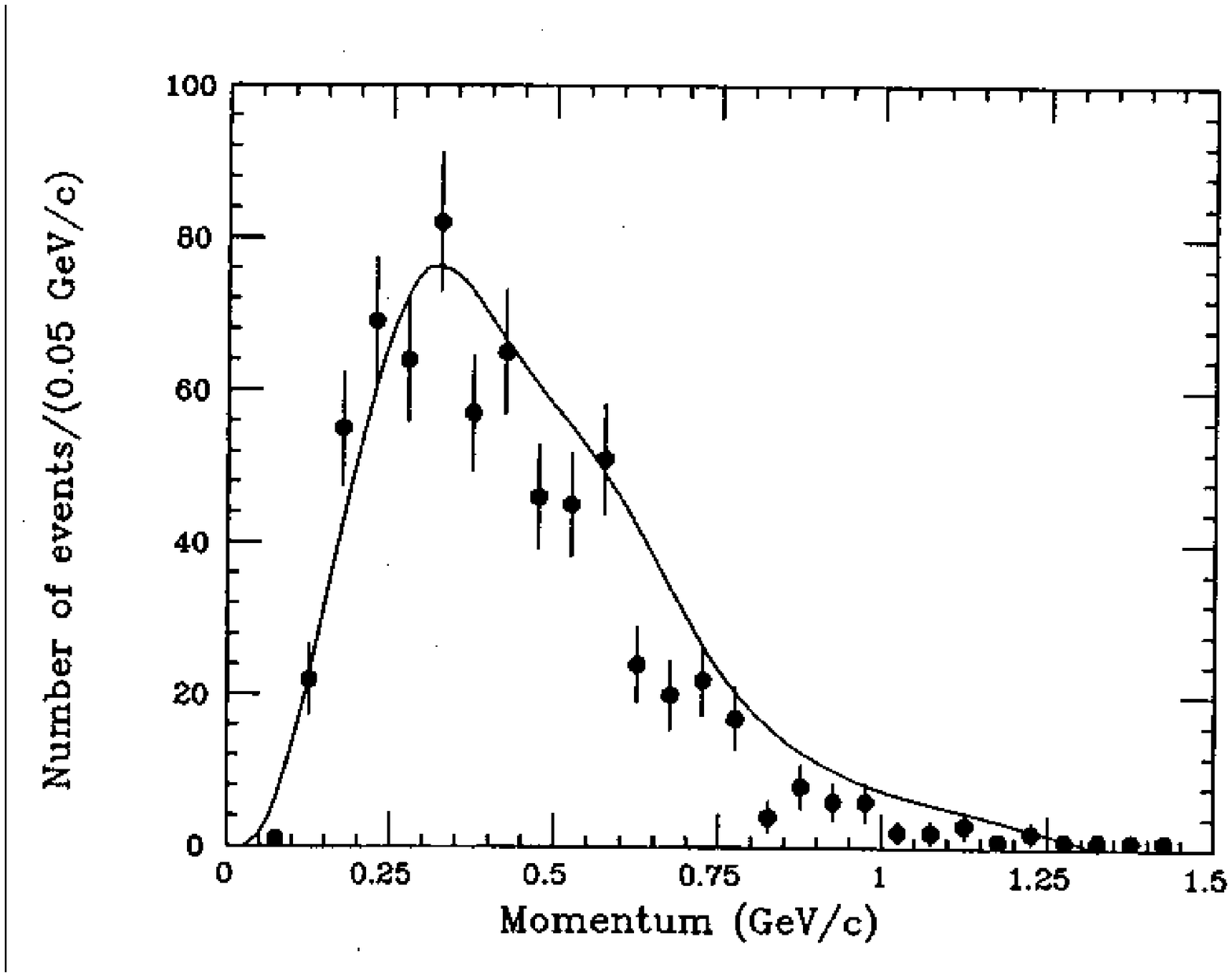}
      {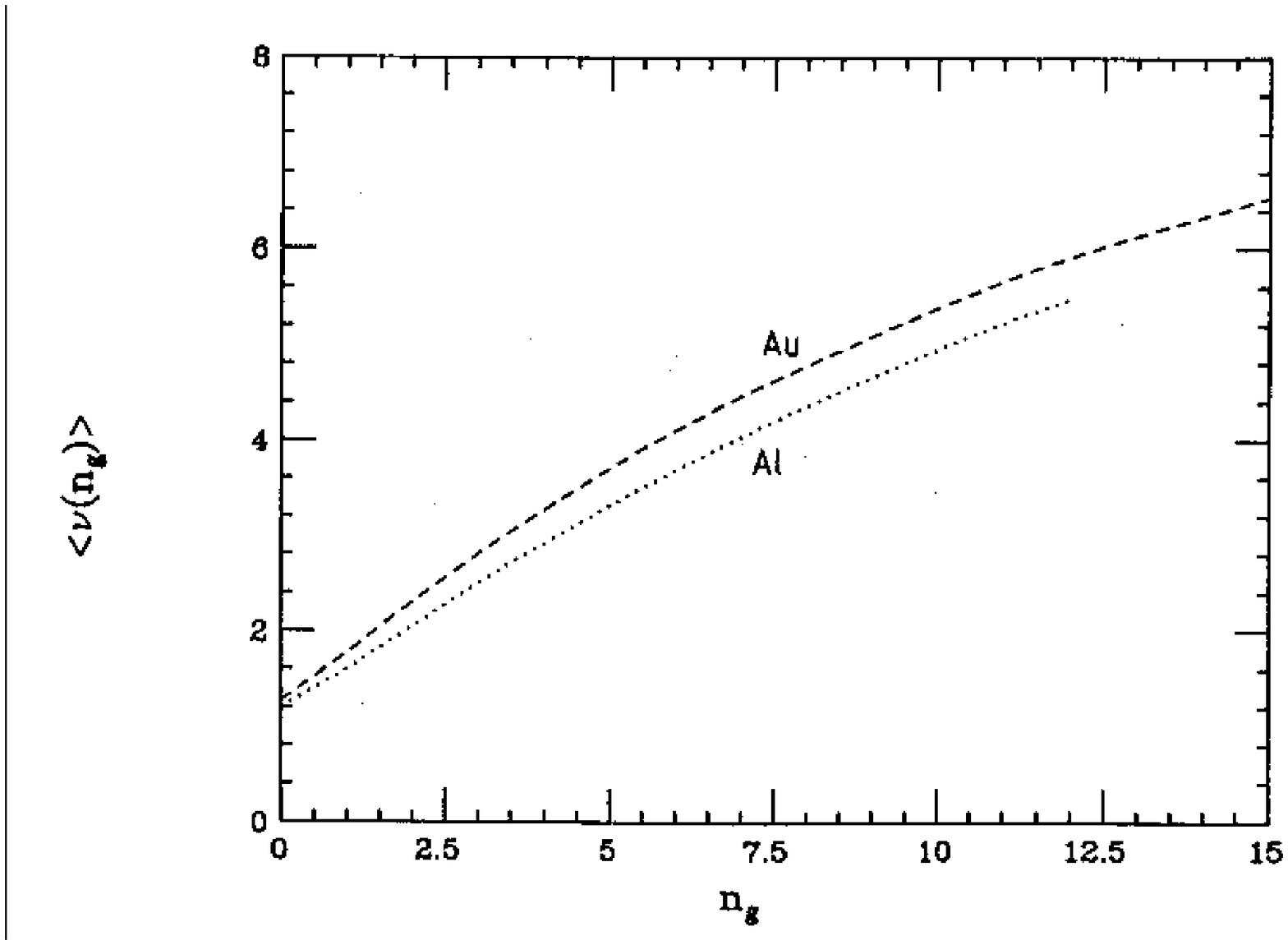}
 
 \caption{\label{fig:bailly} a) Grey tracks momentum distribution. The
solid curve gives the prediction of the multi-chain model. b) Average
number of collisions between the incident protons and the target nuclei as
a function of the number of grey tracks calculated by the multi-chain
model. Both from Ref.~\cite{Bailly:1987vv}.}

\end{figure}

\paragraph{CERN-EHS.} Interactions of 360 GeV/$c$ protons with Al and Au
targets were studied using the European Hybrid Spectrometer equipped with
the rapid cycling bubble chamber at CERN \cite{Bailly:1987vv} and compared
with calculations based on the multi-chain and Lund models
(Fig.~\ref{fig:bailly}). The former is in good agreement with data, while
the latter is not, probably because it does not include cascading.

The multi-chain model assumes that at each collision the projectile loses
a fraction of its momentum according to a probability, and a hadronic chain
is stretched between the projectile and the target nucleon. The projectile
finally fragments into hadrons, the hadronic chain also hadronizes into
pions and the recoil nucleon. The hadrons may come on-shell only after a
characteristic formation time (formation zone concept). The model can
predict the average number of collisions $\langle \nu \rangle$ and
function of the number of knocked out gray particles $N_g$.

\begin{figure}
 \figs{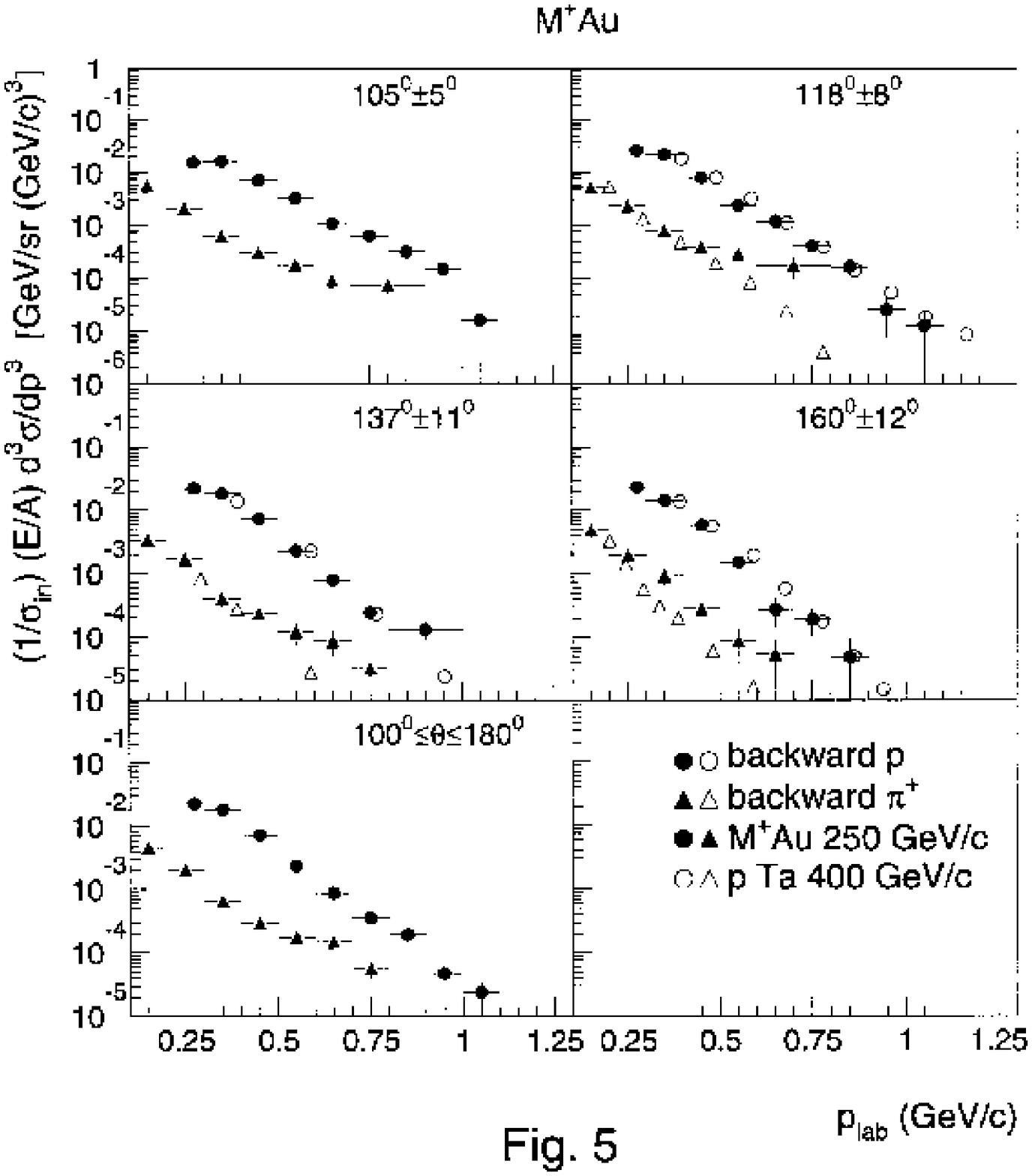}
      {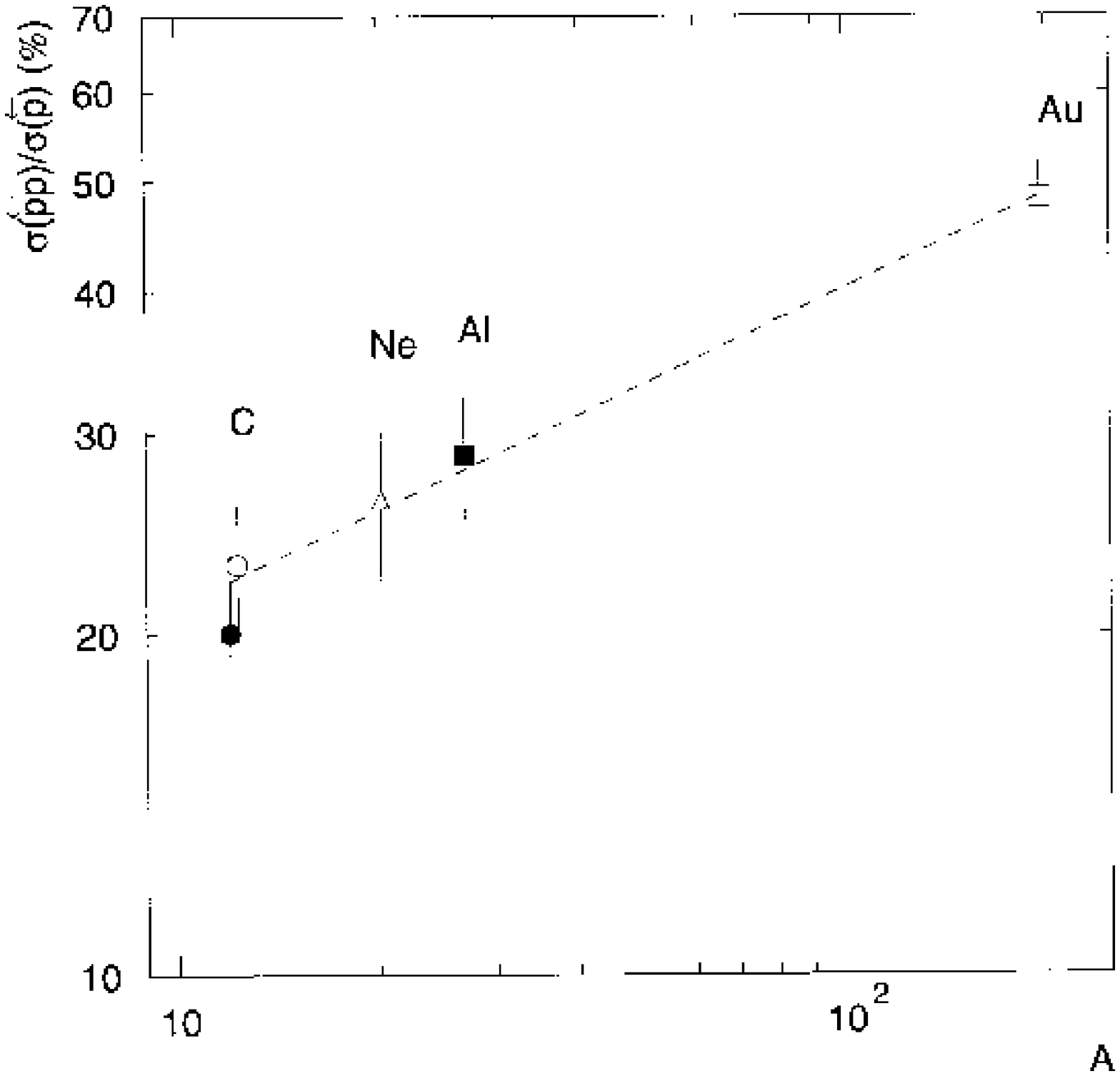}

 \caption{\label{fig:agababian} a). The normalized invariant inclusive
cross-section per nucleon for backward protons and $\pi^+$ mesons in
different angular ranges, as a function of $p_{lab}$, for $M^+$Au
interactions. b) The contribution of secondary pion absorption to backward
proton production, as a function of mass number $A$. Both from
Ref.~\cite{Agababian:1995wy}.}

\end{figure}

\paragraph{CERN-NA22.} The same equipment has been used by the CERN-NA22
experiment, studying backward proton production in $\pi^+$ and $K^+$
collisions with Al and Au nuclei at 250 GeV/$c$ \cite{Agababian:1995wy}.
Inclusive spectra are in agreement with the higher energy FNAL-E592
results (Fig.~\ref{fig:agababian}a). From two-proton correlations it is
found that significant part of backward proton production -- even as high
as 70\% for heavy nuclei -- can be attributed to secondary pion absorption
by a nucleon pair in the nucleus, predicting a low contribution from
double color-charge-exchange mechanism (Fig.~\ref{fig:agababian}b).

\paragraph{FNAL-E154.} Multiparticle production in the interactions of 200
GeV/$c$ protons $\pi^+$ and $K^+$ mesons with nuclei of Au, Ag and Mg have
been studied with bubble-chamber hybrid spectrometer by the FNAL-E154
experiment \cite{Brick:1989dm}. Using the net charge $Q$ of the event
and the number of collisions $\nu$ (obtained via calculation from the
number of grey protons), the number of secondary collisions can be
calculated. Their results suggest that secondary collisions arise from
rescattering of recoiling nucleons rather than produced particles.

\begin{figure}
 \figs{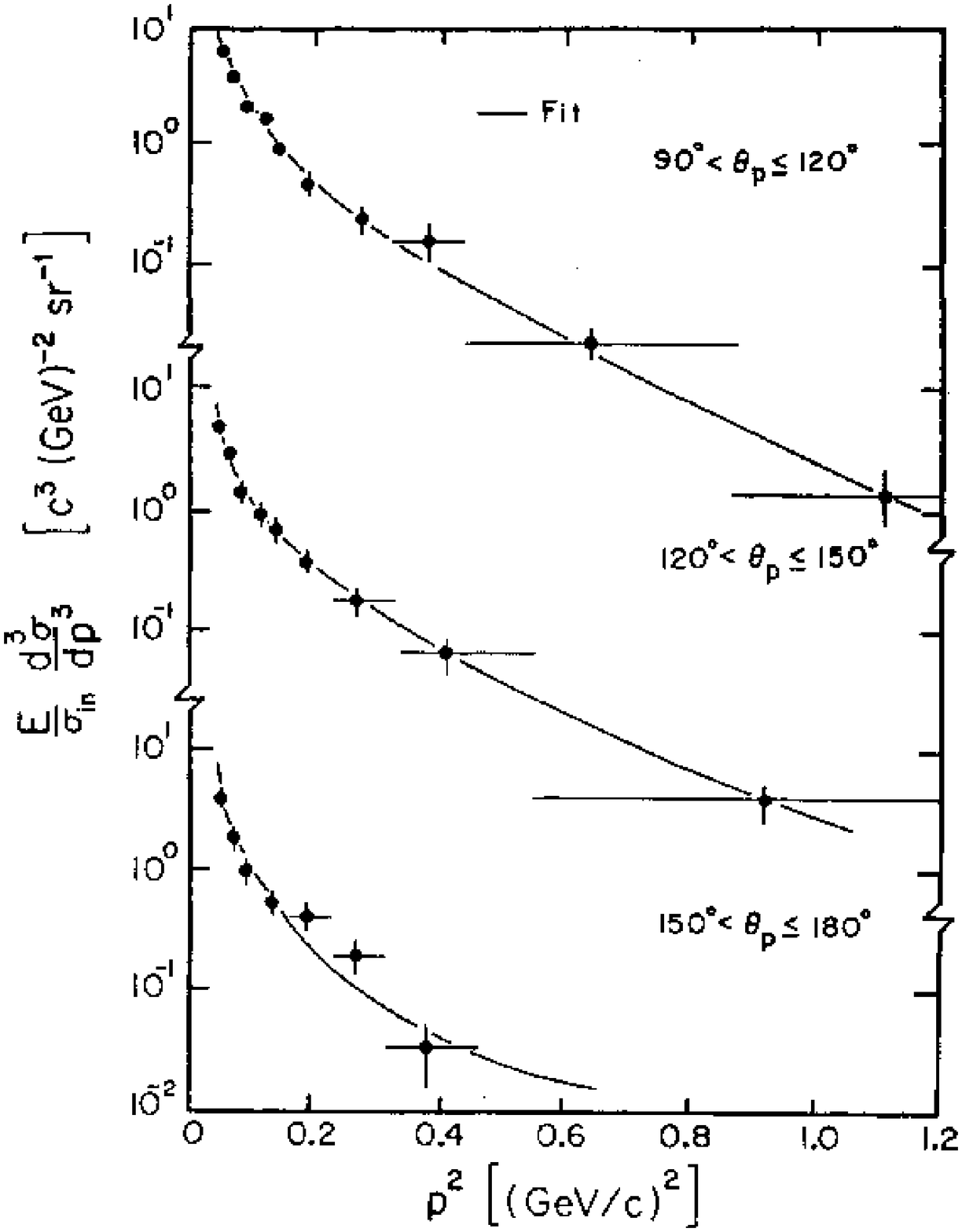}
      {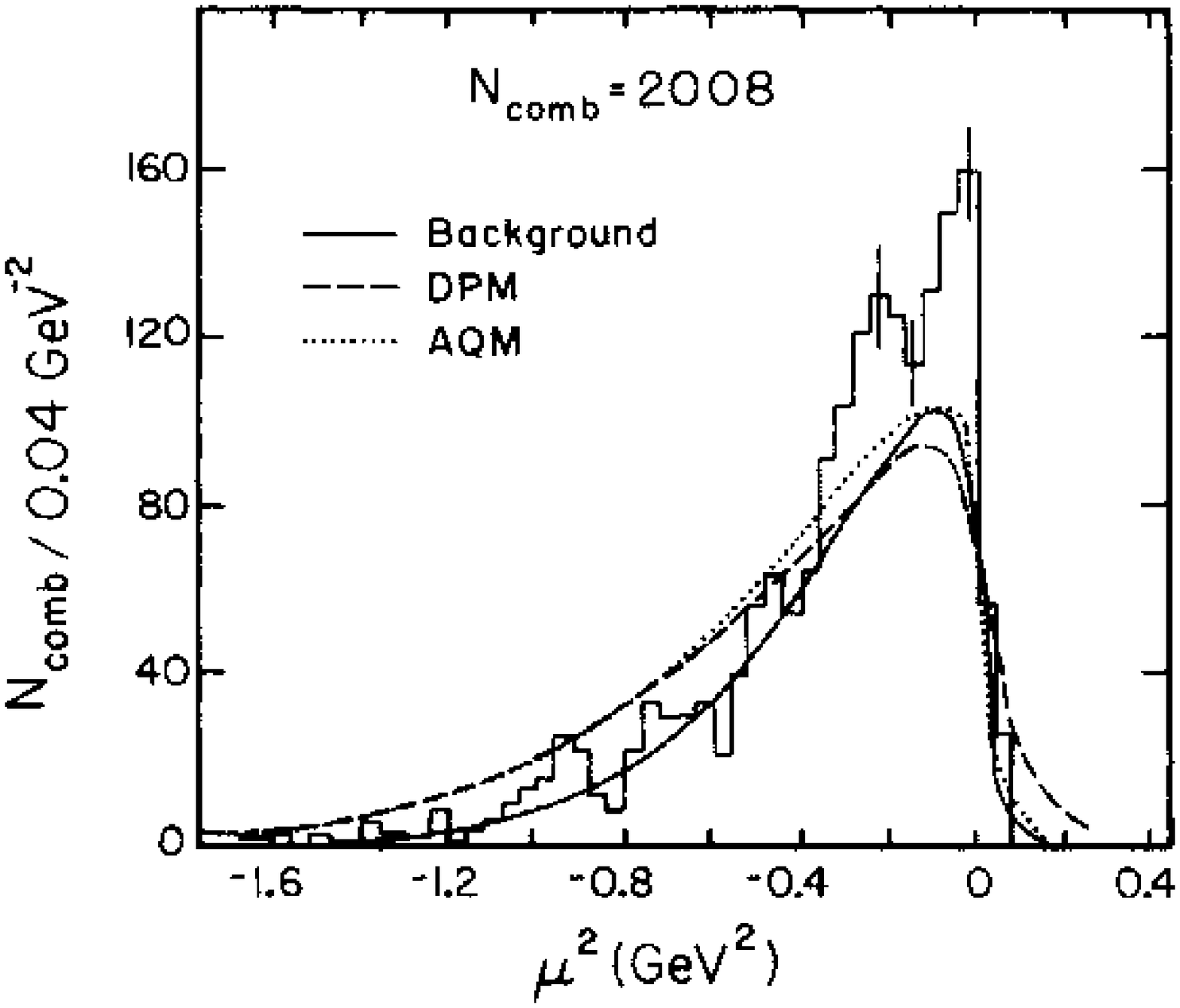}

 \caption{\label{fig:yuldashev} a) The invariant inclusive cross-sections
for protons with different angular ranges as function of $p^2$. b) The
distribution of missing mass square $\mu^2$ corresponding to the reactions
$\pi+"b"\rightarrow$ p+p and $\pi+"b"\rightarrow$ $\Delta$+p. The solid
line represents random background. The excess events are from two peaks at
$\mu^2\sim 0$ and $\mu^2\sim -0.22 (\mathrm{GeV}/c)^2$. Both from
Ref.~\cite{Yuldashev:1992aw}.}

\end{figure}

\paragraph{FNAL-E343.} Cumulative particle production in p $^{20}$Ne
interactions at 300 GeV have been studied by the FNAL-E343 experiment
\cite{Yuldashev:1992aw}. Strong correlation between multiplicities of
forward- and backward produced protons is observed. The inclusive
cross-section for protons can be described by sum of two exponentials
plotted as a function of $p^2$. The slope is an increasing function of
angle (Fig.~\ref{fig:yuldashev}a). Strong signals from $\Delta$ and $N^*$
resonances are observed. Evidence for large contribution of cumulative
protons from absorption of pions by quasi-two-nucleon systems is found,
which can produce up to 40\% of all protons emitted backward
(Fig.~\ref{fig:yuldashev}b). Comparison with models like additive quark
model, Lund and dual parton model shows that one cannot neglect
rescattering (cascading) processes.

%%%%%%%%%%%%%%%%%%%%%%%%
% Electronic detectors

\subsection{Electronic detectors}

\begin{figure}
 \figs{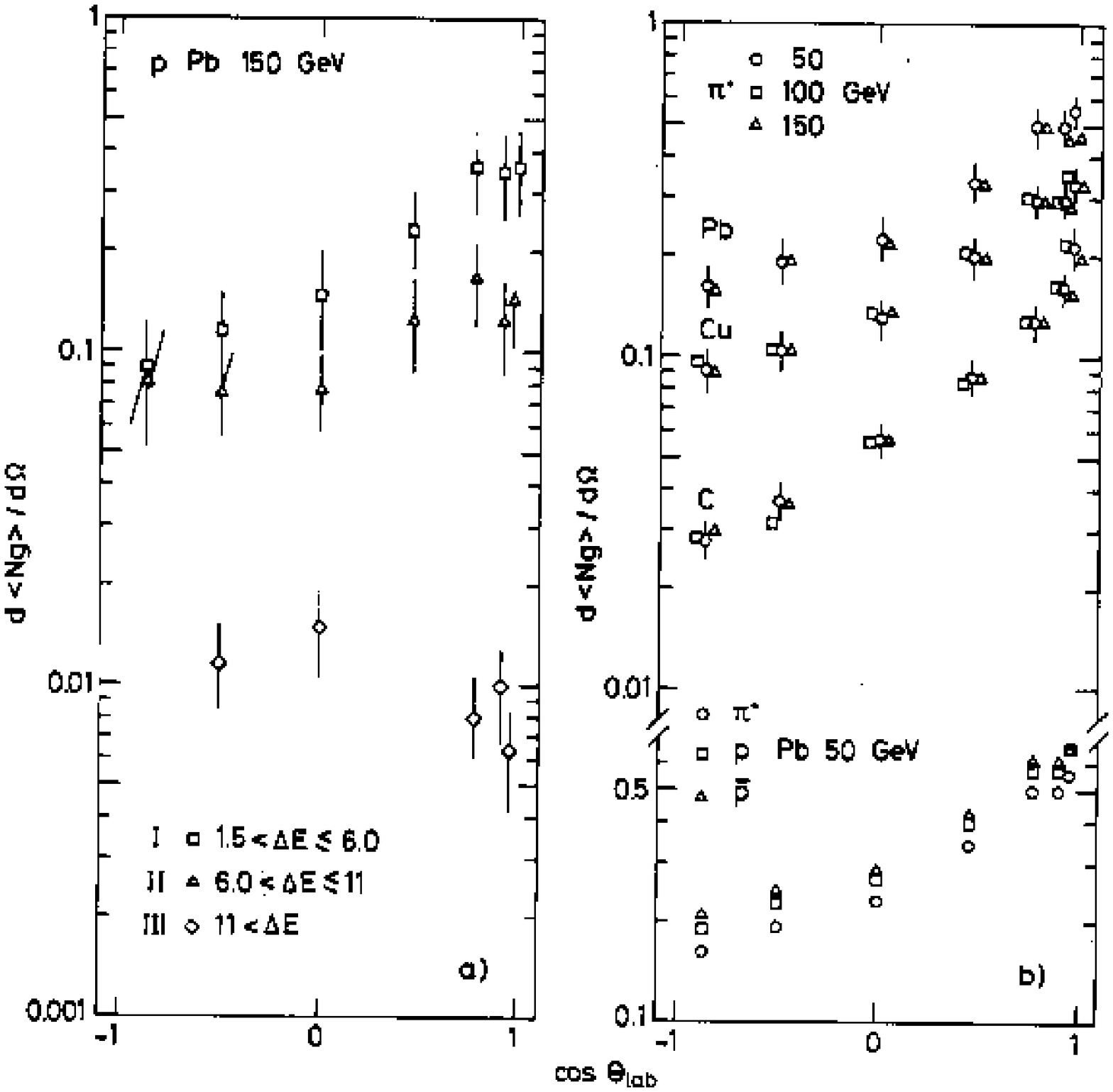}
      {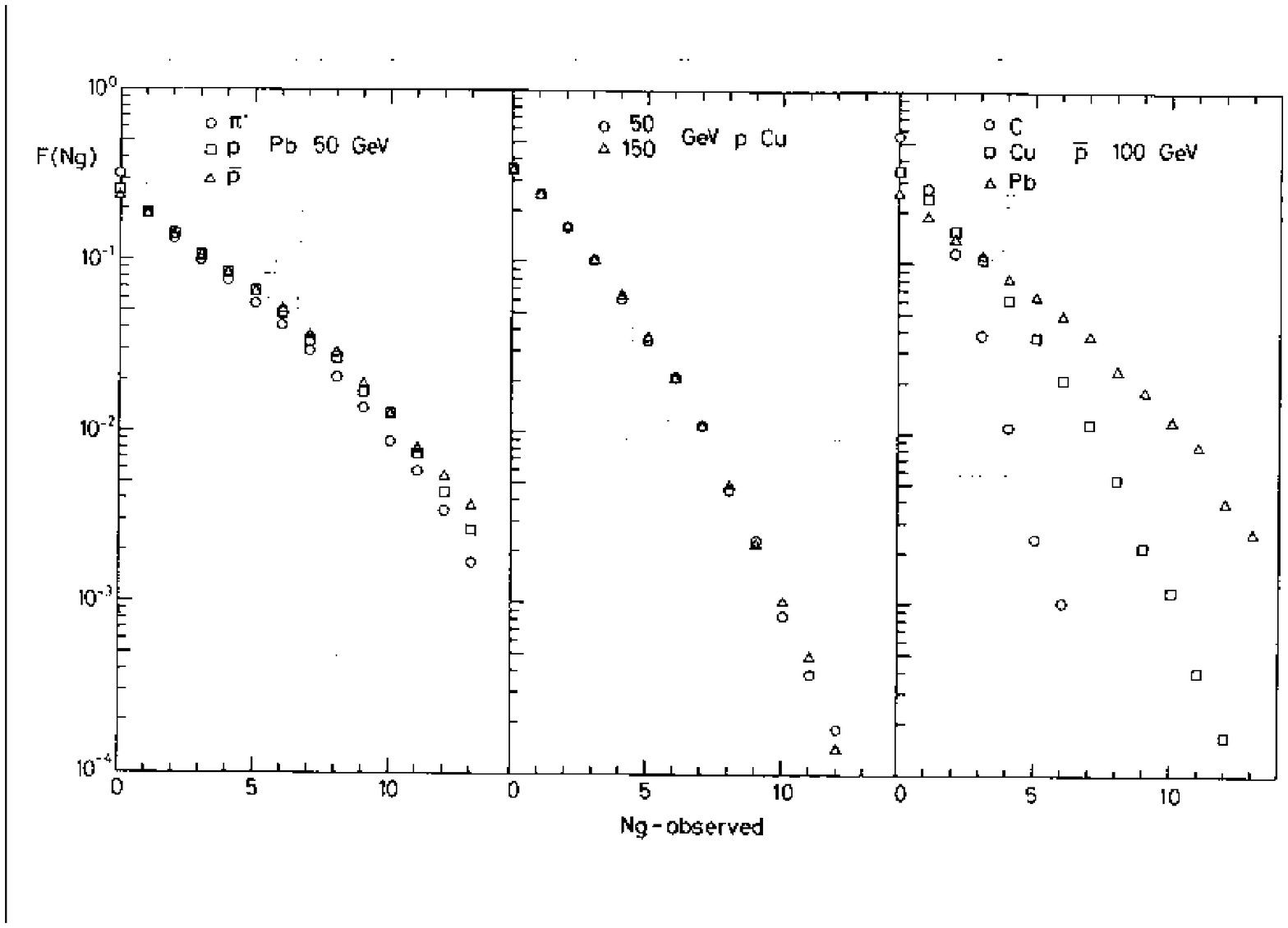}

 \caption{\label{fig:braune1} a) Angular distributions of slow
particles for different intervals of energy loss, $\Delta E$ is given in
units of the energy loss of minimum ionizing particles. b)
Multiplicity distributions of slow particles: projectile, energy and
target dependence. Both from Ref.~\cite{Braune:1982jb}.}

\end{figure}

\paragraph{CERN-WA35.} Slow particles from collisions of 50, 100 and 150
GeV pions, protons and antiprotons on C, Cu and Pb targets were measured
and thoroughly analyzed by CERN-WA35 at the SPS
\cite{Faessler:1979sn,Braune:1982jb,Braune:1983kp}. The slow particles
were separated and their energy loss was measured by many CsI scintillator
counters covering 52\% of the total solid angle, while the forward
direction is covered by lucite hodoscope. Angular distributions for higher
energy particles are strongly forward peaked, the lower energy ones
approach isotropy, reflecting the transition from direct, knock-out
protons to thermal ones. Angular distributions show a significant
dependence on $A$, they are stronger forward-peaked for lighter targets
than for heavier ones, while they do not depend on incoming energy
(Fig.\ref{fig:braune1}).

\begin{figure}
 \figs{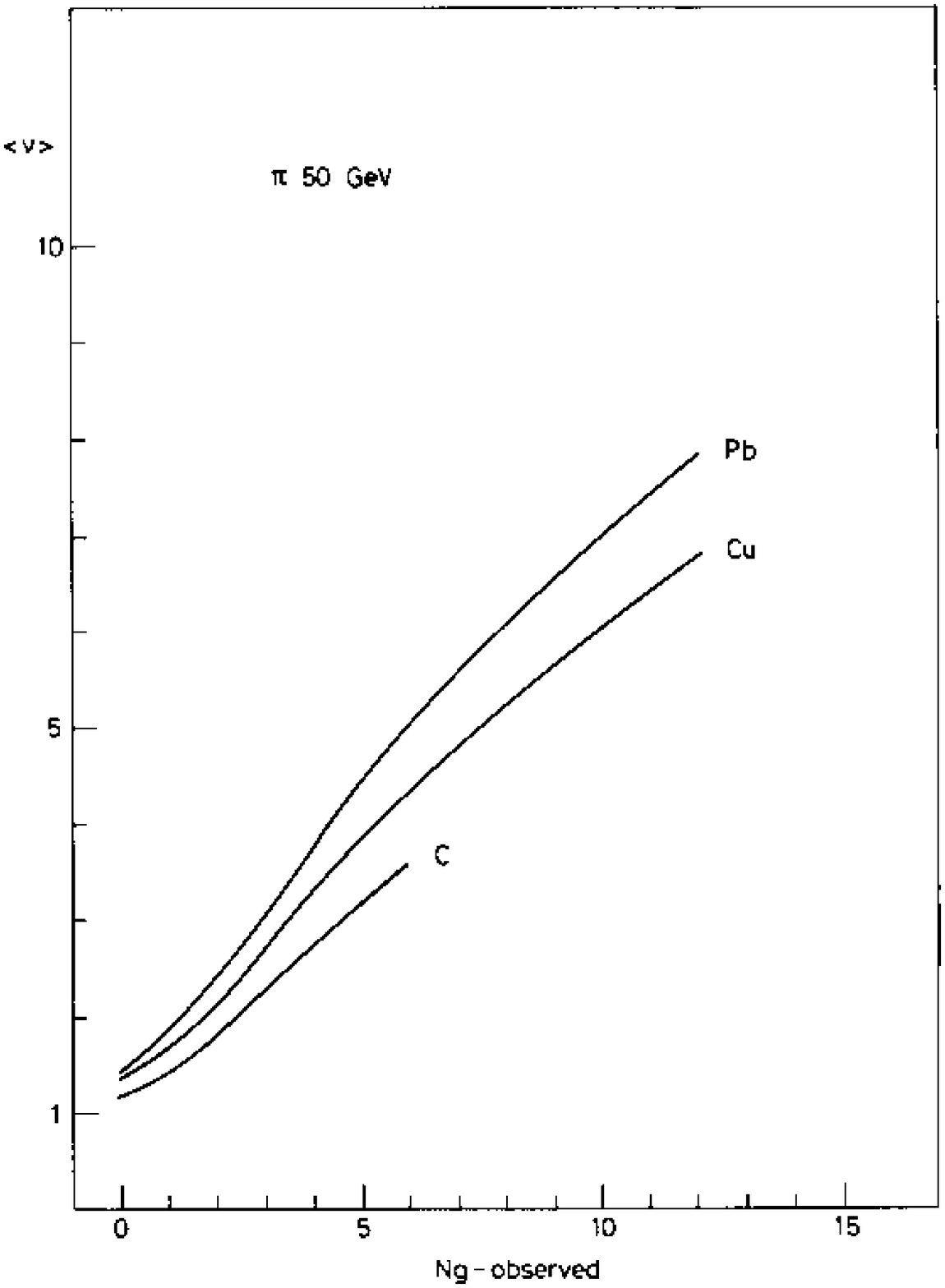}
      {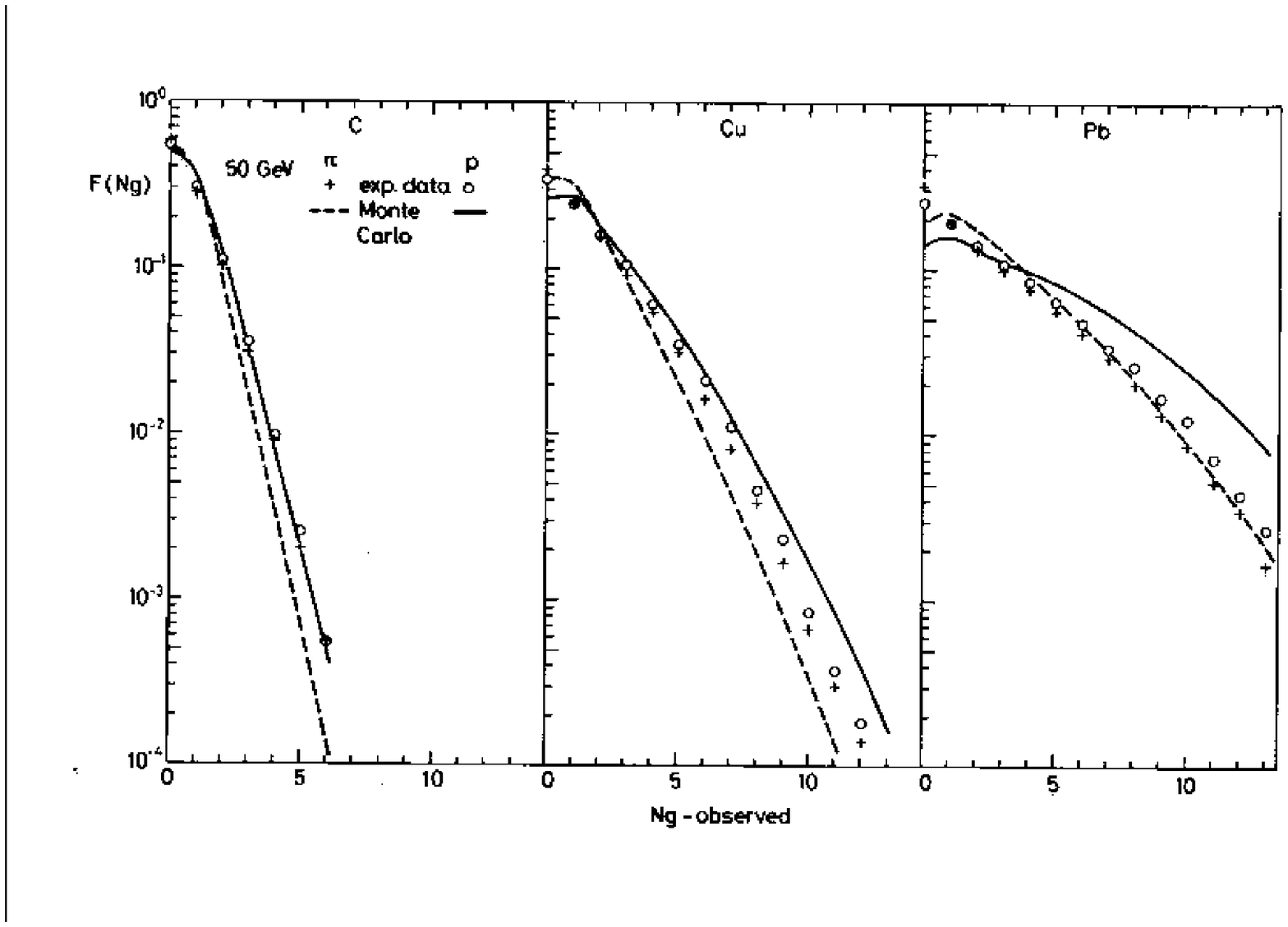}

%left) Calculated probability that an incident proton collides $\nu$ times 
%in C, Cu, and Pb nuclei

 \caption{\label{fig:braune2} a) The mean number of collisions
$\langle\nu\rangle$ as a function of the number of observed slow
particles. b) Comparison of experimental and calculated multiplicity
distributions of slow particles for incident pions and protons on C, Cu,
and Pb. Both from Ref.~\cite{Braune:1982jb}.}

\end{figure}

In their model the incoming projectile collides successively with nucleons
moving on a straight trajectory, each recoiling nucleon initiates an
internuclear cascade leading to slow nucleons. At a given impact parameter
the number of collisions $\nu$ are assumed to be Poisson-distributed with
the mean value obtained from simple Glauber-calculation with Woods-Saxon
density distribution. The distribution of slow particles $N_g$ emitted as
result of the cascade is assumed to be Poissonian with mean value fitted
to multiplicity distribution. From the model relation between $\langle \nu
\rangle$ and $N_g$ can be derived (Fig.~\ref{fig:braune2}a).

The dependence on projectile type is weaker than predicted by models, in
agreement with earlier emulsion experiments (Fig.~\ref{fig:braune2}b).
This could be cured if recoil nucleons and secondary pions can act as
independent hadrons in the nucleus, giving weight to the first collision. The
down-stream nucleons which may be hit by projectile are also likely to be
hit by the slow particles from the first interaction. For the extreme
assumption there could be only one large cascade. It is concluded that the
large cascade does not depend directly on the number of primary
collisions, but on the average depth of the first interaction and on the
remaining thickness of the nucleus at given impact parameter. Hence the
number of slow particles would measure the peripherality or centrality of
the collision.

\begin{figure}
 \includegraphics[width=\textwidth]
                 {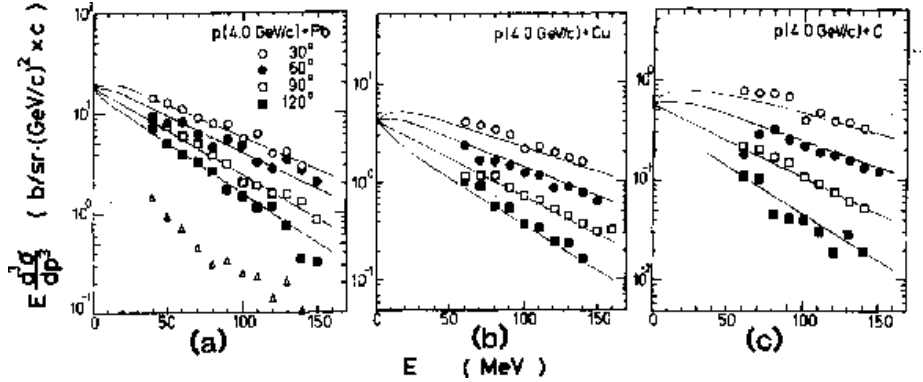} 

 \caption{\label{fig:nakai} Examples of proton spectra from hadron-nucleus
reactions. solid curves are spectra calculated assuming isotropic
emissions in a moving frame, from Ref.~\cite{Nakai:1983ut}.}

\end{figure}
 
\begin{figure}
 \figs{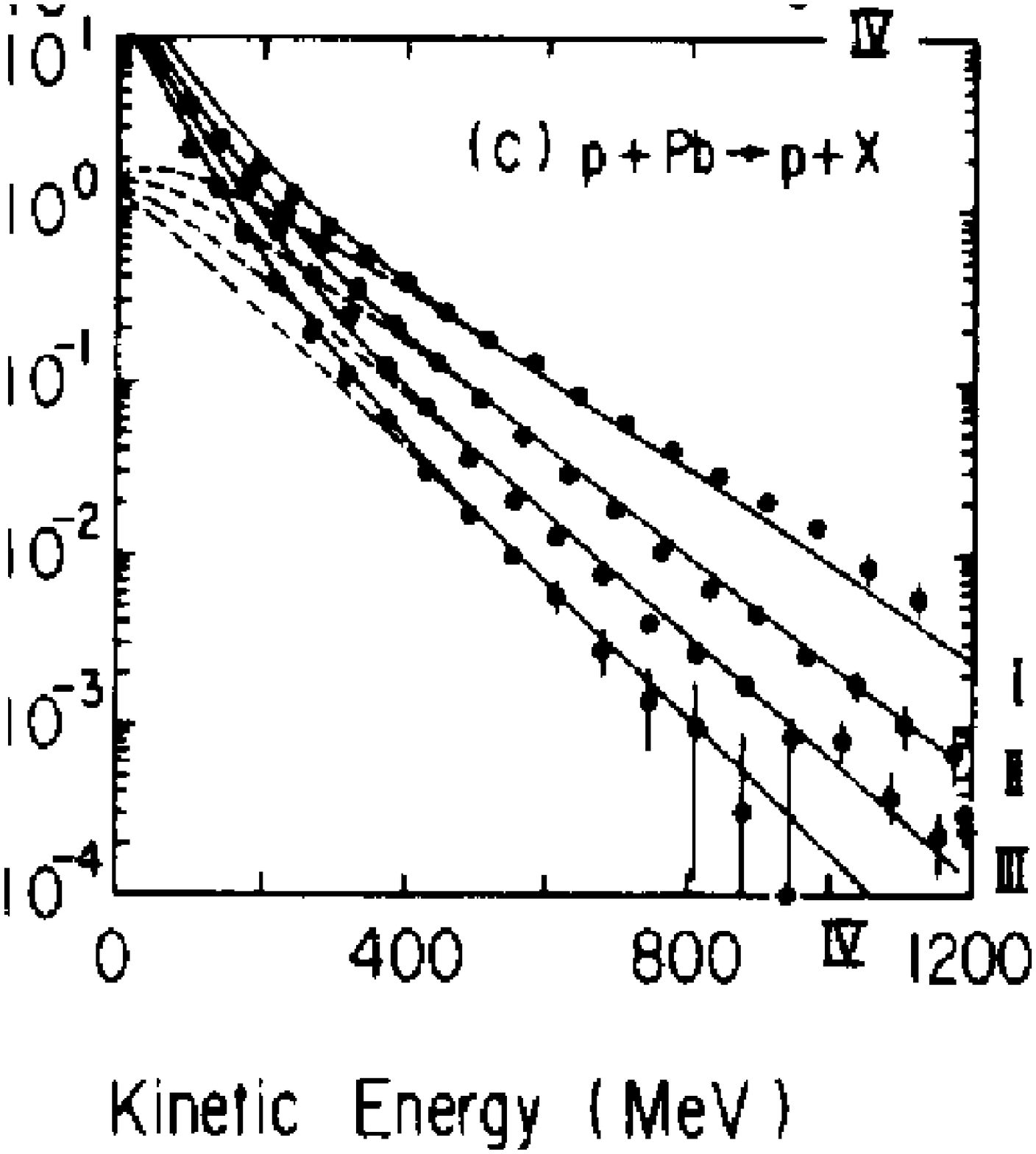}
      {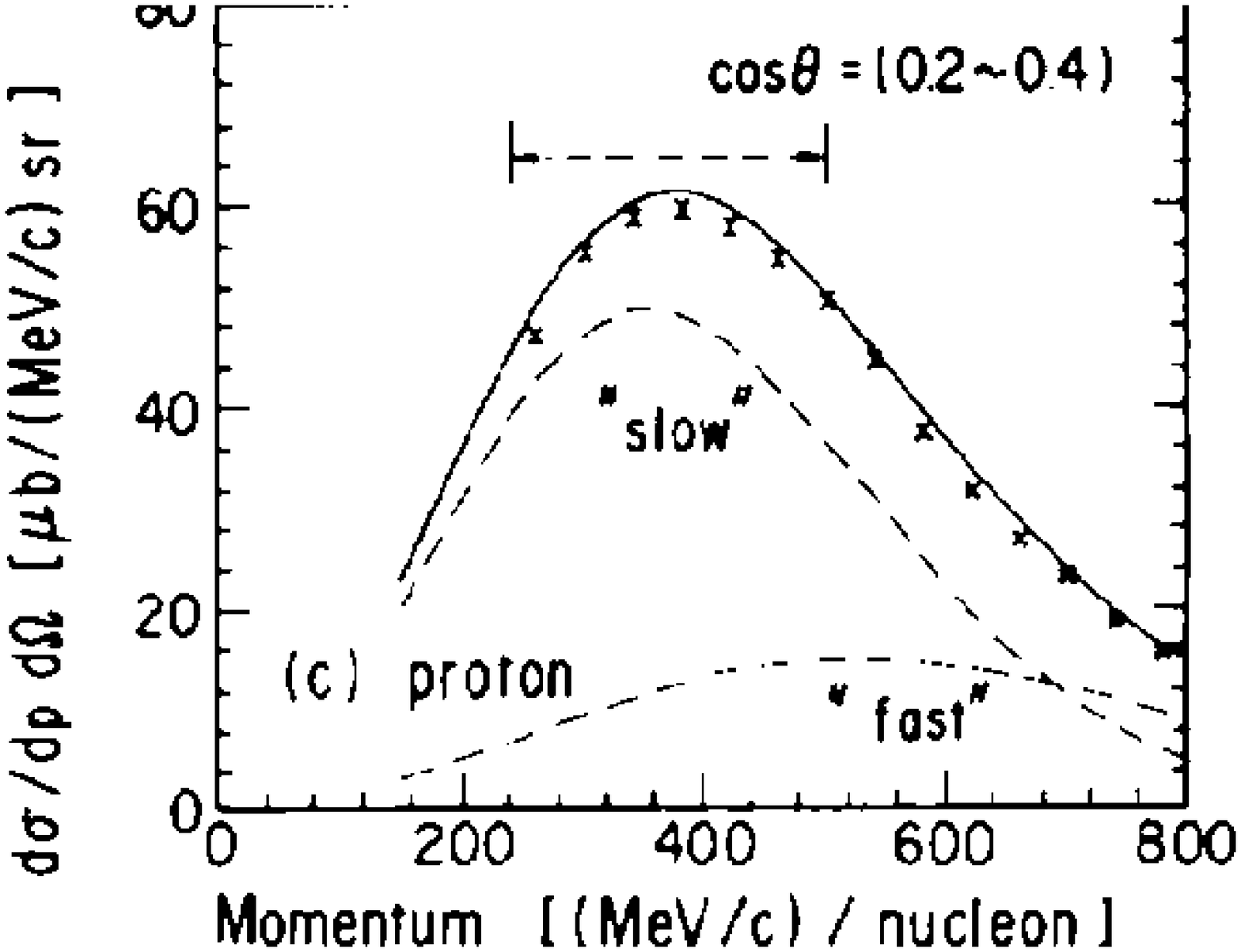}

 \caption{\label{fig:enyo} a). Proton spectra emitted in different angular
ranges. The fits with the two-moving-source model are shown with the solid
line, from Ref~\cite{Enyo:1985ze}. b) Decomposition of proton spectra with
the two-moving-source model, from Ref.~\cite{Tokushuku:1990sp}.}

\end{figure}

\paragraph{KEK-E90.} Spectra of protons and pions emitted in the
target-rapidity region in collisions of 1.5 to 4 GeV/$c$ pions and protons
on C, Cu and Pb targets have been measured with the FANCY spectrometer at
KEK \cite{Nakai:1983ut}. The target was surrounded by a cylindrical
multiwire proportional chamber and a trigger hodoscope. Proton spectra
were measured using $\Delta E$--$E$ scintillation telescopes, calibration
was done by time-of-flight measurements. Proton spectra are fitted
assuming isotropic emission of protons in a frame moving with a velocity
$\beta_s$ and with a spectrum of $E \d^3\sigma/\d p^3 = N_0
\exp(-E/E_0)$, with good agreement (Fig.~\ref{fig:nakai}). The results are
consistent with a model of a single-moving-source formation with
$\beta_s=0.1-0.2$ which decays by emitting protons with characteristic
energy $E_0$ of 40 to 70 MeV. While this experiment is sensitive only to
the gray proton component, they note that previous measurements observed
more low energy components with $E_0$ parameters 6-9 and 1.5-2 MeV,
respectively. From the $\beta_s$ source velocities the number of nucleons
$\nu$ involved in the formation of the source can be estimated with the
relation $\beta_s = p_{inc}/(E_{inc} + \nu M)$, where M is the nucleon
mass.

After pushing the momentum acceptance of the detector to a wider range
data were fitted with a two-moving-source model, thus introducing an
additional, fast source \cite{Enyo:1985ze} (Figs~\ref{fig:enyo}). The
parameters obtained are $E_0 = 50-60$ MeV, $\beta_s = 0.15-0.20$ for slow
protons, $E_0 = 125-165$ MeV and $\beta_s = 0.4-0.5$ for fast ones.
Further analysis concentrated on deuteron data and its understanding with
the coalescence model \cite{Tokushuku:1990sp}.

\begin{figure}
 \figs{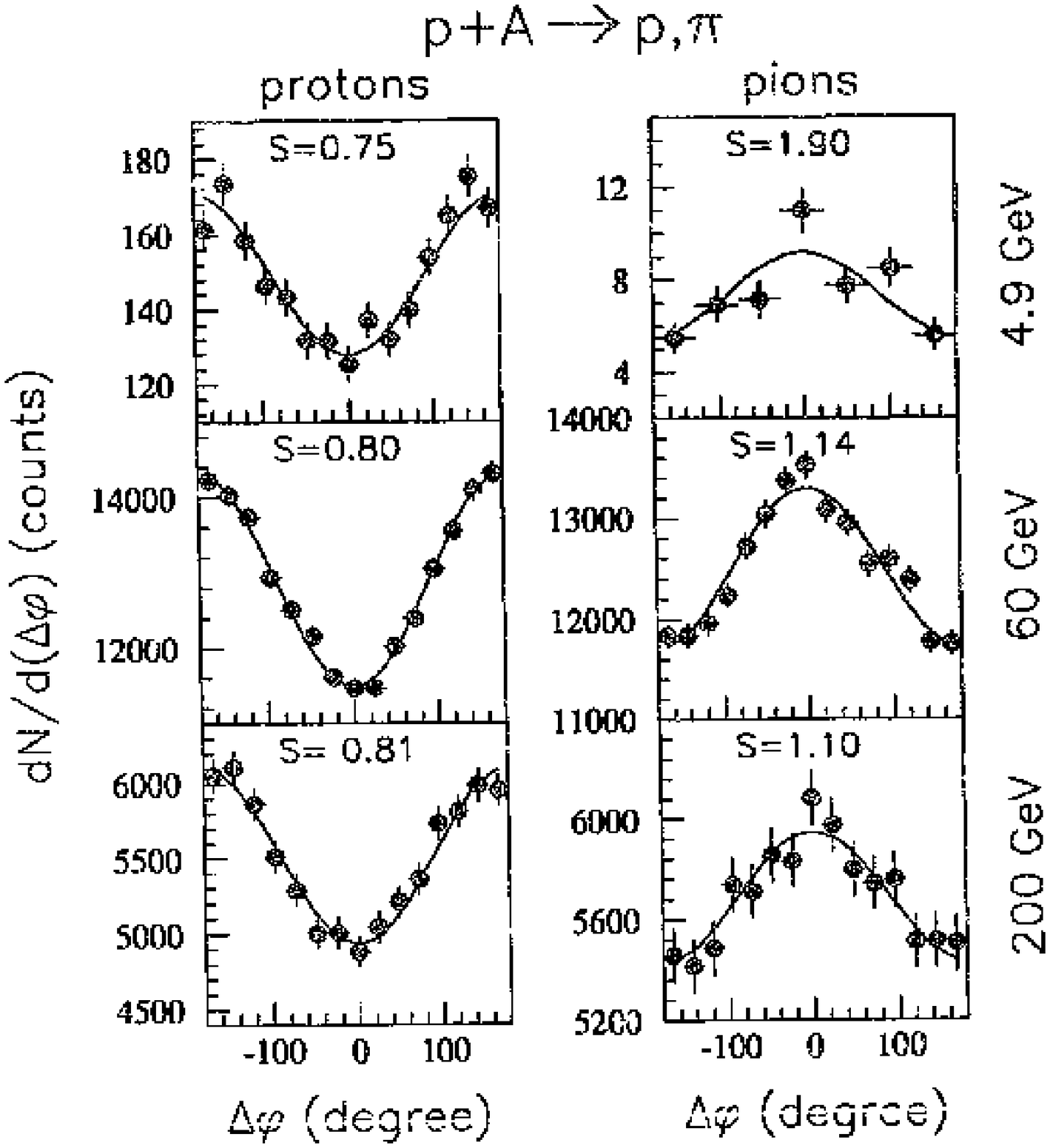}
      {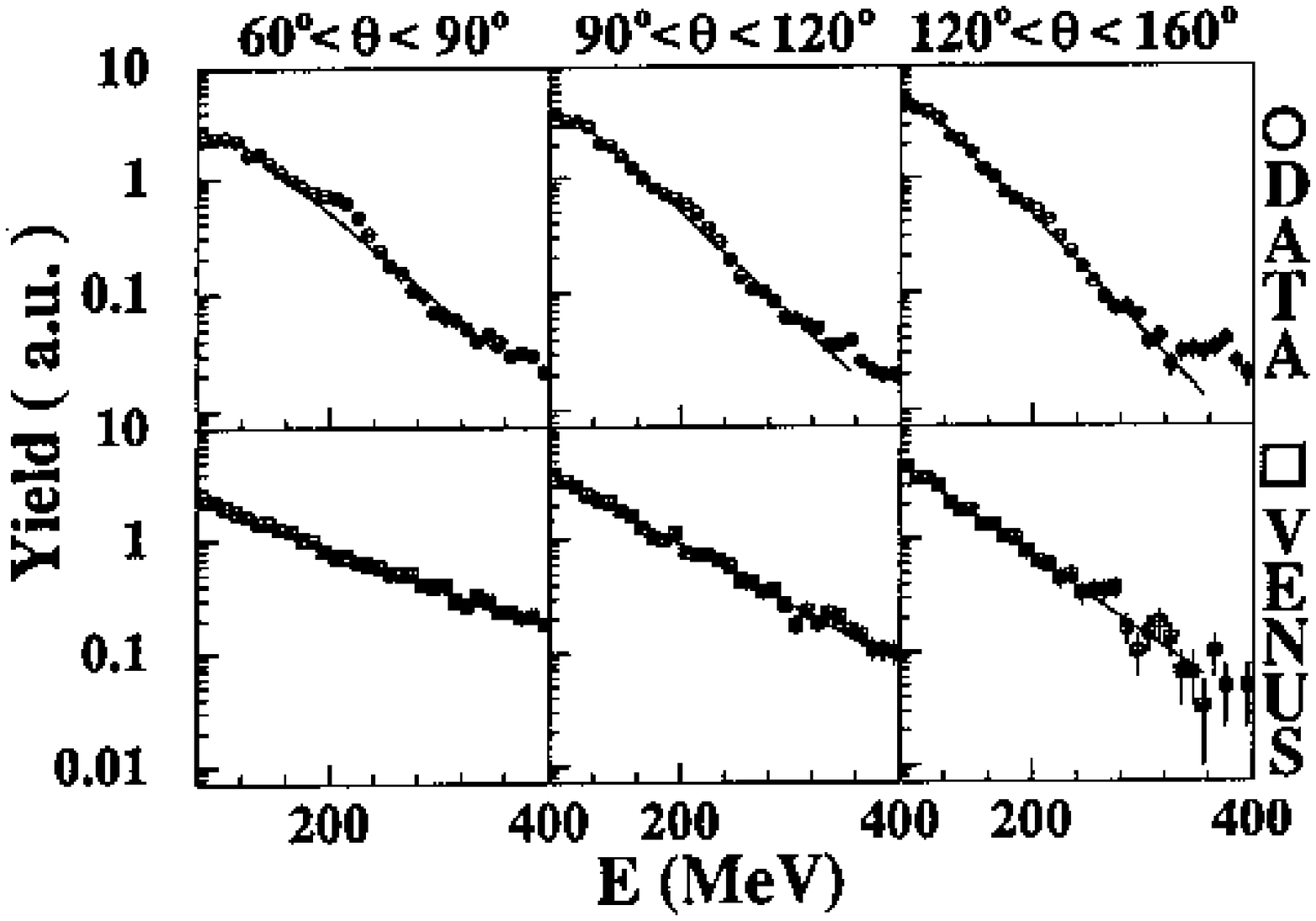}

 \caption{\label{fig:schmidt} a) The correlation function for p+Au
collisions at 4.9 (top), 60 (middle) and 200 (bottom) GeV, from
Ref.~\cite{Schmidt:1992ws}. b) Energy distributions for p+Au reactions in
three different $\theta$ bins. Data and the Venus model are compared. From
Ref.~\cite{Albrecht:1993aj}.}

\end{figure}

\paragraph{CERN-WA80.} Data on proton and pion induced reactions at 60 and
200 GeV/$c$ were taken by the WA80 experiment at CERN SPS with a number of
nuclear targets (C, Al, Cu, Ag and Au) \cite{Schmidt:1992ws}. The results
were obtained using the Plastic Ball spectrometer which consists of 665
particle identifying telescopes with $\Delta E$ -- $E$ detector, working
in the energy range of 30-400 MeV for protons. Comparisons to data taken
at 4.9 GeV bombarding energy at the LBL Bevalac with the same detector
have been done.

Emitted protons are preferentially back-to-back, while pions are emitted
side by side (Fig.~\ref{fig:schmidt}a). This observation is in accord with
pion absorption in the excited target spectator matter. While Venus model
fails to describe pions, the RQMD model seems to reproduce data, which
might be due to the fact that this one uses experimentally measured
cross-sections for $\pi$N reactions, dominated by $\Delta$ excitation.

For the determination of number-, angular- and energy distributions -- in
order to achieve essentially background free sample -- events with high
($E_T > 5$ GeV) transverse energy were selected \cite{Albrecht:1993aj}.
For the study the geometrical cascade model is adopted. The distribution
of number of collisions is calculated using a frozen straight line
geometry and a Woods-Saxon potential. Thus number distributions of slow
singly charged fragments are fitted with one free parameter, giving a good
description. The average number of slow protons produced in an interaction
goes like $\sqrt{A}$ indicating a larger intranuclear cascade.

Neither Fritiof nor Venus produces enough slow protons in comparison with
experimental observations. The energy distribution exhibits angular but not
target mass dependence. The shape of angular distributions is relatively
well described by the form $\exp(\kappa \cos\theta)$. The lighter targets
have a more forward peaked structure, because they are not large enough
for a cascade to become fully developed. Energy distributions are found to
have the form of $\exp(-E/E_0)$ with slope factors 40-60 MeV.

\begin{figure}
 \figs{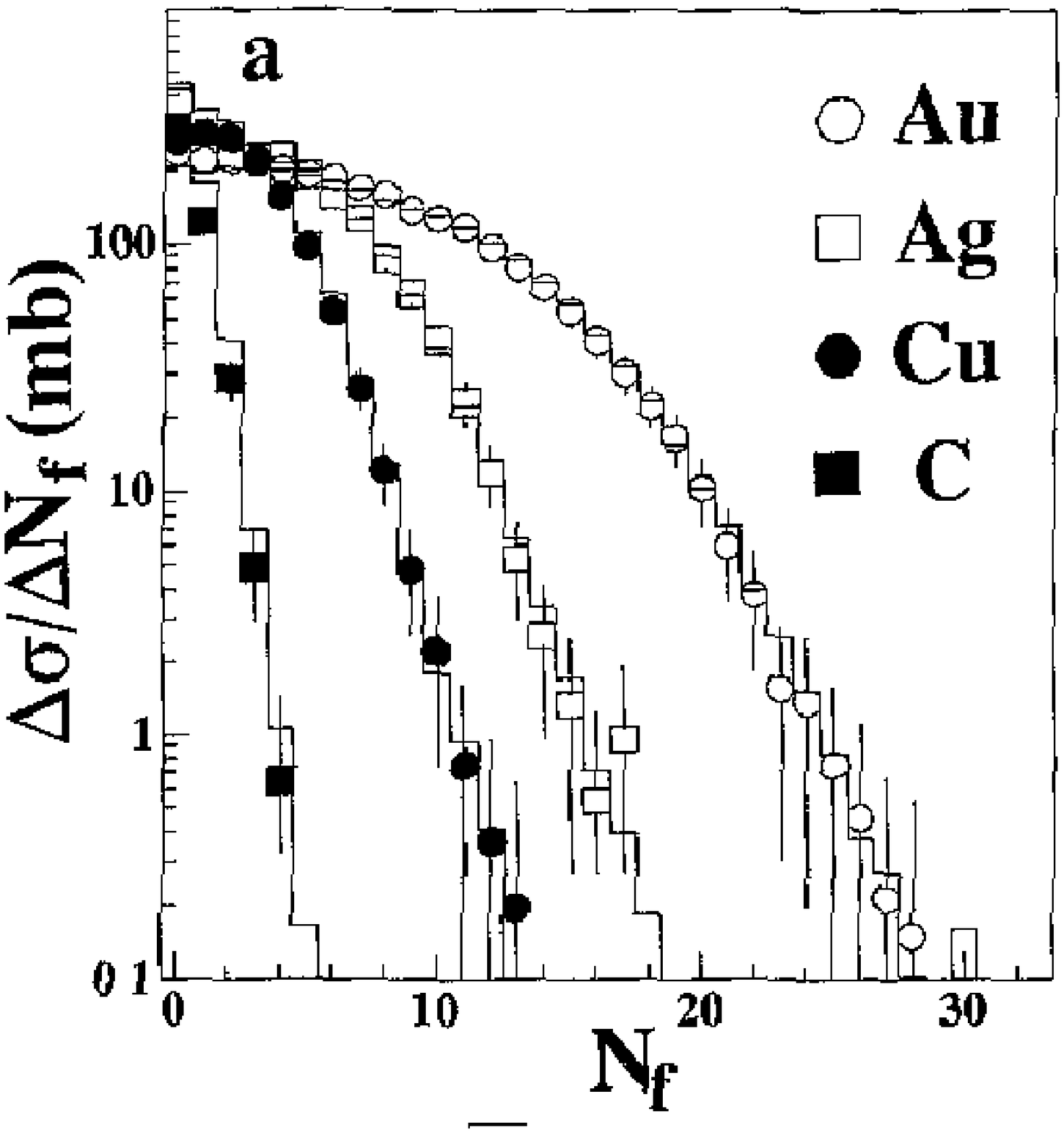}
      {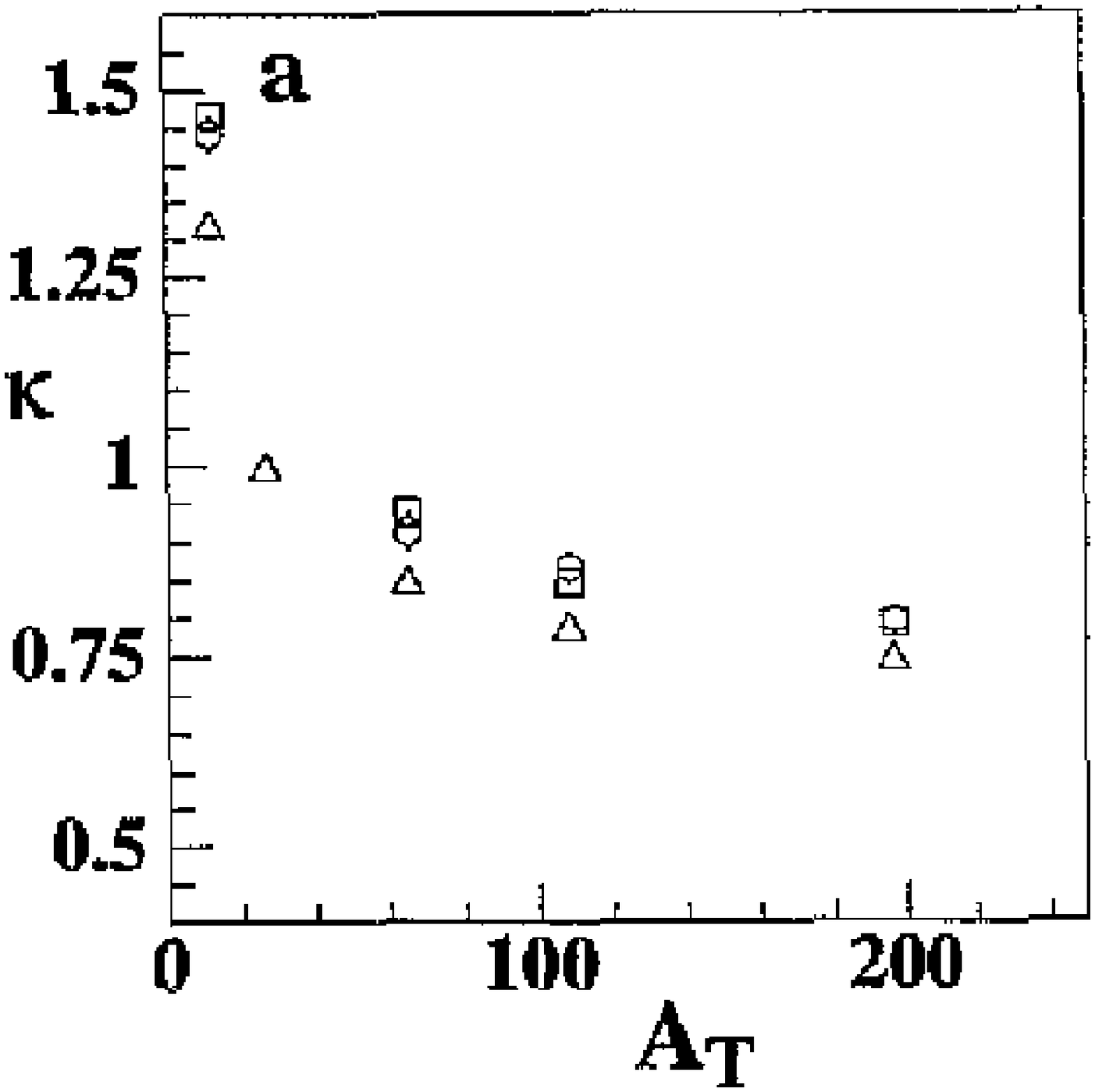}

 \caption{\label{fig:albrecht} a) Multiplicity distribution of slow
fragments in 200 and 60 $A$ GeV $^{16}$O induced reactions on C, Cu, Ag
and Au. b) $\kappa$ dependences in target mass for $^{16}$O-$A$ and p-$A$
collisions. Both from Ref~\cite{Albrecht:1993zi}.}

\end{figure}

Similar studies have been performed using $^{16}$O projectiles
\cite{Albrecht:1993zi} (Fig~\ref{fig:albrecht}).

%%%%%%%%%%%%%%%%%%%%%
% Streamer chambers

\subsection{Streamer chambers}

\paragraph{CERN-NA5.} Interactions of 200 GeV/$c$ protons on H, Ne, Ar and
Xe were studied with a streamer-chamber spectrometer by the CERN-NA5
experiment \cite{DeMarzo:1984eg}. Particles were identified based on
ionization and the charge of tracks. A rather good identification of
knocked out protons in the momentum interval 100 to 600 MeV/$c$ was
achieved.

In the analysis they adopt the geometrical cascade model and calculate the
probability distribution of the number of collisions using the Glauber
model. They find that the dependence of $\overline\nu$ on number of
knocked-out protons is not linear. The dispersion of the centrality
measure is also studied.

The number of collisions in the intranuclear cascade is estimated. Each
collision brings an extra positive charge when colliding with a proton, or
no extra charge when colliding with a neutron. Thus the total number of
collisions inside the nucleus is $\overline{\nu}_{tot} = \langle Q \rangle
\frac{A}{Z}$, where $\langle Q \rangle$ is the average net charge of all
observed secondaries. Having estimated the number of projectile collisions
$\overline{\nu}_p (N_g)$ one can deduce the average number of secondary
collisions $\overline{\nu}_s = \overline{\nu}_{tot} - \overline{\nu}_p$.
It is found that $\nu_s$ increases rapidly, roughly as $\nu_p^2$.

%%%%%%%%%%%%%%%%%%%%%
% Streamer chambers

\subsection{Recent experiments}

\begin{figure}
 \figs{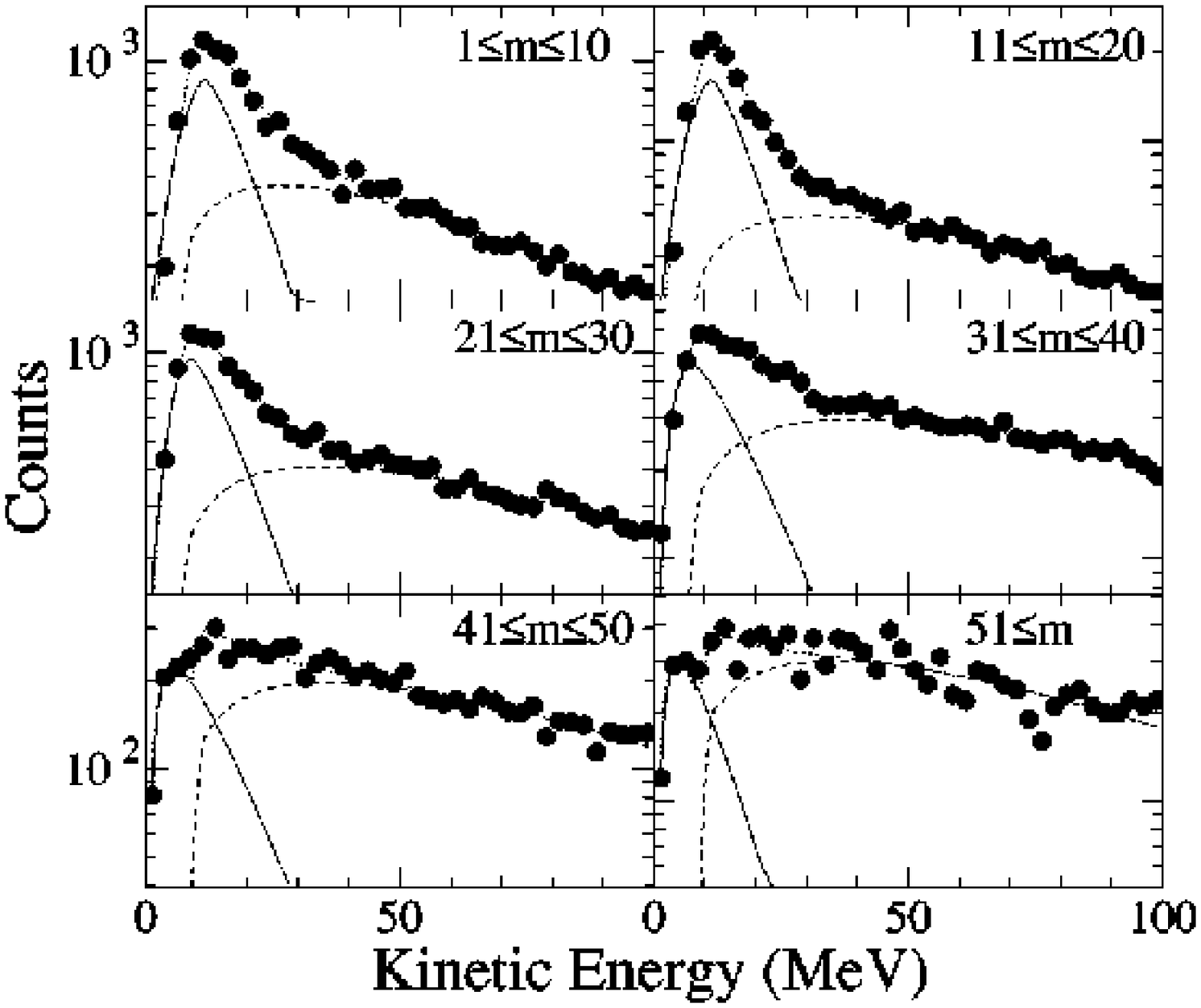}
      {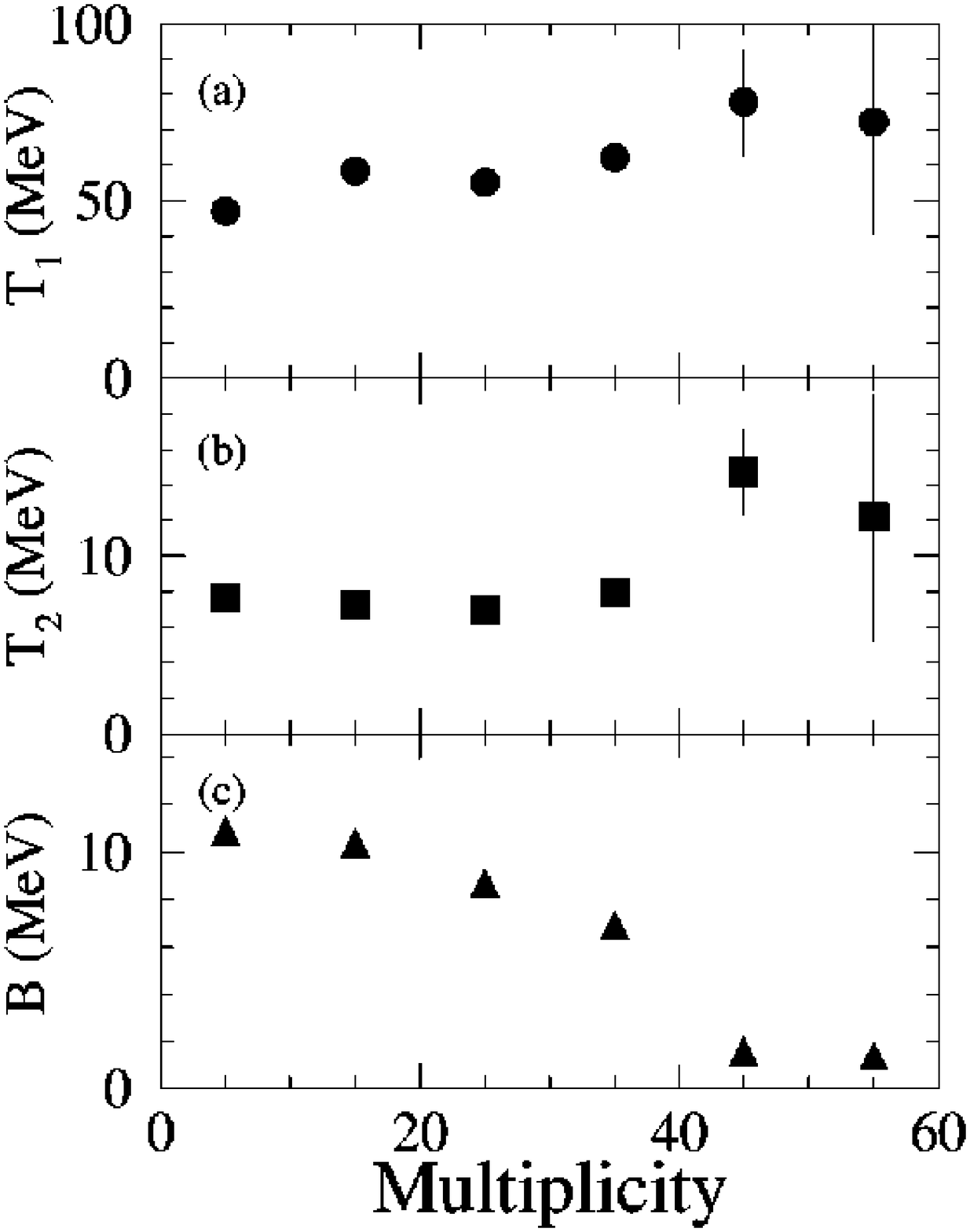}

 \caption{\label{fig:hauger} a) Two-stage fit to the proton kinetic energy
spectrum for indicated multiplicity intervals. b) Multiplicity dependence
of the parameters in the two-stage fit to the proton spectra. Both from
Ref.~\cite{Hauger:1998jg}.}

\end{figure}

\paragraph{LBL-E987.} A high-statistics study of the multifragmentation of
1$A$ GeV Au on C has been performed by LBL-E987 experiment at Bevalac,
using the EOS time projection chamber and a multiple sampling ionization
chamber \cite{Hauger:1998jg}. The proton kinetic energy spectra are fitted
with sum of two Maxwell-Boltzmann-like functions, one for each of the
reaction stages, and studied as function of multiplicity
(Fig.~\ref{fig:hauger}a). Slope parameters of intranuclear cascade with
energetic prompt particles are around 50 MeV, for the emission from an
equilibrated system are at about 8 MeV (Fig.~\ref{fig:hauger}b).

\paragraph{BNL-E900.} Energy deposition in 5-15 GeV/$c$ proton, $\pi^-$
and antiproton induced reactions on Au have been studied by experiments
BNL-E900 and BNL-E900a at the AGS \cite{Lefort:2001pa}. Results were
obtained with the ISiS silicon sphere $4\pi$ charge-particle detector
array. In order to perform the separation between the thermal-like and
nonequilibrium particles, two-component moving source fits, as function of
angle, are performed. For the former one source velocity of $\beta\le
0.01$ is found.

\paragraph{BNL-E910.} Slow protons and deuterons from collisions of 18
GeV/$c$ protons with Be, Cu and Au targets have been measured by BNL-E910
experiment at the AGS, performing a detailed analysis of the data
\cite{Chemakin:1999jd}. Charged particles were tracked by the EOS time
projection chamber in magnetic field supplemented by three drift chambers
and time-of-flight walls.\footnote{The targets were rather thick, 4 mm
for the Au, which means that most of slow particles are stopped in the
target material. The effect is clearly visible in the momentum distributions
and the very steep angular distributions.}

\begin{figure}
 \figs{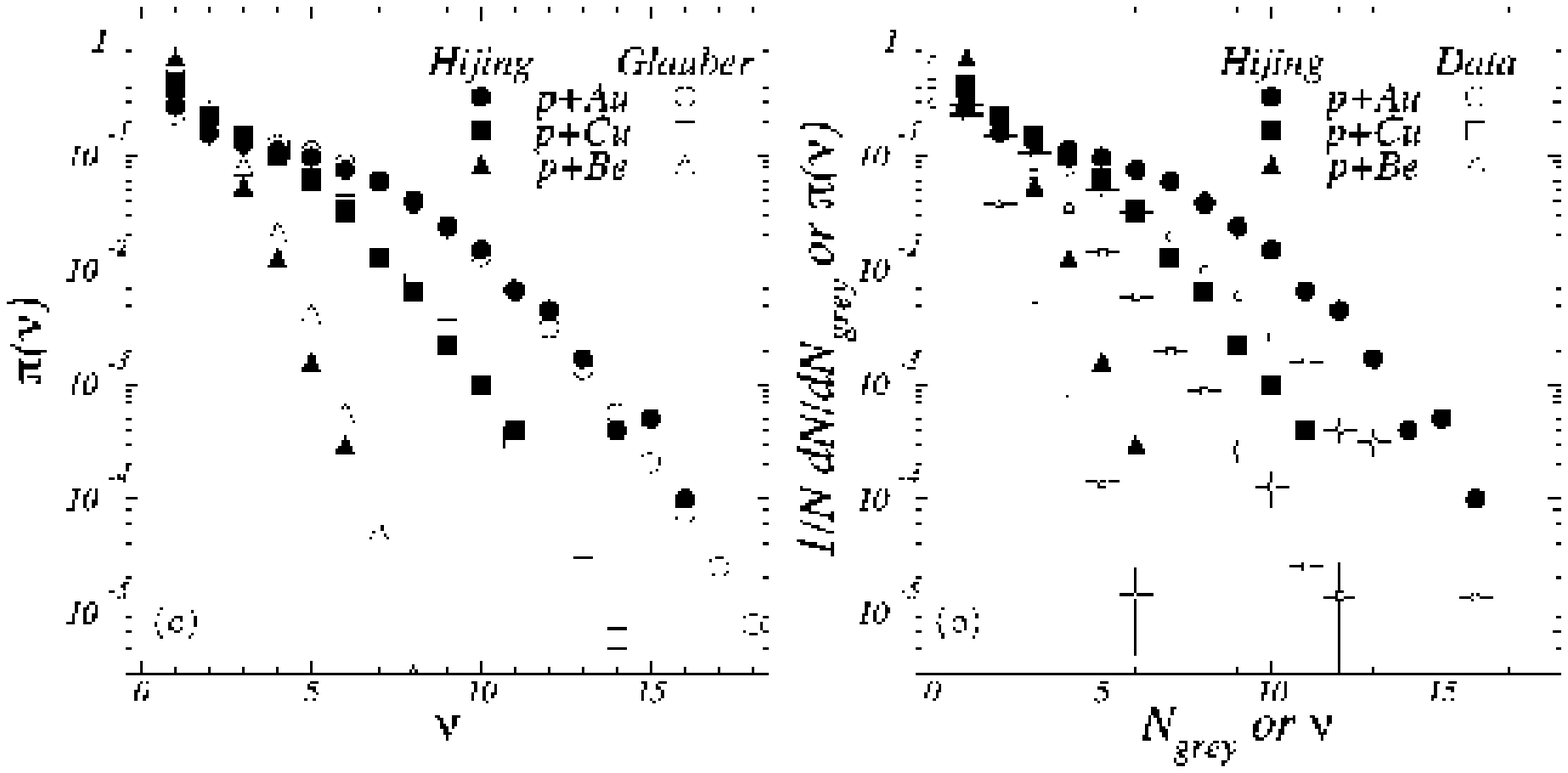}
      {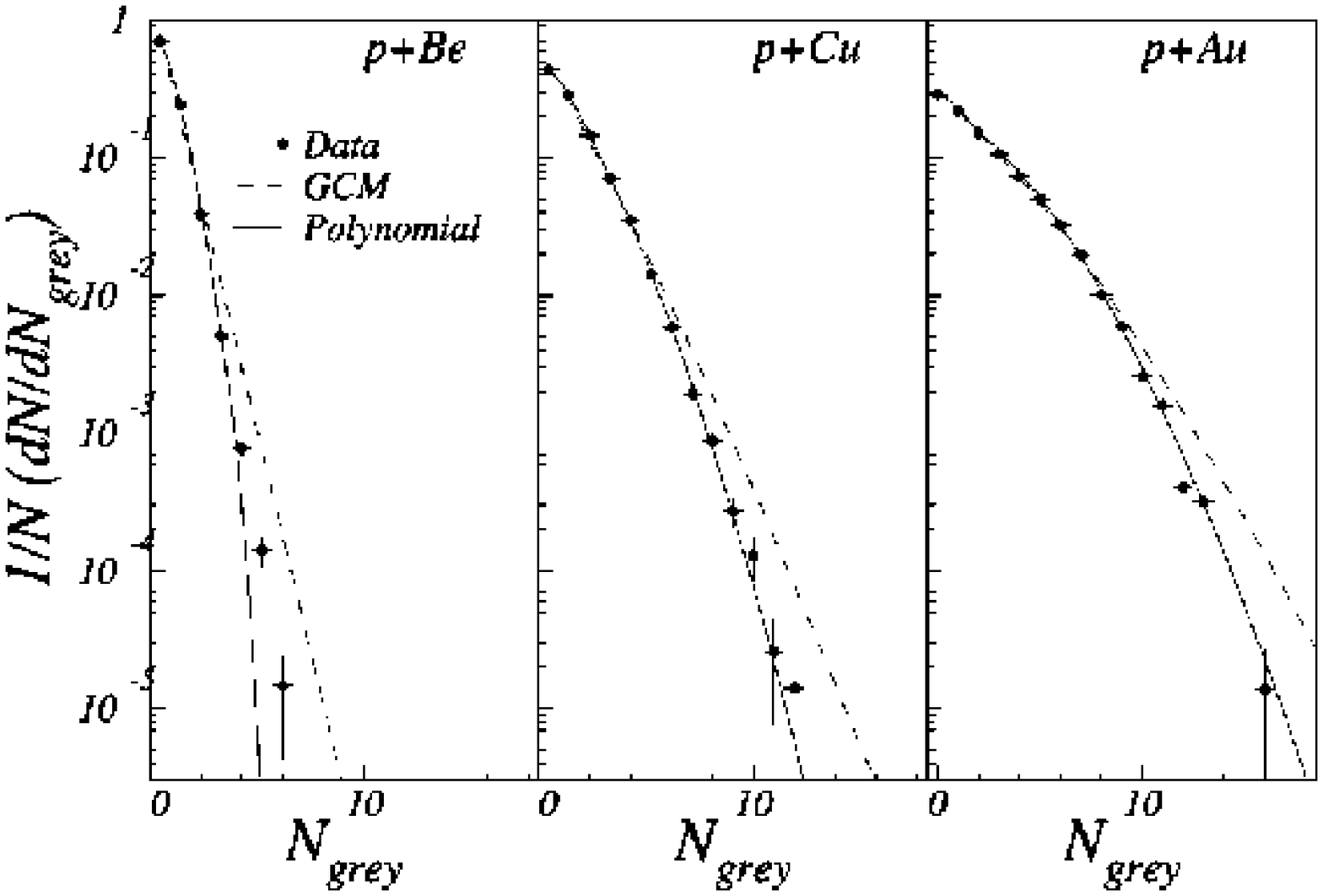}

 \caption{\label{fig:chemakin1} a) Probability distributions for the beam
proton to encounter $\nu$ collisions with target nucleons calculated for
p+Be, p+Cu, and p+Au reactions using Glauber and Hijing models. On the
right, distributions are overlayed with the $N_{grey}$ distributions. b)
Log-likelihood fits to event normalized $N_{grey}$ distributions with the
geometric cascade and the polynomial model. Both from
Ref.~\cite{Chemakin:1999jd}.}

\end{figure}

\begin{figure}
 \figs{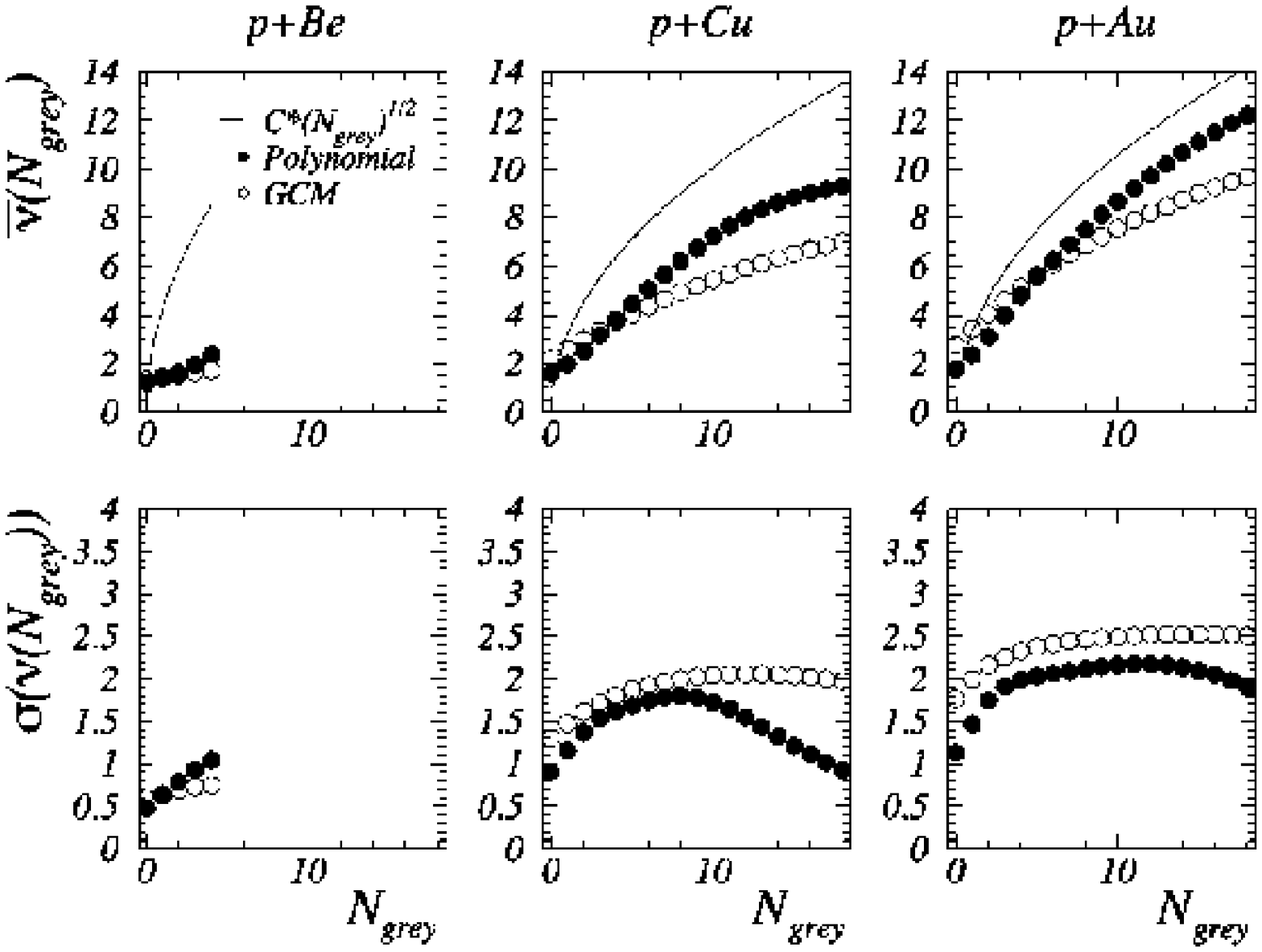}
      {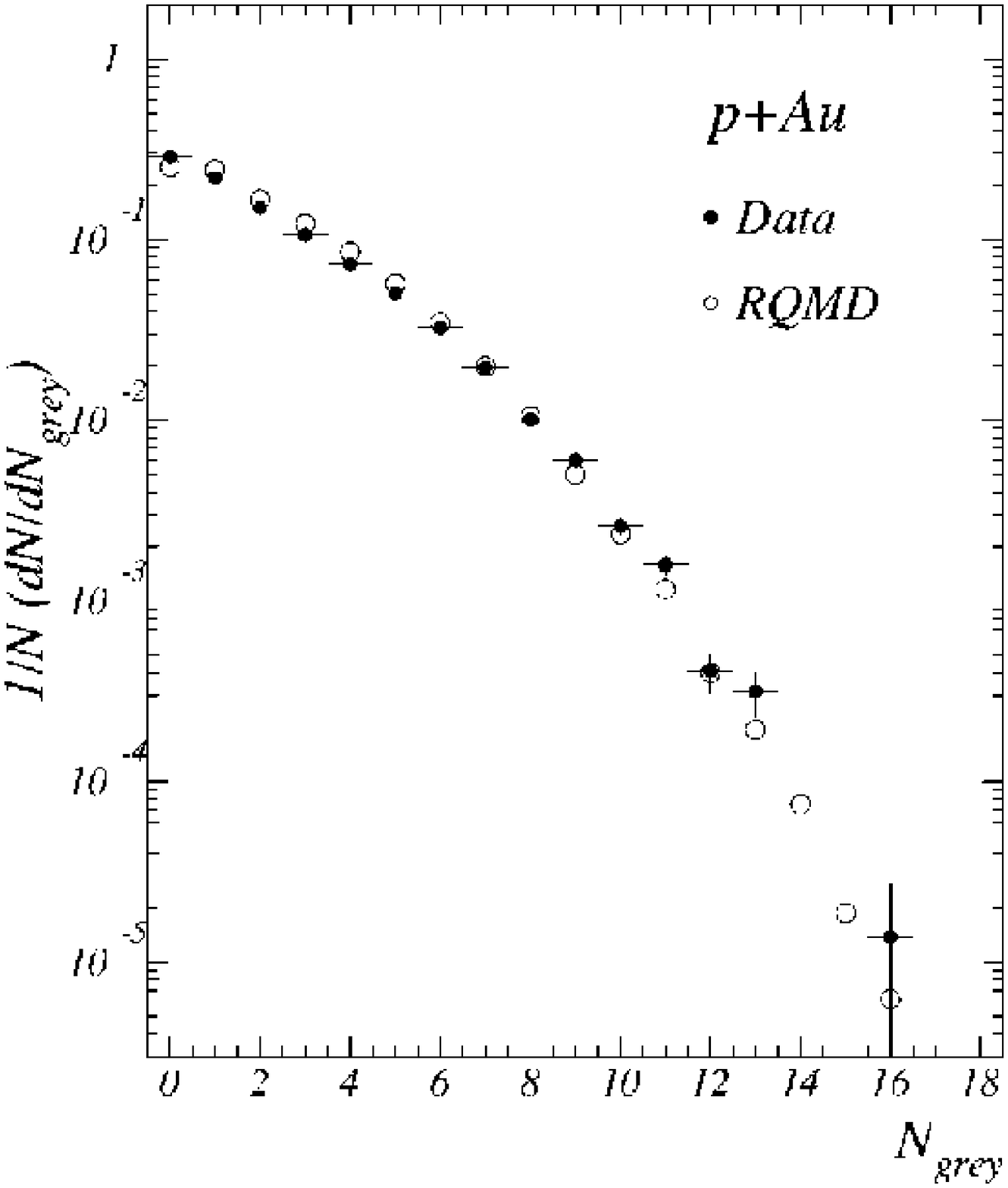}

 \caption{\label{fig:chemakin2} a) $\overline\nu(N_{grey})$ and $\sigma$
generated from the polynomial and geometric cascade models and according
to the $\overline\nu^2$ ansatz. b) Comparison between event normalized
slow fragment multiplicity distributions for p+Au reactions obtained from
data and RQMD calculations. Both from Ref.~\cite{Chemakin:1999jd}.}

\end{figure}

Very detailed studies on the distribution of number of collisions are
given by comparing the simple Glauber-calculation to Hijing model
(Fig.~\ref{fig:chemakin1}). For the description of number distribution of
grey protons a new polynomial model is proposed which incorporates both
geometrical and intranuclear cascade models: a strong linear dependence of
$\overline{N_g}$ on $\nu$ is found. RQMD reproduces distributions and
exhibits similar dependence (Fig.~\ref{fig:chemakin2}).

\paragraph{CERN-NA49.} Hadron-nucleus collisions at 158 GeV/$c$ have been
studied by CERN-NA49 experiment \cite{Afanasiev:1999}. Gray protons have
been detected with a gas detector consisting of proportional tubes
surrounding the target. Data are compared to known angular distributions
and with events from the Venus generator.

\begin{figure}
 \figs{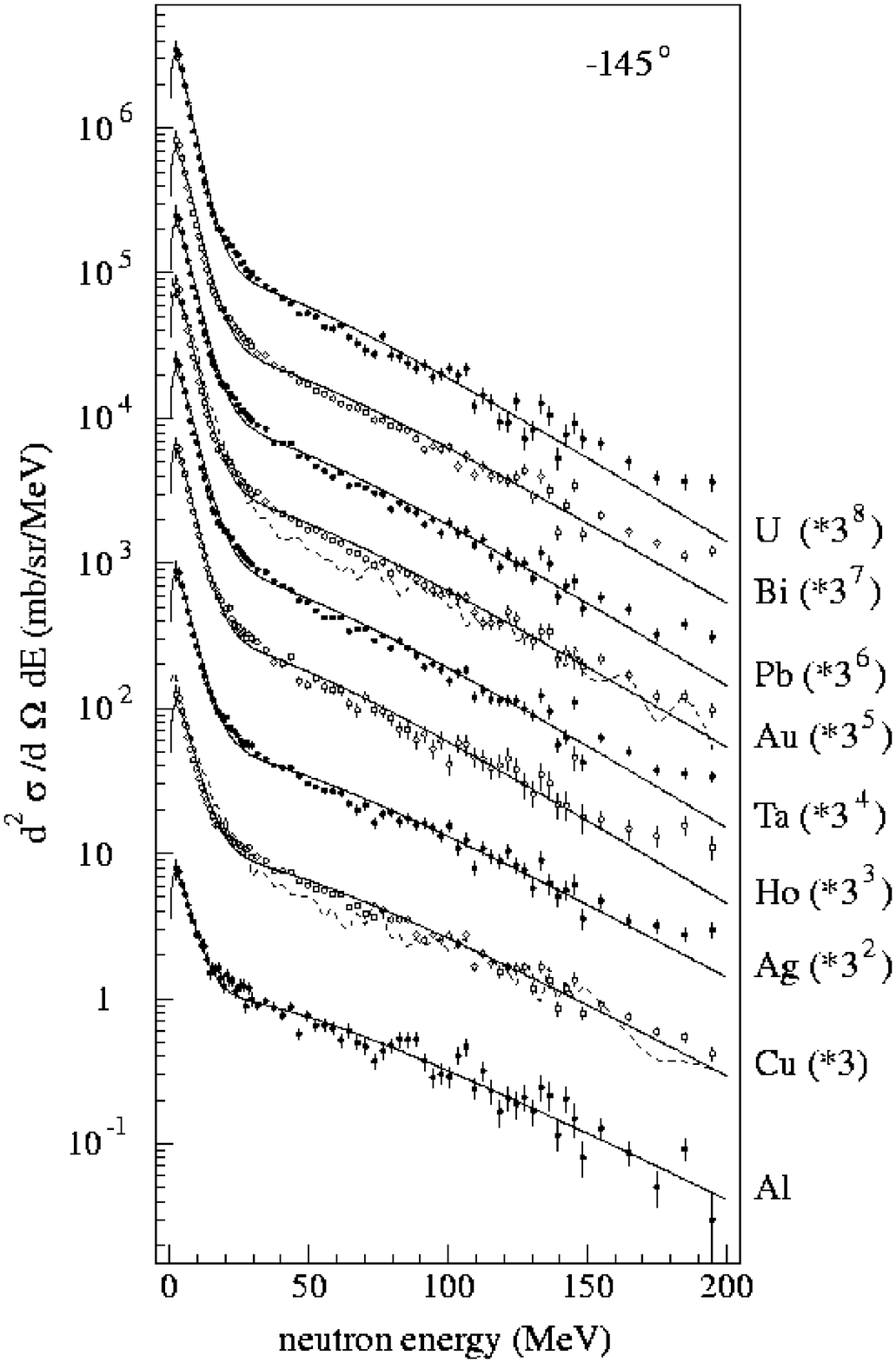}
      {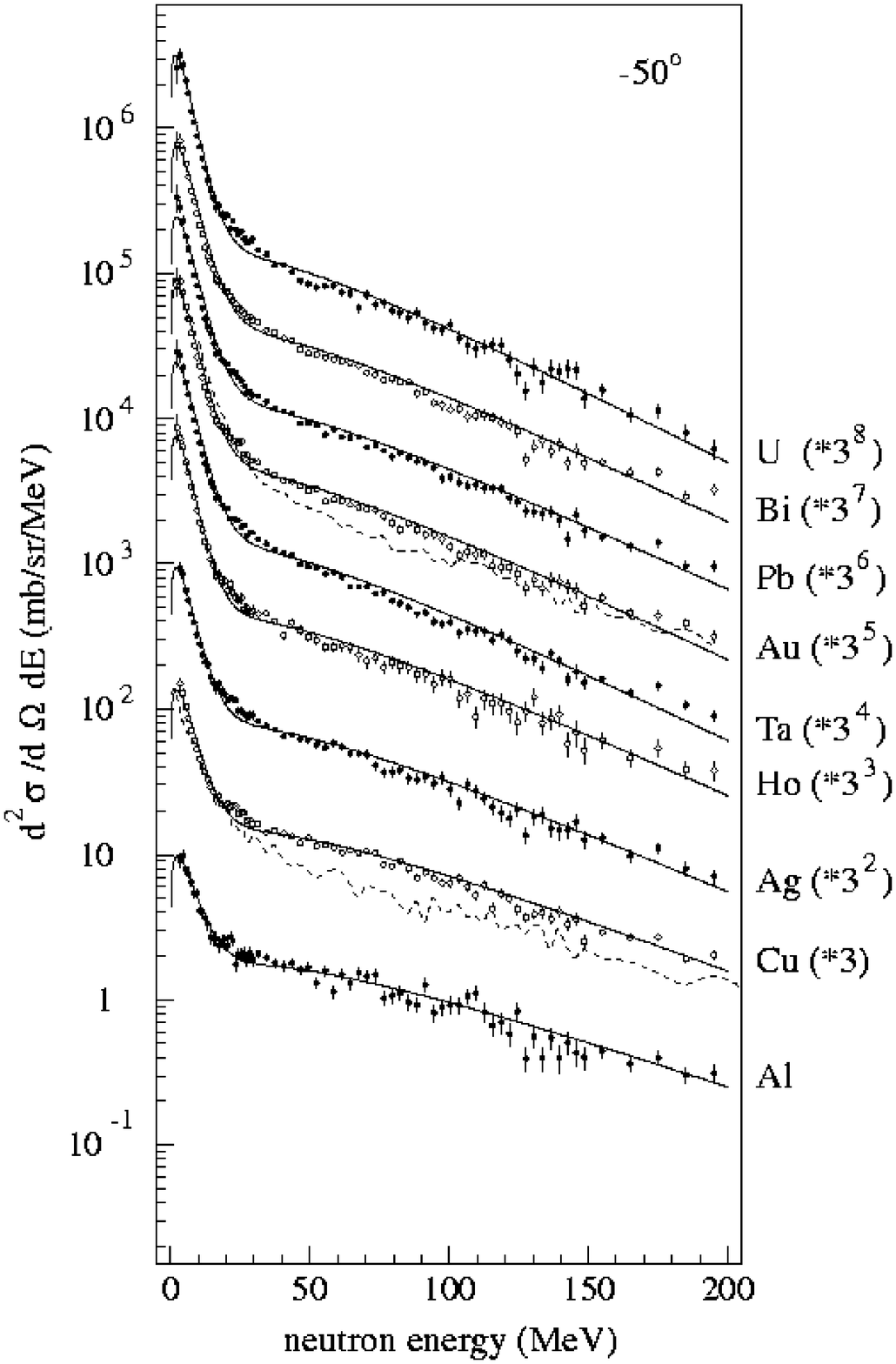}

 \caption{\label{fig:vonegidy1} Double differential inclusive neutron
production cross-sections vs $E$ for Al, Cu, Ag, Ho, Ta, Au, Pb, Bi and U
targets at emission angles of $-145^\circ$ and $-50^\circ$. Maxwellian
fits (solid) and intranuclear calculations (dashed) are also shown, from
Ref.~\cite{vonEgidy:2000kr}.}

\end{figure}

\begin{figure}
 \figs{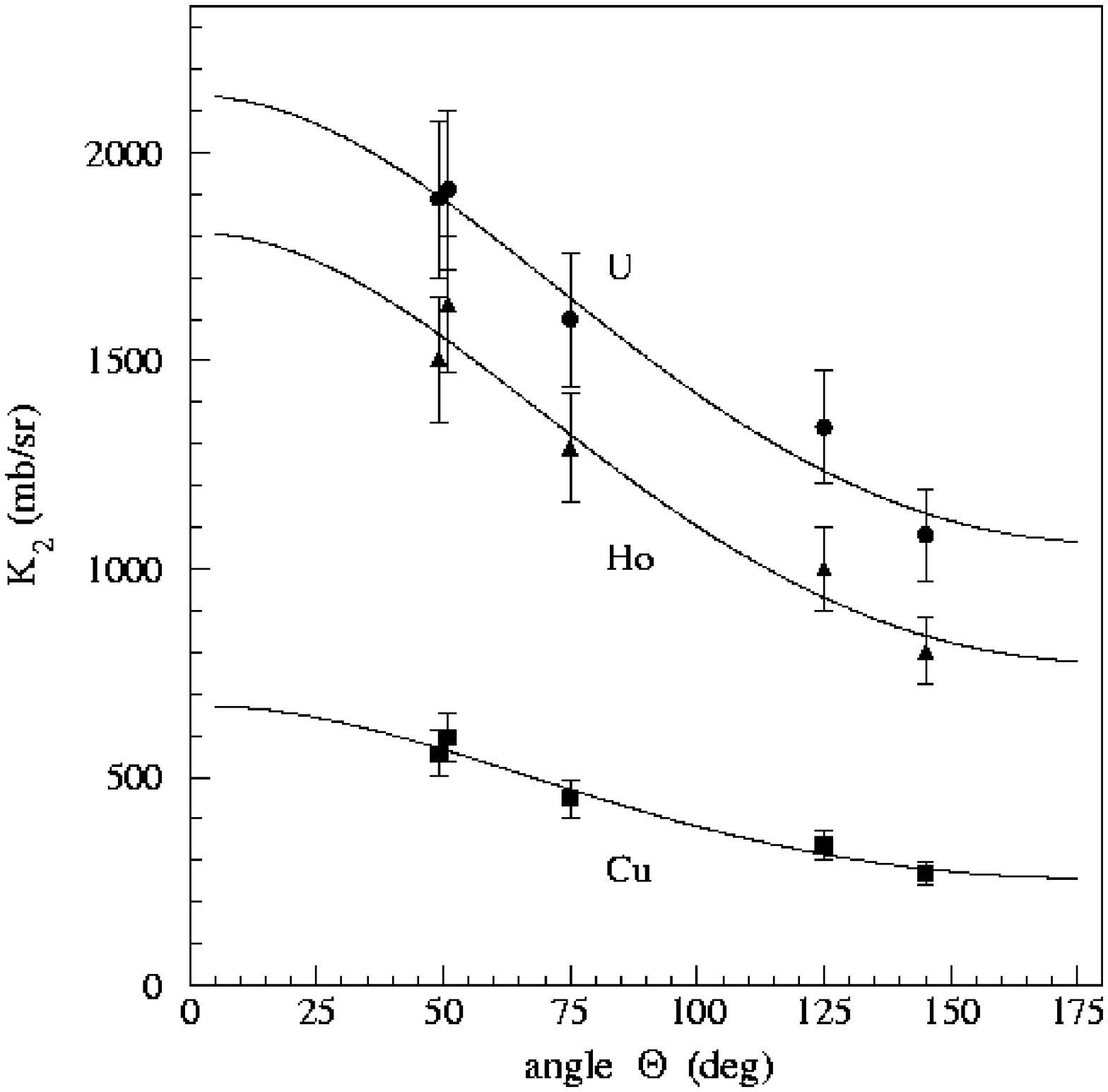}
      {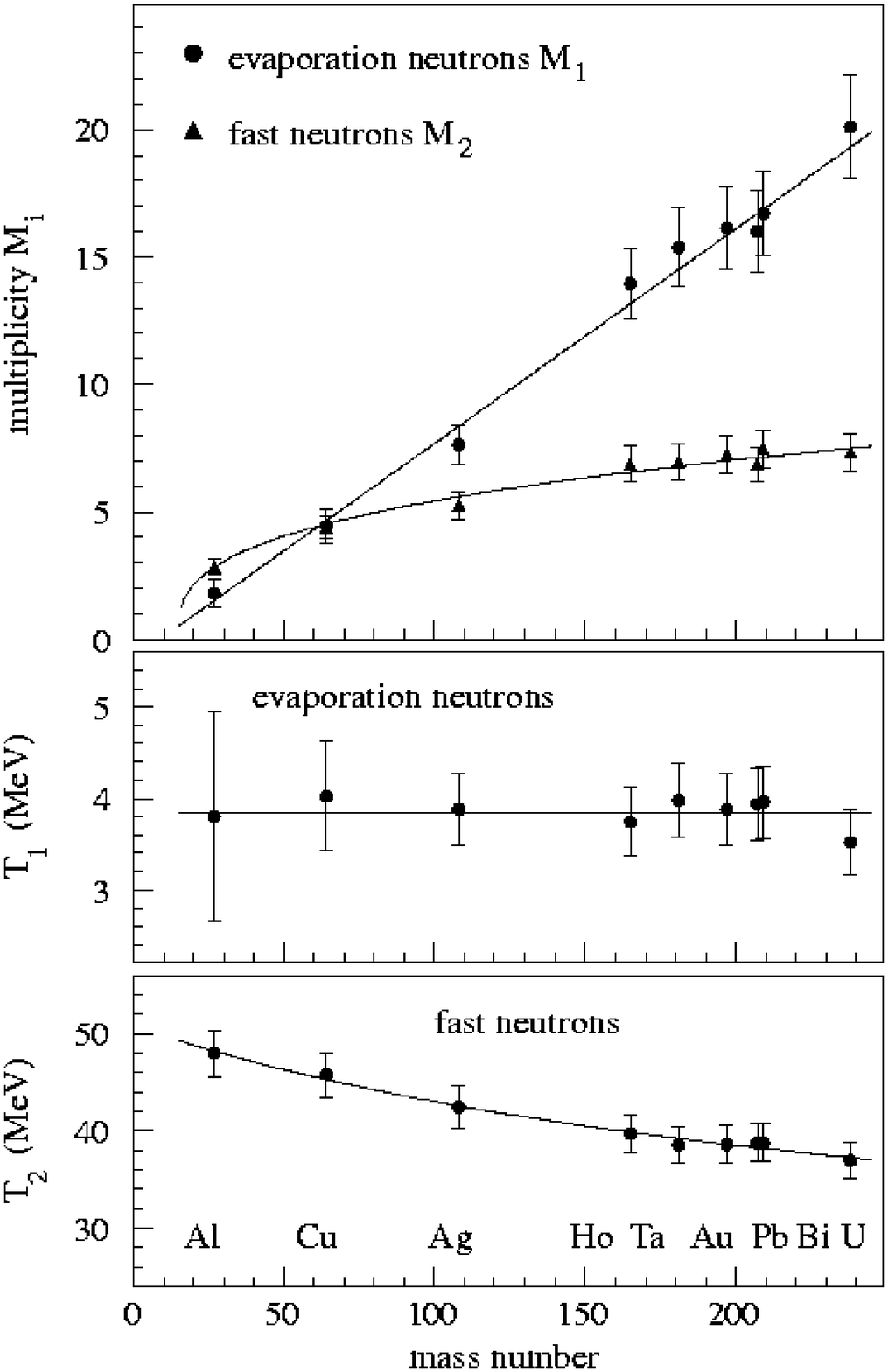}

 \caption{\label{fig:vonegidy2} a) Angular distribution of the fast
neutrons in Cu, Ho and U targets. b) Target mass dependence of the neutron
multiplicities and the inverse slope parameters. Both from
Ref~.\cite{vonEgidy:2000kr}.}

\end{figure}

\paragraph{CERN-PS208.} Inclusive neutron spectra were measured by time of
flight using 1.22 GeV antiprotons from LEAR on variety of targets -- Al,
Cu, Ag, Ho, Ta, Au, Pb, Bi and U -- by the PS208 experiment at CERN
\cite{vonEgidy:2000kr}. Sum of two Maxwellian distributions are fitted to
the spectra obtained at several angles yielding total multiplicities and
slope for the low-energy evaporative and high-energy pre-equilibrium parts
(Fig.~\ref{fig:vonegidy1}). While the former component increases with $A$,
the latter grows with $A^{1/3}$, proportional to the nuclear radius. The
behavior indicates that the path length of the pions or fast nucleons in
the nucleus is responsible for neutron emission. The slope parameters are
nearly independent of $A$: for all targets the evaporation temperature
remains constant around 4 MeV. The slope parameter of the fast neutron
spectra is decreasing from Al ($\approx 48$ MeV) to Ag and then remains
constant near 39 MeV for heavier targets (Fig.~\ref{fig:vonegidy2}). The
agreement with intranuclear cascade calculations is good.

\begin{figure}
 \figs{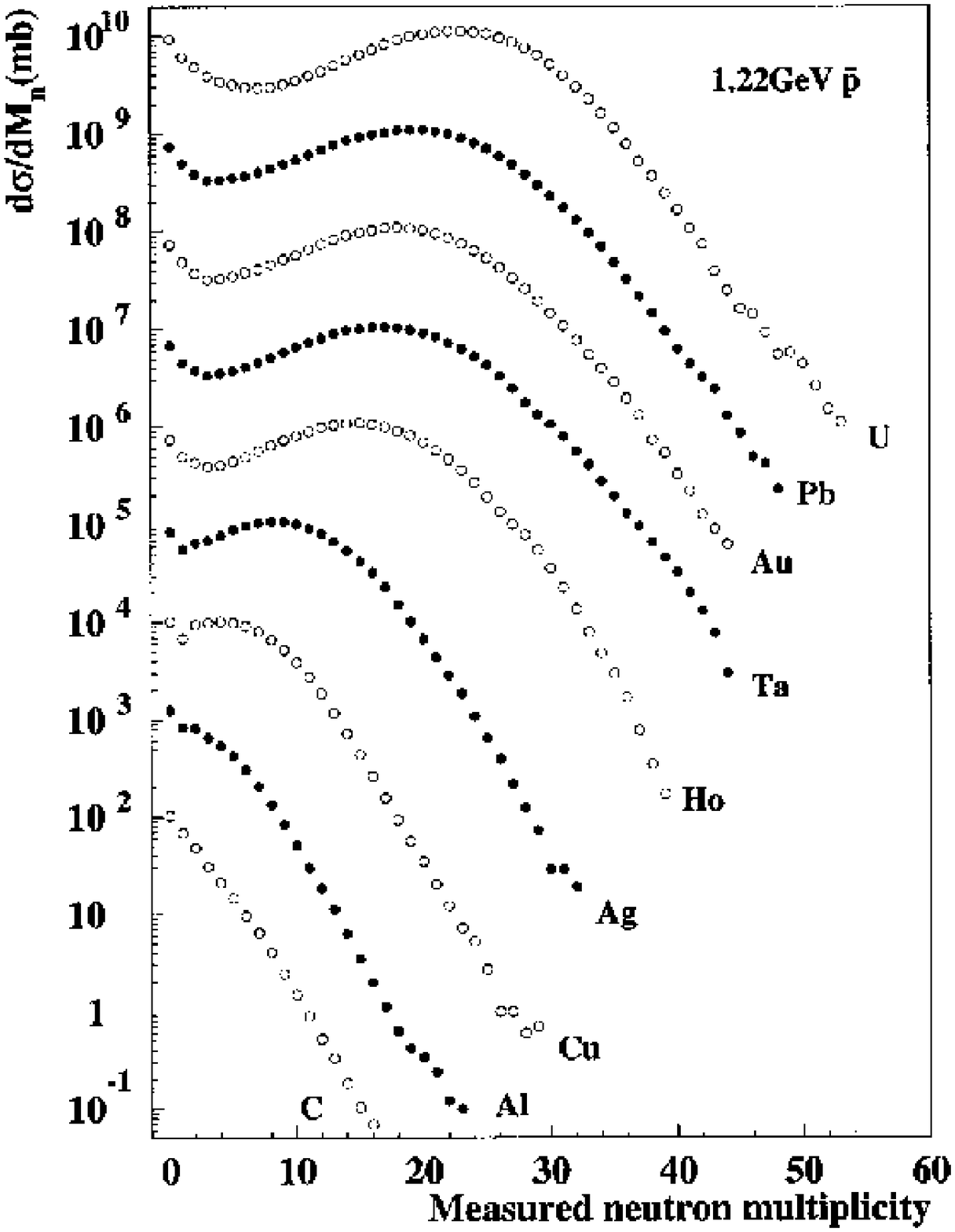}          
      {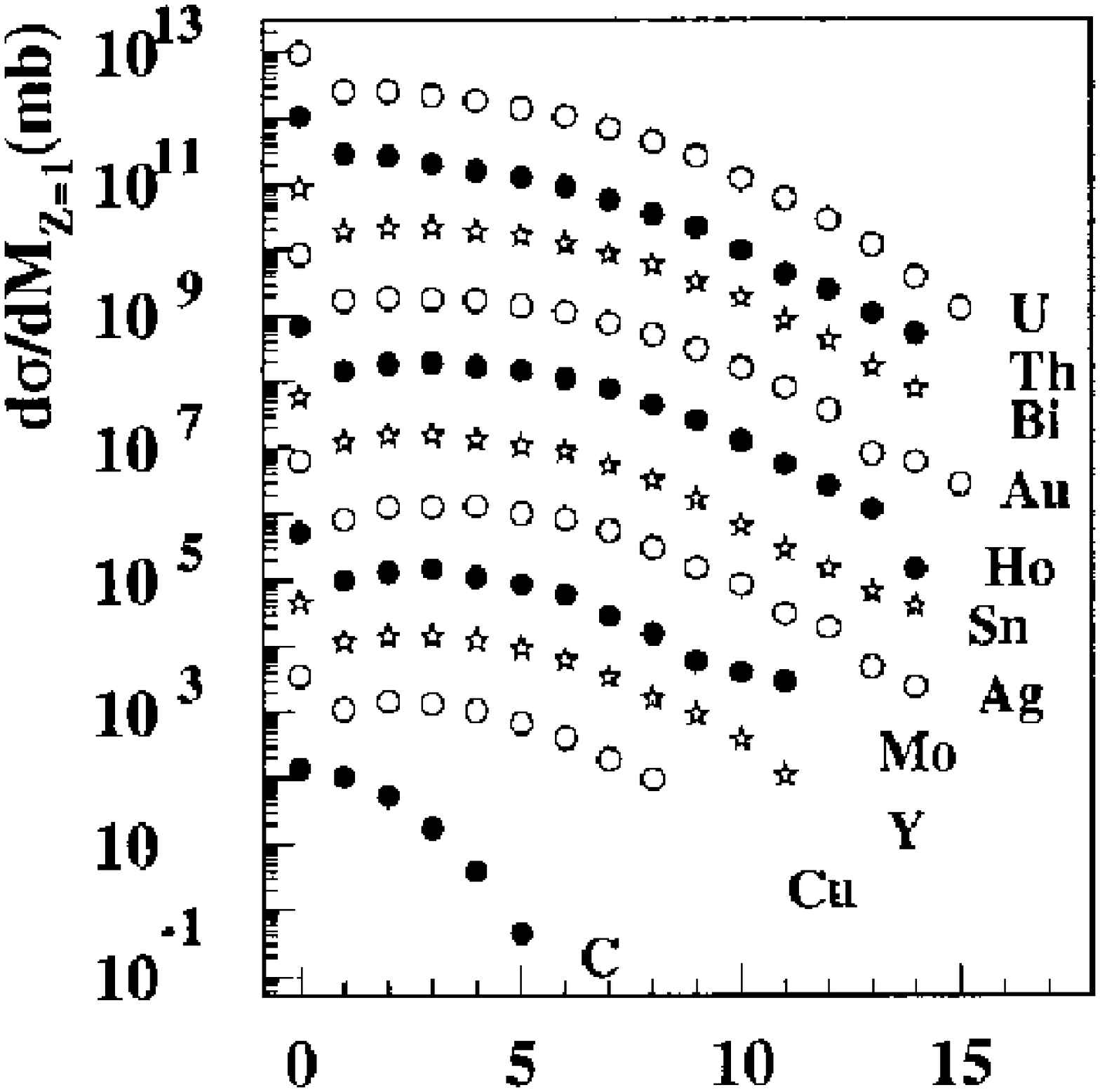}

 \caption{\label{fig:lott} a). Experimental inclusive neutron multiplicity
distributions for different targets. The distributions are multiplied by
successive factors of 10 for clarity. b) Experimental distributions of
multiplicities of $Z=1$ particles for different targets (multiplied by
successive factors of 10 for clarity). Both from Ref.~\cite{Lott:2001mf}.}

\end{figure}

Both neutrons and charged products were detected over a solid angle of
$4\pi$ by means of the neutron ball (spherical tank filled with liquid
scintillator, favoring evaporation-like neutrons) and the silicon ball
(mostly sensitive to intranuclear cascade products), respectively
\cite{Lott:2001mf}. Multiplicity distributions of neutrons and
protons are given, a good agreement with intranuclear cascade model is 
found (Fig.~\ref{fig:lott}).

%%%%%%%%%%
% Models
%%%%%%%%%%

\section{Models}

%%%%%%%%%%%%%%%%%%%%%%%
% Glauber-calculation

\subsection{Number of projectile collisions}

\paragraph{Glauber calculation.}
  
This is the conventional method used by all experiments when presenting
their data. At a given parameter $b$, the average number of hadron-nucleon
collisions is given by the integral

\begin{equation}
 \overline\nu(b) = \int\limits_{-\infty}^\infty \sigma \rho(z,b) \d z
\end{equation}

\noindent where $\sigma$ is the elementary hadron-nucleon cross-section.

For larger nuclei the nuclear density $\rho$ is often described by a Woods-Saxon
distribution in the form

\begin{equation}
 \rho(r) = \frac{\rho_0}{1+e^{\frac{r-R}{a}}}
\end{equation}

\noindent with the mean radius $R \approx 1.1 A^{1/3}$fm and the surface
diffuseness $a \approx 0.6$fm.

The probability of $\nu$ collisions for a given impact parameter is
assumed to follow a Poisson distribution. (In some variants binomial
distribution is used reflecting the finite number of nucleons, maximum
thickness, available in the nucleus.) After integration over the impact
parameter the probability distribution for number of collisions is
obtained,

\begin{align}
 \pi(\nu\vert b) &= \frac{{\overline\nu(b)}^\nu
                          e^{-\overline\nu(b)}}{\nu!} &
 \pi(\nu)        &\propto \int \pi(\nu\vert b) 2 \pi b \d b
\end{align}

\paragraph{Hijing model.}

The distribution $\pi(\nu)$ was also obtained by experiment BNL-E910
\cite{Chemakin:1999jd} with the Hijing generator which in this context is
equivalent to the Lund geometry code. This gives distributions similar to 
Glauber-calculation.

\paragraph{Criticism, alternative descriptions.}

\label{criticism}

The validity of assuming sequential projectile-nucleon collisions in the
nucleus is already questionable at the energy range ($E_{lab}$ from 1 GeV
to 1 TeV) studied in this note. The collision is so fast that the
projectile cannot reach its asymptotic final state, only after having left the
nucleus. This also renders the notion of constant hadron-nucleon cross
section unsure.

The projectile independence of the distribution of slow nucleons shows
that secondary particles acting as independent hadrons can give weight to
the first collision (e.g. \cite{Braune:1982jb}), in the extreme case
producing a single large cascade. Here the production of slow nucleons
would measure the {\it thickness} of the nucleus, thus the peripherality
or centrality of the collision and not the number of projectile
collisions.

The inelasticity of a hadronic interaction is the fraction of energy not
carried off by the fragments of the incoming particle, being available for
particle production. The multiple scattering model can be employed to
study hadron-nucleus interactions, allowing for changes of the
inelasticity of the projectile as it participates in more and more
collisions, when passing through the nucleus. The data are fitted using
known hadron-nucleon spectra and with the inelasticity as function of
collision number. It is found that second and higher interactions of the
excited projectile are relatively elastic: they altogether release only
20\%~(!) of the energy freed in the first collision. The results are
checked at SPS energies (e.g. \cite{Frichter:1997wh}), but there are
analyses using high-energy cosmic ray (exceeding $10^{14}$~eV) cascades
with the same conclusion \cite{Wilk:1998ra,Augusto:1999db}. These findings
are thus consistent with "low-energy" data and coincide with the
string-type model predictions.

%%%%%%%%%%%%%%%%%%%%%%%%%%%%%%%%%%%%%%%%
% Number distribution of gray nucleons

\subsection{Number distribution of gray nucleons}

The analysis is based on the knowledge of two distributions: the
distribution of number of projectile collisions in the nucleus $\pi(\nu)$
and the conditional probability $P(N_g\vert\nu)$ of emitting $N_g$ grey protons
in case of $\nu$ collisions.

The number distribution of grey particles is given by

\begin{equation}
 P(N_g) = \sum\limits_\nu P(N_g\vert \nu) \pi(\nu)
\end{equation}

If $N_g$ gray particles are emitted the average number of collisions is

\begin{equation}
 \overline{\nu}(N_g) = \frac{\sum\limits_\nu \nu P(N_g\vert\nu) \pi(\nu)}
                            {\sum\limits_\nu     P(N_g\vert\nu) \pi(\nu)}
 \label{eq:joint}
\end{equation}

The accuracy of $\overline\nu$ measurement is limited by the dispersion

\begin{equation}
 \sigma = \sqrt{\overline{\nu^2}(N_g) - \overline{\nu}^2(N_g)}
 \label{eq:accur}
\end{equation}

\paragraph{Geometric cascade model.}

According to this simple model, each collision of the incoming hadron in
the nucleus corresponds to the same distribution of grey prongs,
independent of the projectile \cite{Andersson:1978wg}. Different
encounters give independent contributions. Looking at experimental data
showing $N_g$ distributions at $\nu=1$, for a single encounter a
normalized geometric distribution is proposed

\begin{align}
 P(N_g\vert \nu=1) &= (1-X) X^{N_g}, & 
 X &= \frac{\overline{N_g}(\nu=1)}{1+\overline{N_g}(\nu=1)}
\end{align}

\noindent where the constant $X$ depends on target nucleus and can be
adjusted with a fit to data. From the assumption of independence, for
$\nu$ collisions

\begin{equation}
 P(N_g\vert \nu) = \binom{N_g+\nu-1}{\nu-1} (1-X)^\nu X^{N_g}
\end{equation}

\noindent which is a negative binomial distribution. This also means that

\begin{equation}
 \overline{N_g} \propto \overline\nu
\end{equation}

To be able to fit different data sets, it is assumed that the distribution
of number of grey prongs is the same whether the incident hadron is a pion
or a proton. Furthermore, model imposes no maximum on the number of
protons that can be emitted from a nucleus.

\paragraph{Intranuclear cascade model.}

The cascade can also described from a microscopic point of view
\cite{Hegab:1981iq,Hegab:1982ud}. The hadron-nucleus collision is viewed as a
sequence of three steps. The hadron passes through the nucleus on an
almost straight line and collides with $\nu$ target nucleons. The struck
primary nucleons recoil, travel mostly in forward direction, about half of
them detected as grey particles. Their elastic scatterings yield the
knocked-out second generation, all of them counted as grey particles. If
the primary nucleon is excited they can emit mesons which are too slow and
too light to produce grey particles, therefore they are left out of the
description. The secondary nucleons collide with other target nucleons to
produce the third generation and so on, but they contribute to black
tracks and will also be left out.

The following relation for the average number of grey protons can be 
derived:

\begin{equation}
 \overline{N_g} = \frac{Z}{A} \left[p\langle\nu\rangle + 
        \frac{1}{2} \frac{\sigma_{NN}}{\sigma_{hN}} 
        \overline{\nu(\nu-1)}\right]
\end{equation}

Here the first term represents the contribution of the primary protons
which will be detected as grey with probability $p$. Secondary greys are
produced with an effective cross-section $\sigma_{NN}$. If a primary
nucleon is produced in the $\mu$th collision it will scatter
$(\nu-\mu)\sigma_{NN}/\sigma{hN}$ times before it escapes. Summing up
these contributions leads to the second term. Using fits to data
\cite{Faessler:1979sn}, $p=0.55$ and $\sigma_{NN}=29$mb is obtained. The
two terms can be approximately combined to

\begin{equation}
 \overline{N_g} \approx \frac{1}{2} \frac{\sigma_{NN}}{\sigma_{hN}}
                             \frac{Z}{A} \overline{\nu^2} 
 \label{eq:hegab}
\end{equation}

Two fundamental problems are also discussed. While one is usually
interested in the number of inelastic collisions, elastic and inelastic
hadron-nucleon collisions both lead to recoiling primary nucleons.

After the first inelastic collision the hadron is excited and becomes a
different object: its cross-section for the collisions to come may be
different from the elementary hadron-nucleon cross-section. (One should
note that it is defined for asymptotic states only.) It can be assumed
that its cross-section changes continuously along the path: the hadron
expands. However from the observation of the grey particles one cannot
decide between this description and the usage of a constant effective
hadron-nucleon cross-section.

By neglecting the fluctuations in Eq.~(\ref{eq:hegab}) 

\begin{equation}
 \overline\nu(N_g) \propto \sqrt{N_g}
\end{equation}

The dispersion of this estimator is proportional to
${\overline\nu}^{-1/2}$, thus decreases with increasing number of
collisions.

\paragraph{Polynomial model.}

Despite fundamental differences both geometric and intranuclear cascade
models successfully reproduce distributions for a number of experiments. A
new model is proposed by BNL-E910 experiment \cite{Chemakin:1999jd} that
draws elements from both models. The main assumption is that for a given
nucleus the mean number of grey tracks $\overline{N_g}$ is a second order
polynomial of the number of primary interactions $\nu$

\begin{equation}
 \overline{N_g}(\nu) = c_0 + c_1\nu + c_2\nu^2
\end{equation}

It is further assumed that the distribution is binomial, each target
proton can be emitted with probability $p=\overline{N_g}(\nu)/Z$,

\begin{equation}
 P(N_g\vert\nu) = \binom{Z}{N_g} p^{N_g} (1-p)^{Z-N_g}
\end{equation}

The coefficients $c_i$ are derived from fits to the data, which show that 
the quadratic component is negligible and a simple proportionality holds. 
This is consistent with the geometric cascade model.

%%%%%%%%%%%%%%%%%%%%%%%%%%%%%%%%%%%%%%%%%
% Number distribution of black nucleons

\subsection{Number distribution of black nucleons}

Although multifragmentation, breakup of the nucleus is known since cosmic
rays physics, it became popular again when it was found to occur in
high-energy nuclear reactions. It is a phenomenon involving excitation
energies of the order of the nuclear binding energy. In recent years
observations pointed to its thermal nature: the remnant undergoes
equilibration before breakup. Hence thermodynamic or statistical
interpretations should be appropriate (e.g.
Refs.~\cite{Hauger:1998jg,Beaulieu:1999jv,Lefort:2001pa,Beaulieu:2001pc}).
The resulting remnant can be characterized by a few global variables, such
as charge, excitation energy and temperature, which may depend on the
centrality of the collision.

Due to the above mentioned nature of fragmentation the number distribution
of black nucleons can be well described in the form of a binomial distribution,

\begin{equation}
 P(n) = \binom{m}{n} p^n (1-p)^{m-n}
\end{equation}

\noindent where $m$ is the number of available nucleons, $p$ is the
elementary probability, which has the form $\exp(-B/T)$ with $B$ being the
emission barrier.

One can assume that the average black nucleon multiplicity -- hence the
target excitation -- depends linearly on the number of projectile
collisions. This can be deduced from the observation that $\overline{N_b}$
is proportional to $N_g$
(Refs.~\cite{Babecki:1973fe,Anzon:1975wr,Dabrowska:1993pj,Cherry:1994})
and from the success of the geometric cascade model which provides the
proportionality between $N_g$ and $\nu$. In other words, each collision
provides independent and identical production of prompt gray nucleons with
identical excitation of the nucleus, leading to the emission of black
nucleons.

The saturation of $\overline{N_b}$ for $N_g\le 7$ may simply be the
manifestation of the steep distribution of the number of collisions
$\pi(\nu$). (Anyway, the region above this is already rarely populated.)

\paragraph{Average values.}

Number distribution of slow nucleons are often presented but almost always
subject to biases: the fixed target experiments are mostly insensitive to
low energy black nucleons. This makes predictions for the ratio of gray to
black nucleons somewhat difficult. A low energy measurement of CERN-PS208
shows that the average multiplicity of grays is 7 and 16 for blacks for Pb
nucleus \cite{vonEgidy:2000kr}, yielding a black/gray ratio of 2.2.
According to Refs.~\cite{Babecki:1973fe,Anzon:1975wr} $2.6\pm0.1$ gray and
$5.0\pm0.2$ black protons have been observed for 200 GeV proton collisions
in nuclear emulsion, yielding a ratio of 1.9.

%%%%%%%%%%%%%%%%%%%%%%%%%%%%%%%%%%%%%%%%%%%%
% Modified Maxwell-Boltzmann distribution

\subsection{Momentum distribution of slow nucleons}

\paragraph{Modified Maxwell-Boltzmann distribution.}

\label{sec:modified_maxwell_boltzmann}

Assuming that the observed system is large enough to be considered
statistically, distributions may be parametrised in the form of the
Maxwell-Boltzmann distribution \footnote{The notion of statistical
emission of particles from moving frames is most elaborated at
Hagedorn's bootstrap model \cite{Hagedorn:1965st,Hagedorn:1968ua}.}: the
particles are emitted isotropically, but from a source moving with
velocity $\beta_\parallel$. The invariant cross-section can be written as

\begin{equation}
 E\frac{\d^3\sigma}{\d p^3} \propto \exp(-E_{kin}/E_0)
\end{equation}

\noindent where $E_{kin}$ is the kinetic energy, $E_0$ is the
characteristic energy per particle, both in the moving system. In the
laboratory frame this translates to

\begin{equation}
 \frac{\d N}{\d p \d\cos\theta \d\phi} \propto
  \frac{p^2}{\sqrt{p^2+m^2}}
  \exp\left[-\frac{\gamma(\sqrt{p^2+m^2} - \beta_\parallel p \cos\theta) - 
m}{E_0}\right]
 \label{eq:mb}
\end{equation}

\noindent where $m$ is the particle mass, $\theta$ is the laboratory angle
between the emitted particle and the initial projectile, the beam
direction. Experiments are mostly sensitive to some momentum range only,
thus for the angular distribution

\begin{align}
 \frac{\d N}{\d\cos\theta} &\propto
  \exp(\kappa \cos\theta) &
 \kappa &= \frac{\beta_\parallel \overline{p}}{E_0}
\end{align}

\begin{figure}
 \figs{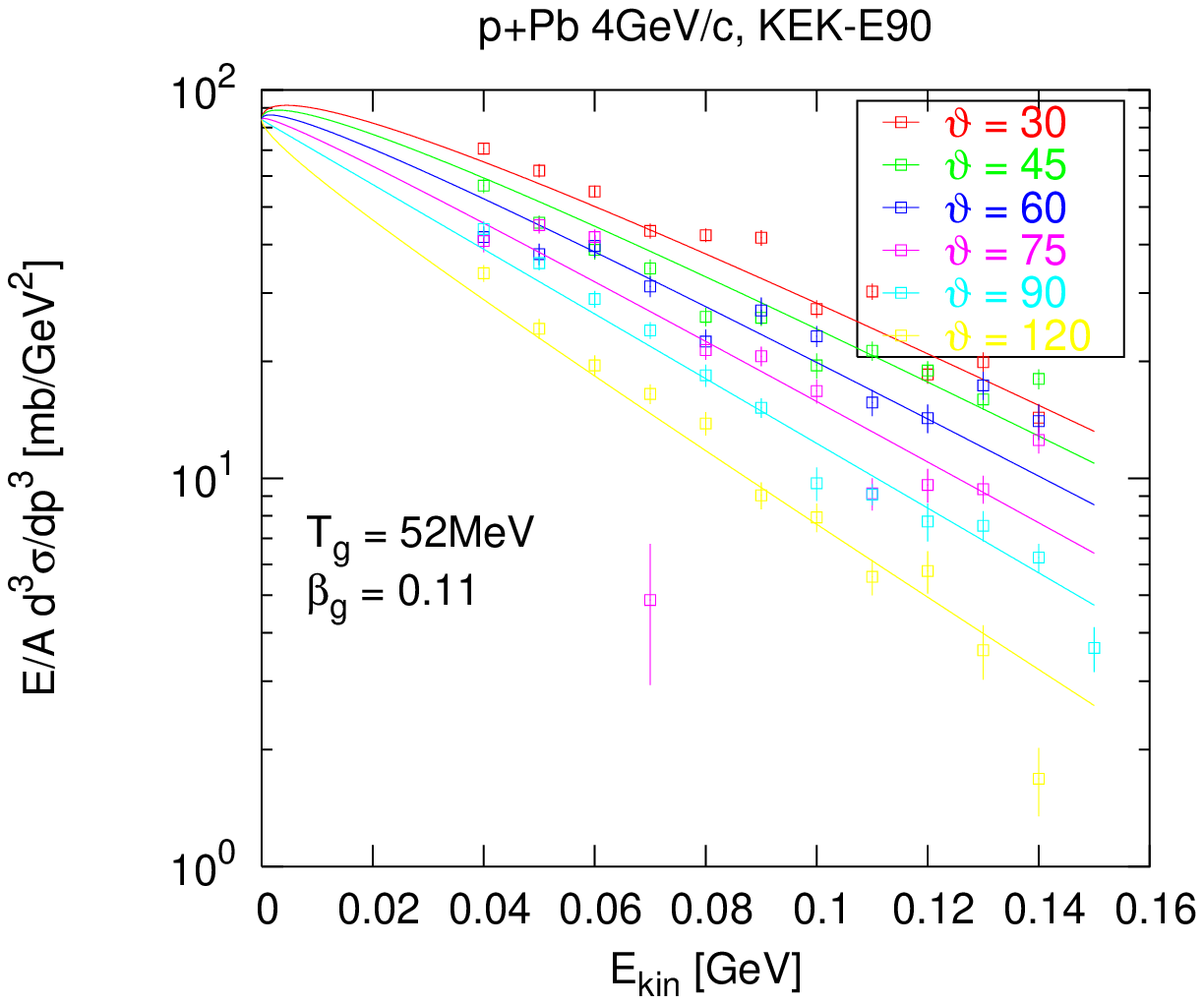}
      {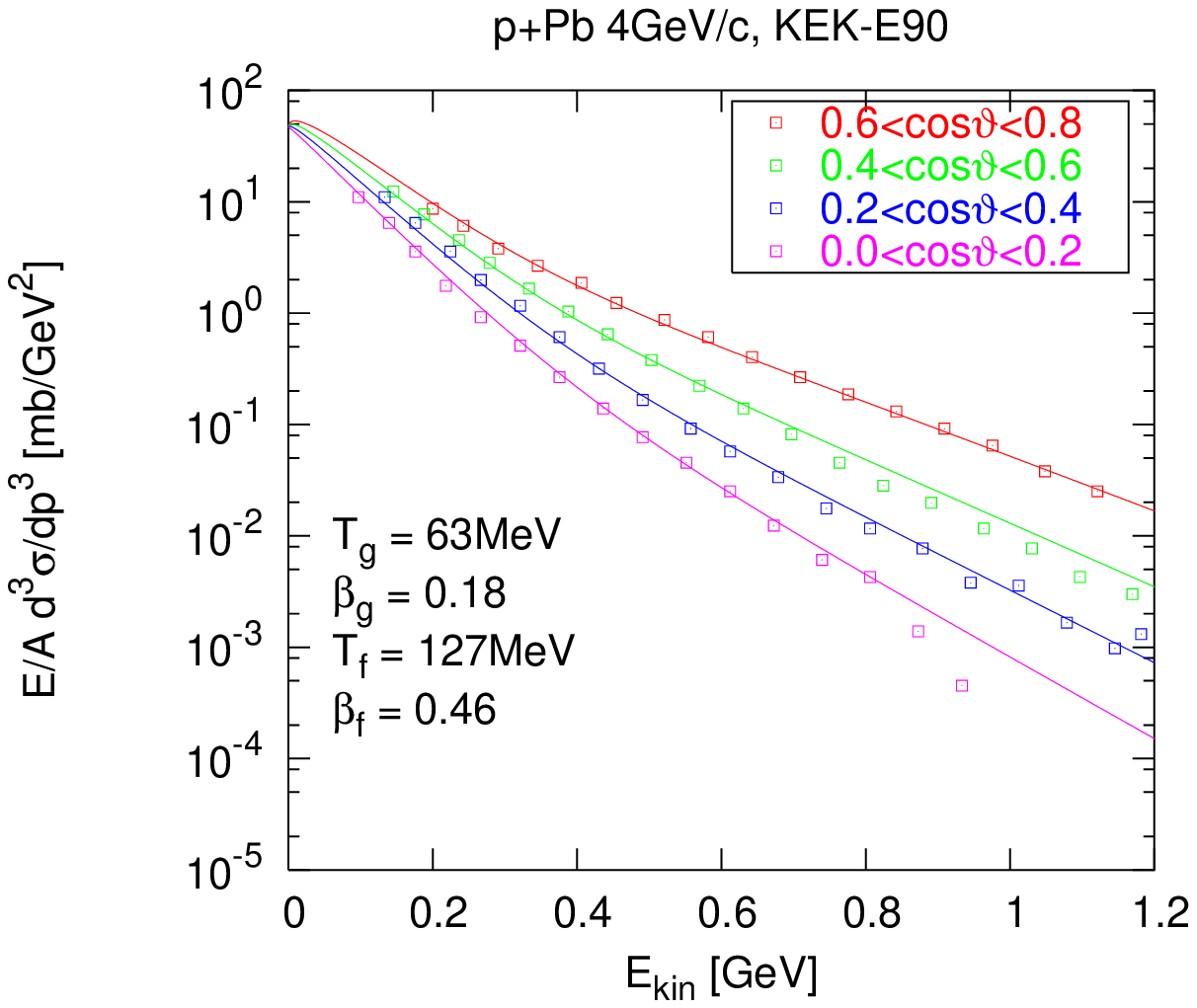}

 \figs{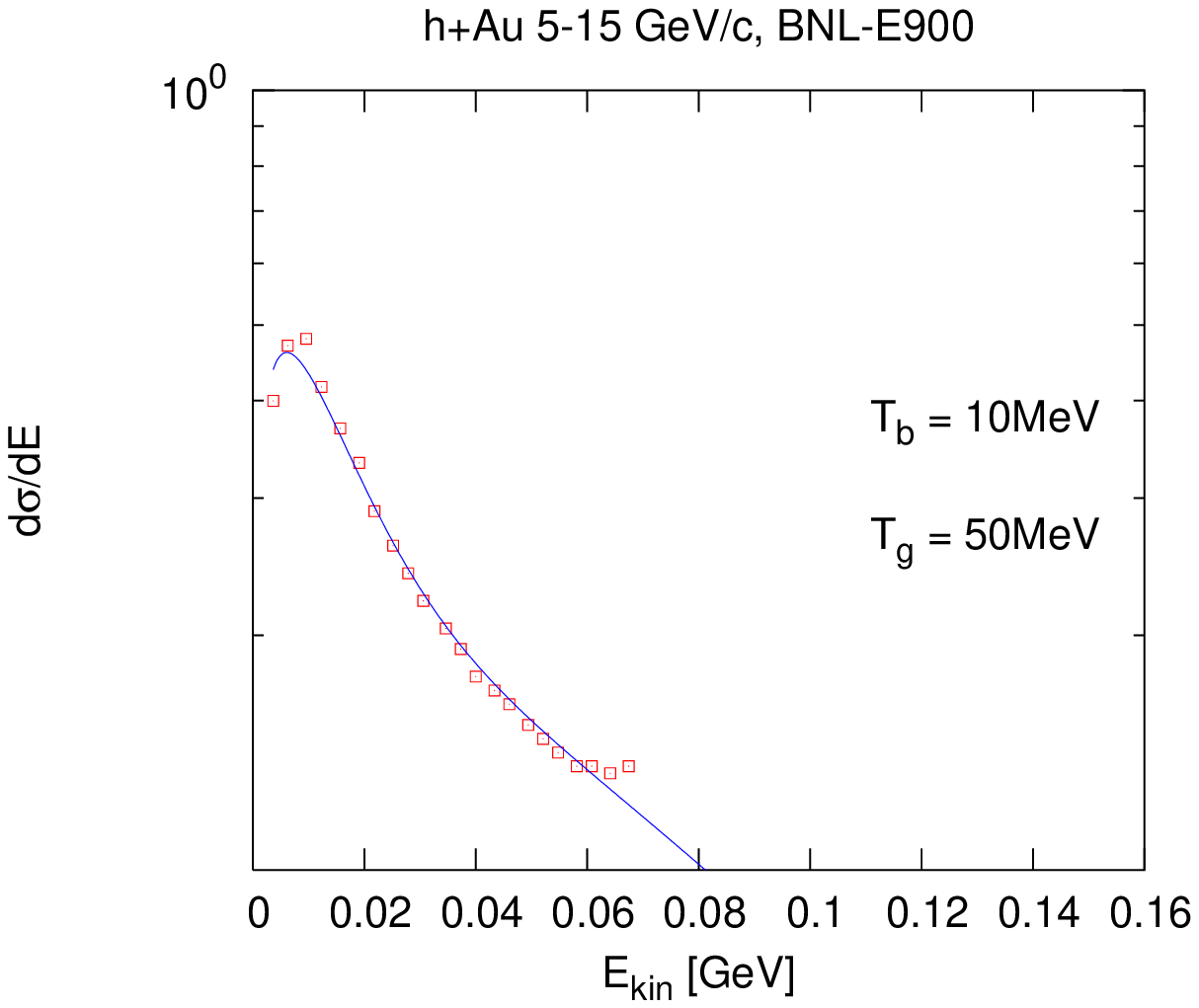}
      {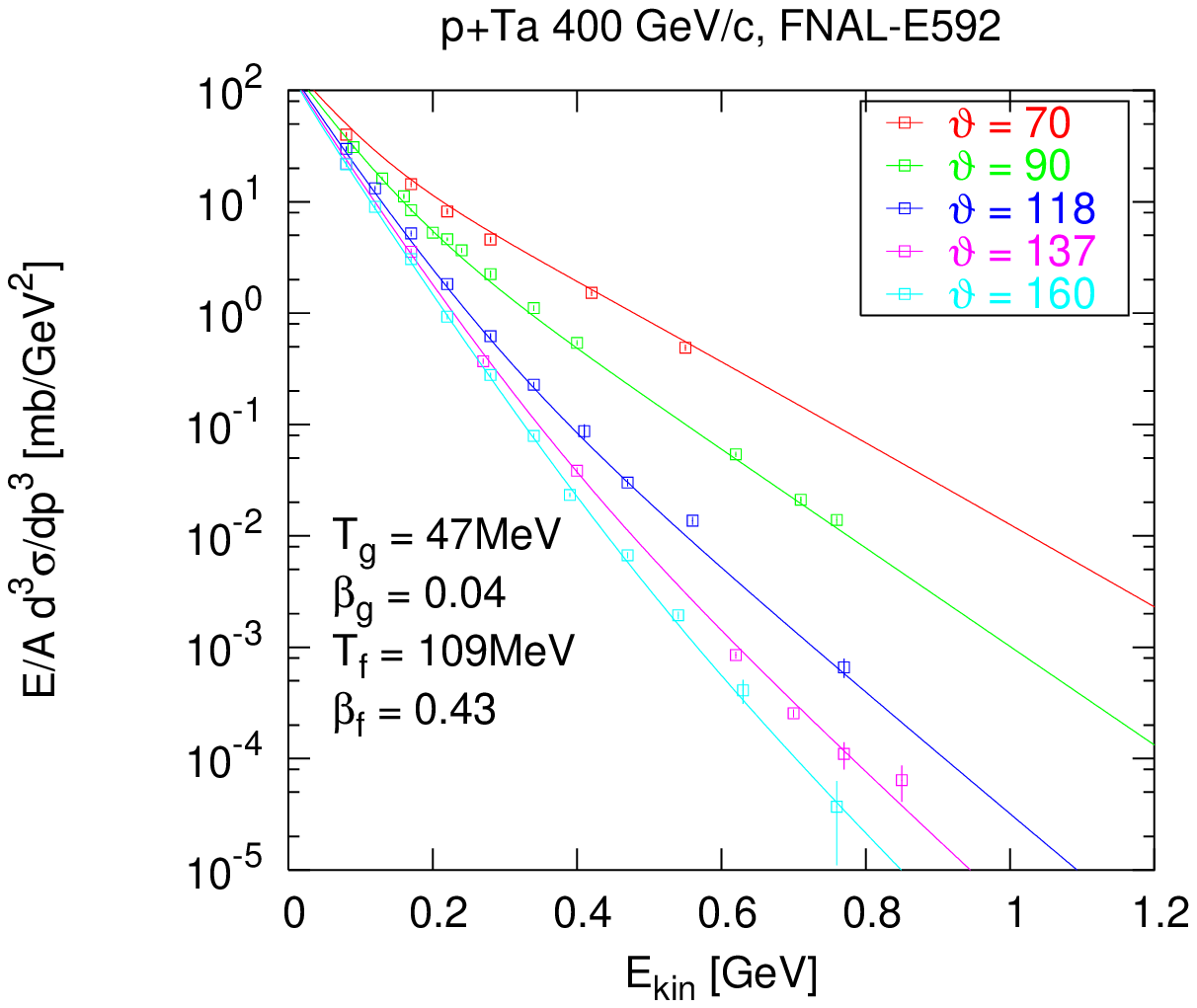}

 \caption{\label{fig:refits} Refits of some of the most complete angle
dependent proton cross-sections from collisions of hadrons and heavy nuclei.
Sum of Maxwell-Boltzmann distributions were used (b -- black, g -- gray, f
-- fast).}

\end{figure}

Some results of fits to slow nucleon angular and momentum distributions are
given in Table~\ref{tab:average}. Although the numbers are rather
scattered (Fig.~\ref{fig:refits}), an average from the more reliable ones
can be drawn.

\begin{table}
\begin{center}
 \begin{tabular}{lccccc}
 \hline\hline
 & $p_{lab}$ [GeV/$c$] & $T_b$ [MeV] & $T_g$ [MeV] & $\beta_g$ \\
 \hline
 CERN-PS208 & 1.22 & 4  & 40 & 0.05 \\
 LBL-E987   & 1    & 8  & 50 & \\
 KEK-90     & 3-4  &    & 50-60 & 0.1-0.2 \\
 BNL-E900   & 5-15 & 10 & 50 & $\le$ 0.01 \\
 FNAL-E592  & 400  &    & 45 & 0.04 \\
 \hline
 "average"  &      & 5 & 50 & 0.05 \\
 \hline\hline
 \end{tabular}
\end{center} 
 
 \caption{\label{tab:average} Result of Maxwellian fits to slow nucleon
angular and momentum distributions of several experiments on targets with
atomic number close to that of Pb. In the last line a kind of "average" is
given, assuming energy independence. (Big weight is given to the precise
measurements of CERN-PS208 and FNAL-E592.)}

\end{table}

\paragraph{Momentum correlations.} Correlation studies show that big part
of slow nucleon emission is due to pion absorption in nuclear matter, on
quasi two-nucleon-systems (Refs.~\cite{Yuldashev:1992aw,Schmidt:1992ws}).
This effect may be partially washed out by rescattering.

%%%%%%%%%%%%%%%%%%%%%%%%
% Experimental summary
%%%%%%%%%%%%%%%%%%%%%%%%

\section{Experimental summary}

Although the picture is rather mixed, the important features of the
accumulated data in the $E_{lab}$ range of 1 GeV -- 1 TeV can be
summarized.

The similarities at different energies suggest the idea that the emission
of slow particles is dictated by nuclear geometry, hence supporting the
hypothesis of limiting fragmentation. 

The data are traditionally analyzed in the framework of the Glauber-model
using Woods-Saxon density distribution -- providing the distribution of
number of projectile collisions in the nucleus $\pi(\nu)$ -- with a model
giving the probability distribution of number of slow nucleons
$P(N\vert\nu)$. For this latter the usage of the geometric model is most
popular.

Based on experimental results the average numbers of black and gray
nucleons, in a minimum bias hadron-nucleus collision, are

\begin{align}
 \overline{N_b} &\approx 0.08 A & \overline{N_g} &\approx 1.2 A^{1/3}
\end{align}

\noindent For centrality selected collisions on Pb target this amounts to

\begin{align}
 \overline{N_b} &\approx 4\nu & \overline{N_g} &\approx 2\nu
\end{align}

\noindent per collision.

Both black and gray components of slow nucleons can be described by
independent statistical emission from a moving frame. (It is a bit of
surprise for the prompt gray ones where such equilibrated behavior would
not be expected.) The number distributions follow binomial distributions.

\begin{align}
 P(N\vert\nu) &= \binom{M}{N} p^N (1-p)^{M-N} & p &= \overline{N}/M 
\end{align}

\noindent where $M$ is the maximum available protons/neutrons in the 
nucleus, $p$ is the emission probability.

The estimation of $\overline\nu$ when detecting $N$ slow nucleons is given
by the projection of the joint $P(N\vert\nu)\pi(\nu)$ distribution (see
Eq.~(\ref{eq:joint})), its dispersion can also be obtained (see
Eq.~(\ref{eq:accur})).

However, one should stress again that experimental observations do not
support the analysis described above (see Sec.~\ref{criticism}). It is
often concluded that the production of slow nucleons would measure the
{\it thickness} of the nucleus, thus the peripherality or centrality of
the collision and not the number of projectile collisions. Aware of this,
one should say that the estimator $\overline\nu(N)$ is a very much model
dependent quantity: $N$ itself could be also employed as a measure of
centrality \cite{Afanasiev:1999}.

The momentum distributions are of Maxwell-Boltzmann type with the
introduction of a source velocity (see Eq.~(\ref{eq:mb})). While the black
nucleons are emitted from a stationary source, the gray nucleons are from
a frame moving slowly in the direction of the beam particle (values given
for targets with atomic number close to that of Pb):

\begin{align}
 \beta_b &= 0     & \beta_g &= 0.05 \\
 T_b     &= 5 MeV & T_g     &= 50 MeV
\end{align}

\noindent The values of $\beta$ decrease with increasing $A$: for bigger
nuclei the intranuclear cascade can develop more, yielding a more
isotropic emission of particles.

Some experiments find another very fast components which can be identified
with central production or particle emission associated with the
projectile hadron.

%%%%%%%%%%%%%%%%%%%%%%%%%%%%%%%%%%%%%%%
% Predictions for colliders
%%%%%%%%%%%%%%%%%%%%%%%%%%%%%%%%%%%%%%%

\begin{figure}
 \figs{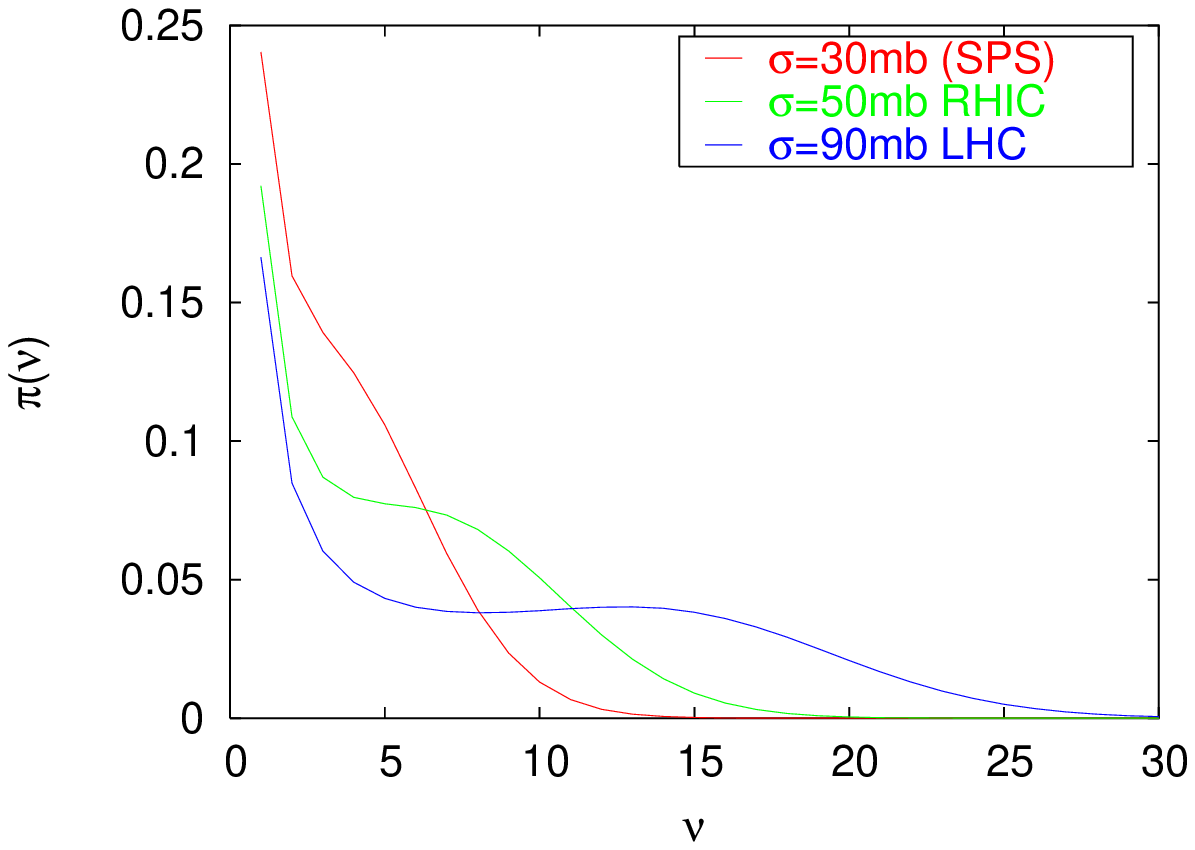}
      {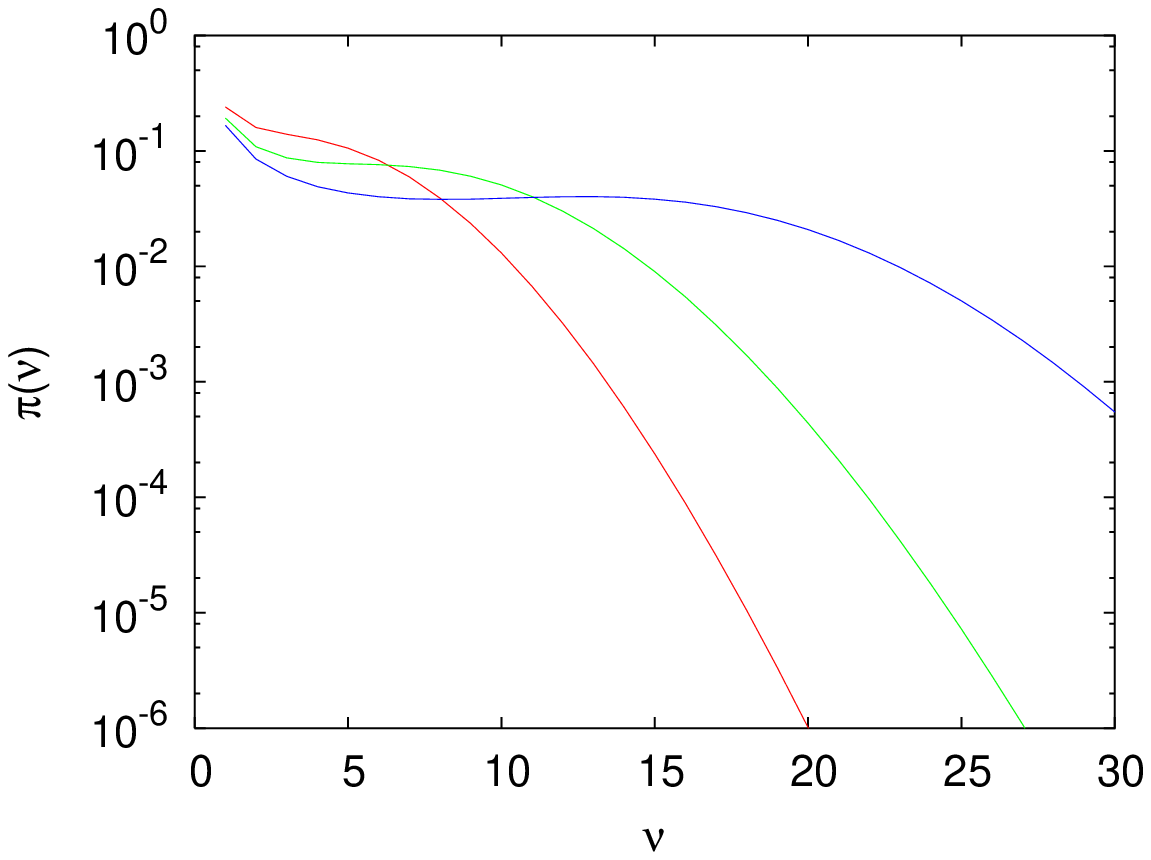}
 \figs{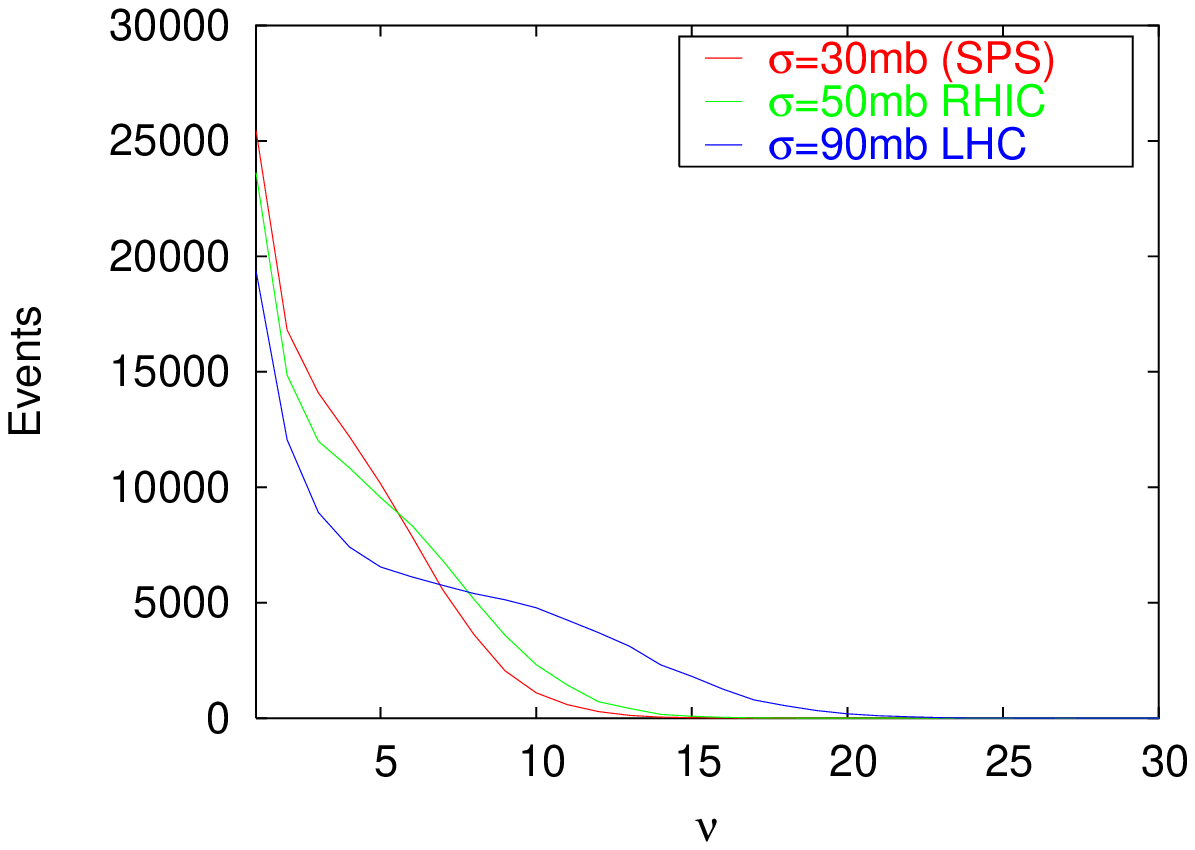}
      {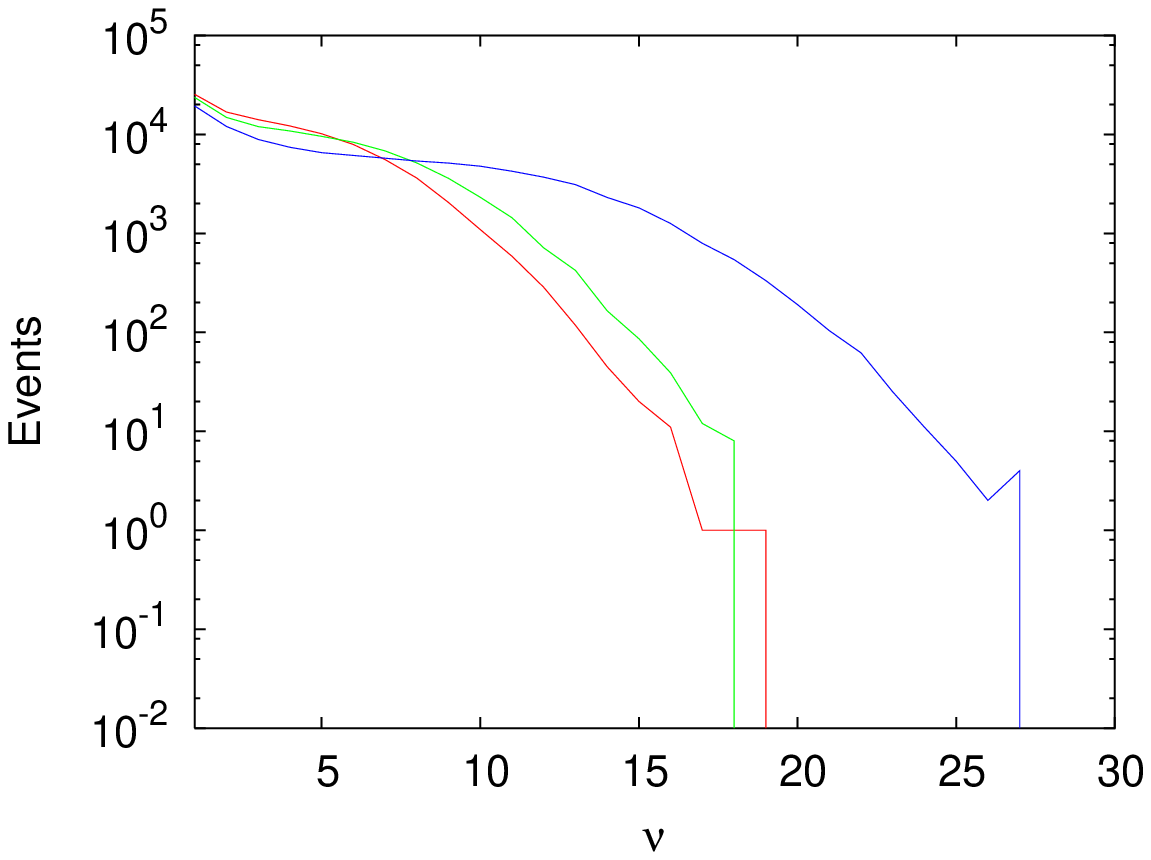}
 
 \caption{\label{fig:prediction1} Probability distribution of number of
projectile collisions in the nucleus at different energies, plotted in
both linear (left) and logarithmic (right) scale: upper) using the Glauber-model. lower)
using $10^5$ events from the Hijing generator, for the three accelerator
energies.}

\end{figure}

\section{Predictions for colliders}

The detection of slow nucleons is possible with the zero degree 
calorimeters (ZDCs), originally planned to measure centrality of
nucleus-nucleus collisions. If the proton ZDC is segmented, the black and
gray protons can be detected separately, because of the different   
velocities of frames they are emitted from. Otherwise -- and for the
neutrons -- only the sum of them, the number of "heavy" particles can be
measured.

Heavier nuclei are expected to go down undetected in the beam pipe. Light
nuclei can already make hits in the proton ZDC, mostly Z/A=2/3
($\mathrm{{}^3He}$) and partly Z/A=1/2 (d,$\mathrm{{}^4He}$) should be
considered. They can be produced by repeated coalescence of prompt gray
protons and neutrons. In this case the eventual detection of these nuclei
(mostly deuterons with some $\mathrm{{}^3He}$) means just the hit of a few
gray nucleons at the same time. However, most of the light nuclei are the
products of the late phase of the interaction: evaporation, fragmentation
of the nucleus (mostly $\mathrm{{}^4He}$ \cite{Hauger:1998jg}). They can
be treated as black nuclei, reporting on the excitation of the nucleus
just like black nucleons.

For planning a collider experiment the good knowledge of the momentum
distribution of the emitted particles is crucial, due to the presence of a
big Lorentz-boost. Data on hadron-nucleus interactions are exclusively
available from fixed target experiments. This is why one has to
concentrate on results where detailed angle dependent momentum/energy
spectra of slow nucleons have been obtained.

The features of the produced slow particles are highly energy independent,
they are very similar in the range of projectile energy from 1 GeV to 1
TeV. This applies to angular, momentum and number distributions as well.
However it is unclear whether this behavior still holds at collider
energies (recall the criticism detailed in Sec.~\ref{criticism}).

\begin{figure}
 \figs{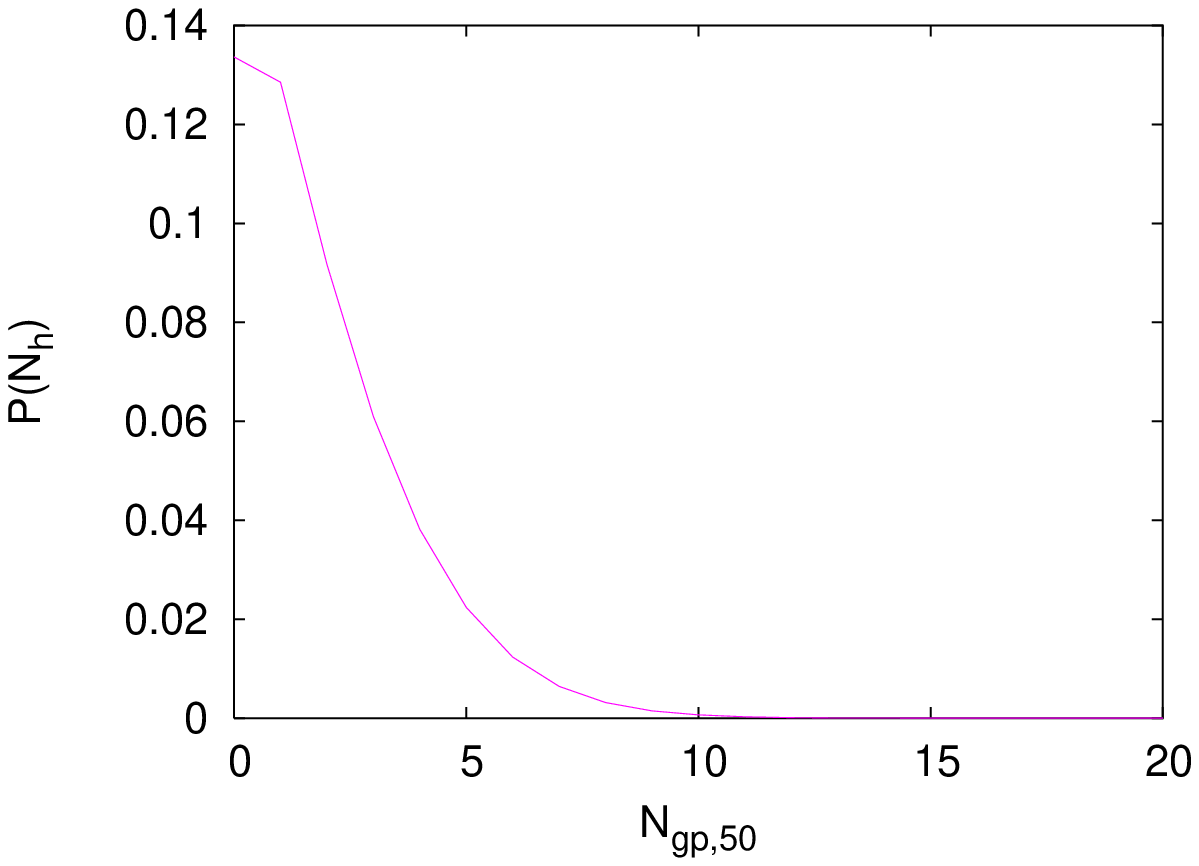}
      {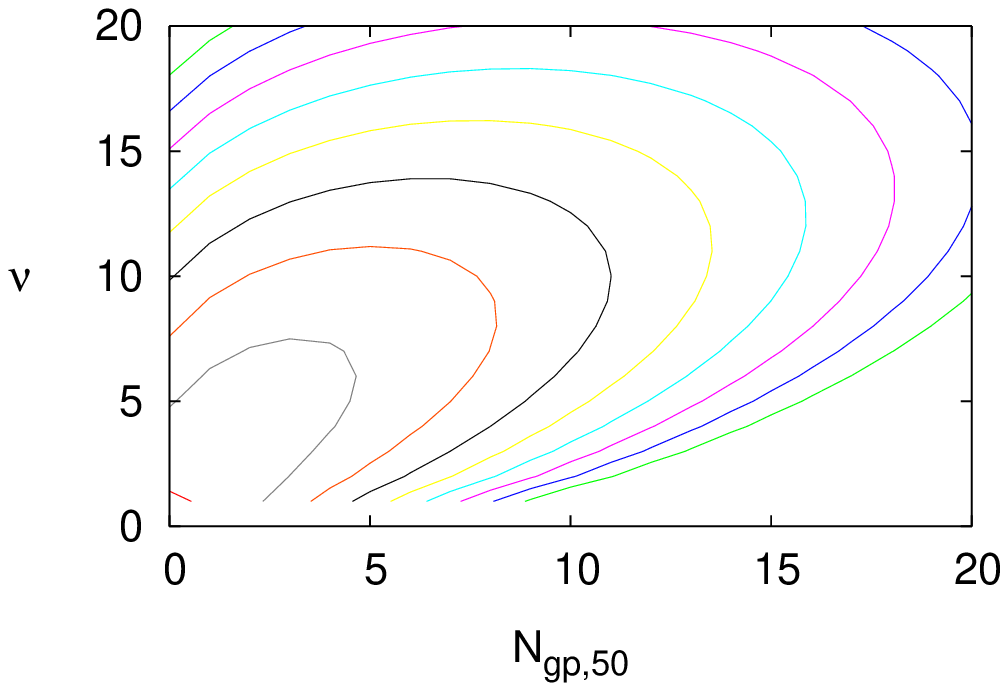}
 \centerline{\includegraphics[width=0.5\textwidth]{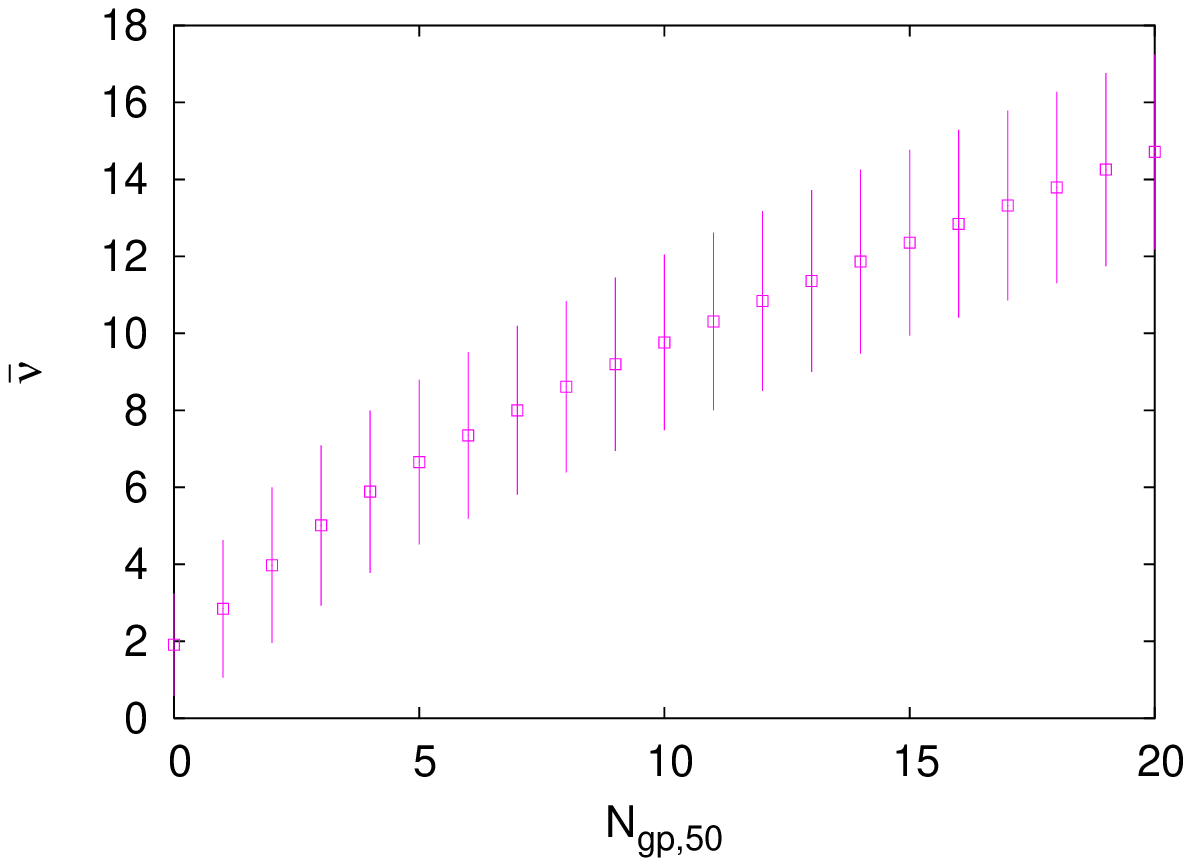}}

 \caption{\label{fig:prediction2} a) Number distribution of gray protons $N_{gp,50}$
assuming 50\% detection efficiency at SPS energy. b) Contour plots of
$N_{gp,50}$ vs $\nu$. Each contours are separated by factors of 10. c) The
$\overline\nu$ estimator: projection of the joint distribution with the
errorbars indicating the resolution.}

\end{figure}

\begin{figure}
 \figs{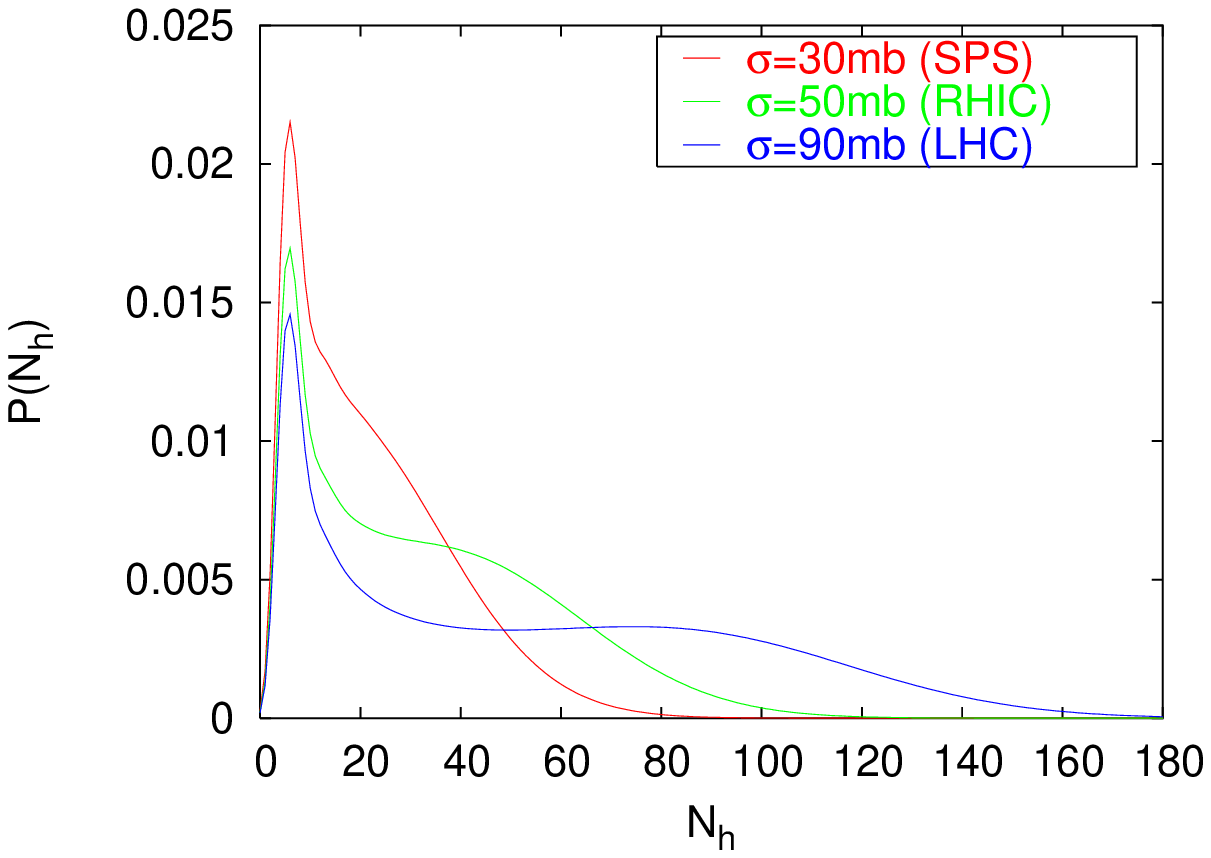}
      {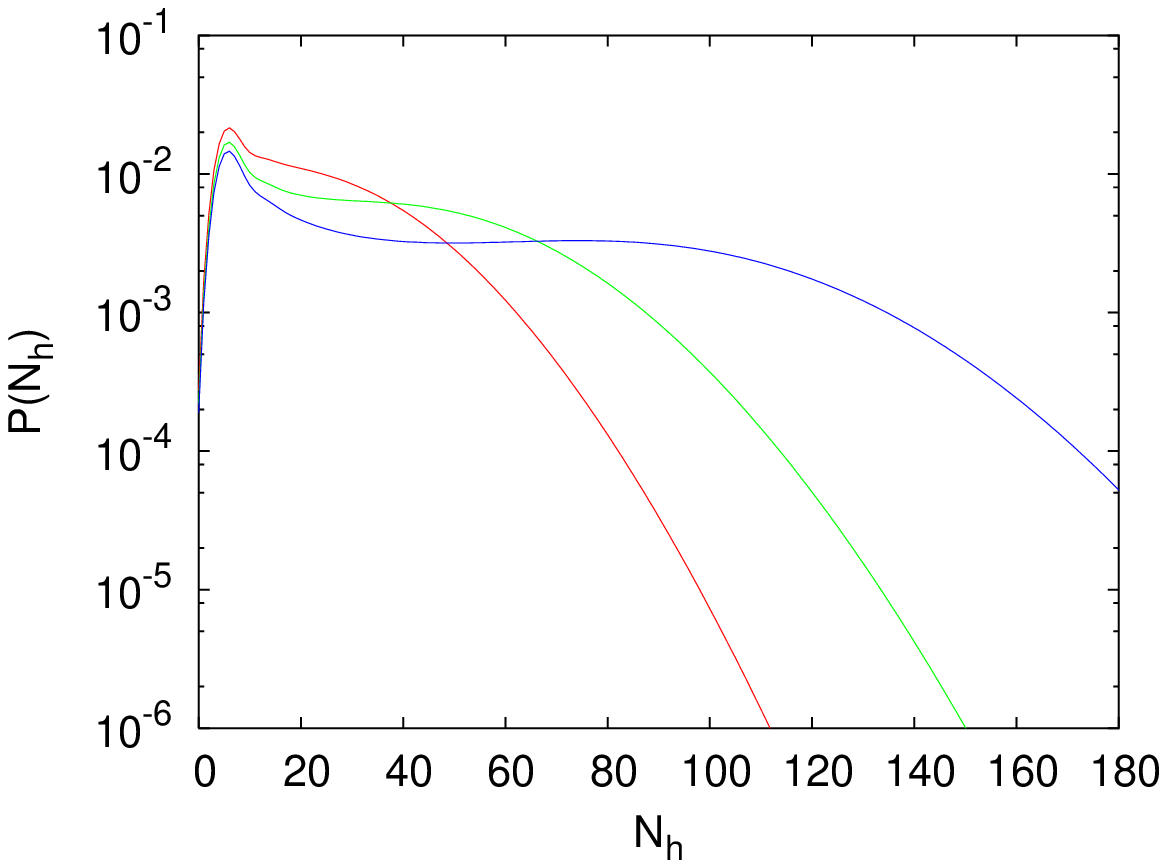}
 \figs{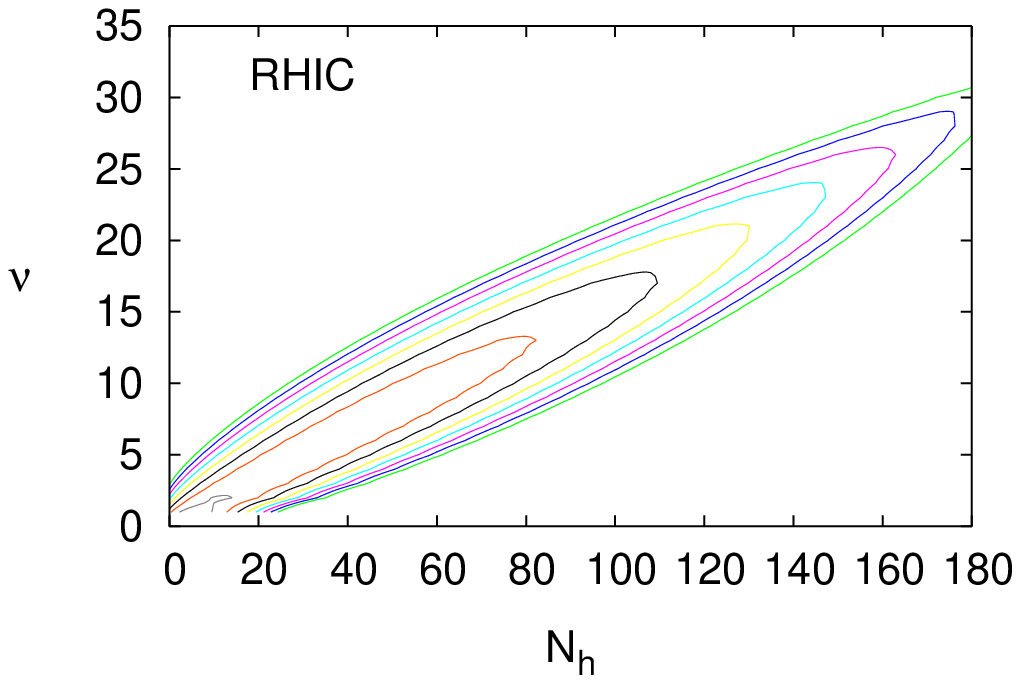}
      {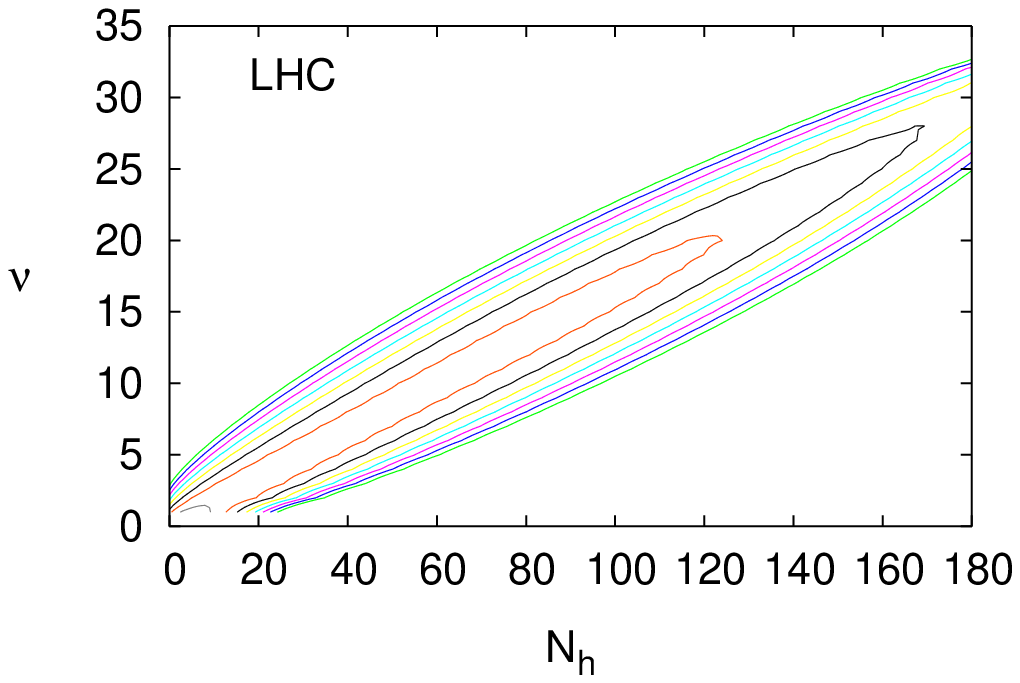}
 \figs{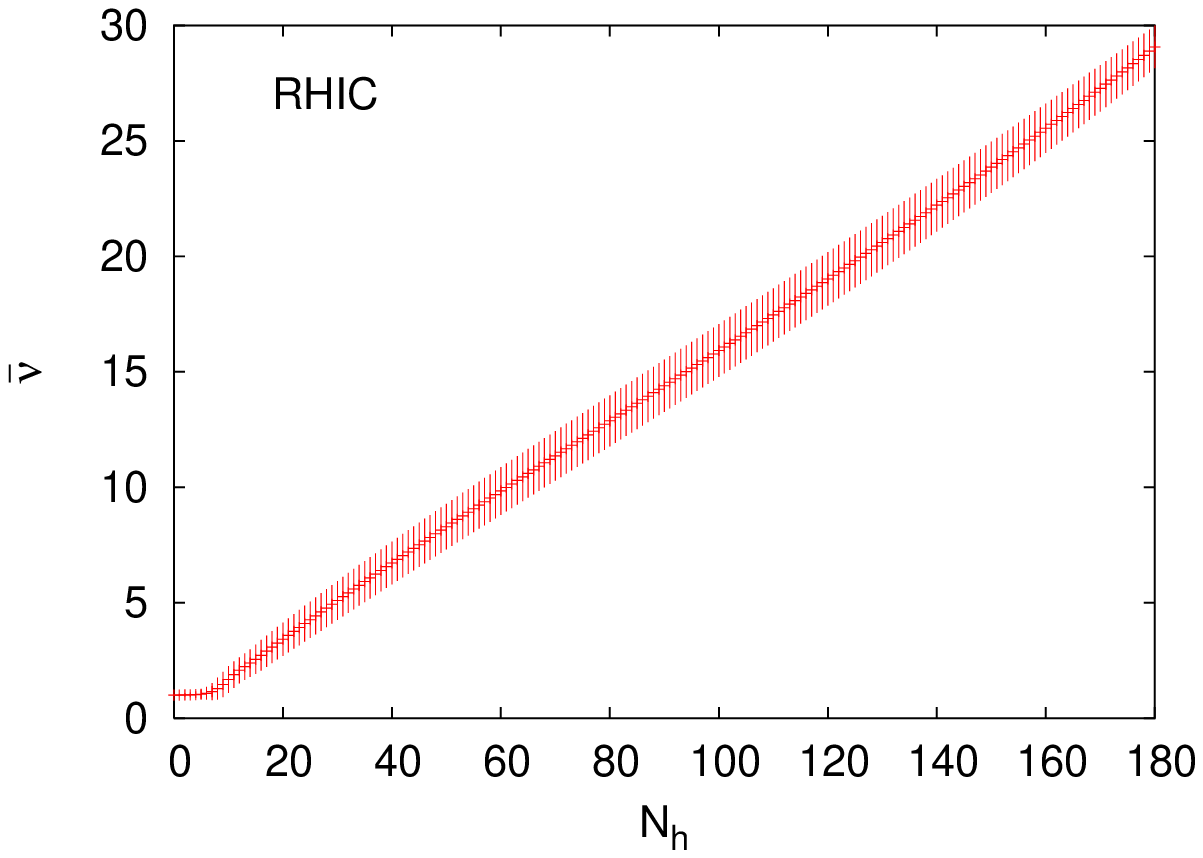}
      {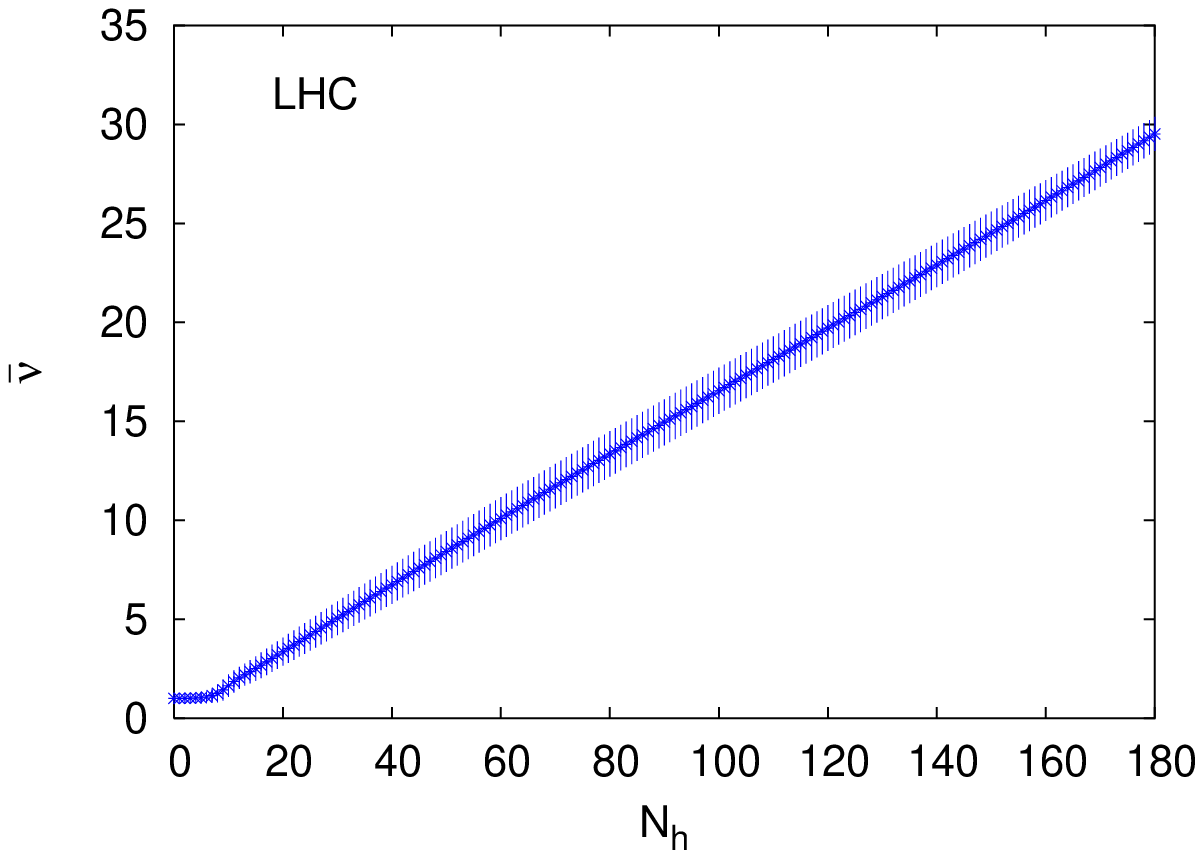}
 
\caption{\label{fig:prediction3} top) Number distribution of slow nucleons
$N_h$ assuming 100\% detection efficiency at different energies. middle)
Contour plots of $N_h$ vs $\nu$. Each contours are separated by factors of
10. bottom) The $\overline\nu$ estimator: projection of the joint
distribution with the errorbars indicating the resolution.}

\end{figure}

Additional complications come about the increase of elementary
hadron-nucleon cross-section, $\sigma_{NN}\approx$ 50 mb for RHIC, 90 mb
for LHC, compared to 30 mb at SPS. Does this mean that a hadron will have
three times more collisions, thus three times more black and gray
nucleons, at LHC than at SPS? Is it only geometry which counts,
irrespective of the cross-section? These are such open questions which can
only be answered after the first results from RHIC are at hand.

Finally, results of the conventional analysis at SPS, RHIC and LHC
energies are shown. The distribution of number of projectile collision is
given in Fig.~\ref{fig:prediction1}, using both Glauber and Hijing model
calculations. Experiments where only gray protons are detected with 50\%
efficiency are shown in Fig.~\ref{fig:prediction2} (typical for fixed target). Results where protons
and neutrons are detected with 100\% efficiency, both black and gray, are
given in Fig.~\ref{fig:prediction3} (typical for collider). Of course all the plots are highly
model dependent.

\section*{Acknowledgments}

The author wishes to thank to D. Barna and S. Hegyi for helpful
discussions and support. This work was supported by the Hungarian National
Science Foundation (F034707) and by the J\'anos Bolyai Research Grant.

\bibliography{review}
\bibliographystyle{apsmod}

\end{document}